\documentclass[%
reprint,
amsmath,amssymb,
aps,
]{revtex4-1}

\usepackage{graphicx}
\usepackage{dcolumn}
\usepackage{bm}
\usepackage{mathrsfs}

\DeclareMathOperator{\tr}{tr}

\begin{document}

    \preprint{APS/123-QED}

\title{\large\bf Strong Coulomb interactions in the problem of Majorana modes in a wire of the nontrivial topological class BDI}

\author{S.\,V.\, Aksenov}%
\email{asv86@iph.krasn.ru}
\author{A.\,O.\, Zlotnikov}
\email{zlotn@iph.krasn.ru}
\author{M.\,S.\, Shustin}%
\email{mshustin@yandex.ru}

\affiliation{%
    Kirensky Institute of Physics, Federal Research Center KSC SB RAS, 660036 Krasnoyarsk, Russia}

\date{\today}

\begin{abstract}

In this study, the problem of strong Coulomb interactions in topological superconducting wire is analyzed by means of the density-matrix-renormalization-group (DMRG) approach. To analyze properties of edge states in the BDI-class structure a quantity called Majorana polarization is used. From its dependence on wire length and an entanglement-spectrum degeneracy topological phase diagrams are obtained. The DMRG calculations for the Shubin-Vonsovsky-type model of the wire show the transformation of phases with Majorana single and double modes (MSMs and MDMs, respectively) under the increase of on- and inter-site correlations. In particular, we demonstrate different scenarios including the possibilities of both induction and suppression of the MSMs and MDMs. It is shown that in the strongly correlated regime the contributions of single-particle excitations to the Majorana-type states significantly decrease at low magnetic fields. Moreover, the $t-J^{*}-V$-model is derived allowing to study the effective interactions and improve the DMRG numerics. It is found out that in the limiting case of the effective Hamiltonian with infinitely strong on-site repulsion, $t$-model, the topological phases are destroyed. Finally, the ways to probe the MSMs and MDMs via the features of caloric functions are discussed.

    \begin{description}
        \item[PACS number(s)]
        71.10.Pm, 
        74.78.Na, 
    \end{description}
\end{abstract}

\maketitle


\section{\label{sec1}Introduction}

Starting from studies \cite{read-00,kitaev-01} the properties of topological superconductors (TSCs) attract considerable attention. Under open boundary conditions such systems host the zero-energy Majorana modes (MMs) which are edge states. MMs are being considered as perspective basic elements for topological quantum computing since they are stable against local perturbations and obey non-Abelian exchange statistics \cite{ivanov-01,kitaev-03}.

Among the systems proposed to observe MMs semiconducting wires, where SC pairing is induced by the proximity effect (in the following we will call them 'SC wires'), are ones of the most intensively investigated \cite{lutchyn-10,oreg-10}. To probe the appearance of MMs InAs and InSb wires characterized by strong spin-orbit interaction and large g-factor values are utilized \cite{mourik-12}. In turn, the SC pairing can be provided by a substrate or Al layer that partly covers the wire \cite{mourik-12,krogstrup-15}.

The advances in epitaxial growth of such low-dimensional hybrid nanostructures allowed to study ballistic transport in tunnel-spectroscopy experiments. The measurements revealed zero-bias conductance peak with the height of $2G_{0}$ ($G_{0}=e^2/h$ - conductance quantum) remaining in a wide range of magnetic fields and gate voltages \cite{zhang-18}. This feature can be accounted for resonant local Andreev reflection on MM when the TSC phase settles down. However, alternative explanations exist such as resonant transport mediated by the Andreev bound state predominantly localized in a normal quantum-dot region between the SC wire and metallic contact or by the one emerging in the SC wire where a spatially varying inhomogeneous potential is present \cite{cayao-15,liu-17b,moore-18a,reeg-18}. Thus, the ongoing disagreements leave a room for further investigations of MM features \cite{valkov-17b,valkov-17c}. 

The majority of studies analyzing the MM formation in the SC wires use the quadratic Hamiltonians without the consideration of Coulomb interactions between fermions. In this approach the classification of topological phases was obtained \cite{schnyder-08,kitaev-09} as well as the quantum-calculation algorithms based on MMs were developed \cite{kitaev-03,dassarma-15}. Simultaneously, it was supposed that the Coulomb correlations in the SC wires are weak. Hence, the stability of these issues against interaction effects is still insufficiently studied. However, it was shown recently that the InAs wires can be driven into the regime of strong electron-electron interactions as the system becomes more depleted due to gate electric field \cite{sato-19}. Thus, it emphasizes the necessity to revisit the problem of description of topological phases and MM detection in the regime of strong Coulomb correlations.

It is worth to note that taking into account strong electron interactions meets fundamental theoretical difficulties related to the significant renormalizations of effective interactions and change of topological classification \cite{fidkowski-10,wang-14}. Additionally, single-particle excitations possessing the features analogous to the ones of MMs in the system of non-interacting (or weakly 'mean-field' interacting) fermions have to be unambiguously defined \cite{katsura-15,kells-15,miao-17}.

To address these problems, we study the impact of strong Coulomb correlations on the 1D wire belonging to the BDI-Hamiltonian-symmetry class with extended s-wave SC pairing mainly by the density-matrix-renormalization-group (DMRG) tool \cite{white-92,white-93}. That is in contrast to the vast majority of works concerning 1D and quasi-1D systems with conventional s-wave superconductivity where the interaction factor was already investigated using DMRG \cite{stoudenmire-11,thomale-13,haim-14,gergs-16}. As it has been shown in \cite{wong-12} the BDI symmetry of the SC wire can be achieved by the presence of SC pairings between nearest neighbors. We consider a BDI-type model of the SC wire with both on- and inter-site s-wave pairings. Such a scenario is able to be achieved if the extended s-wave symmetry in the neighboring d-wave superconductor is realized due to: 1) effective on- and inter-site attraction between electrons leading to Cooper instability \cite{tanaka-95, martin-98}; 2) inner inhomogeneities \cite{belzig-98, hogan-99}.

Many theoretical studies concerning the interaction problem in the 1D and quasi-1D topological systems are based on bosonization and renormalization-group methods \cite{gangadharaiah-11,stoudenmire-11,lutchyn-11,klinovaja-14}. In this article we propose an alternative analytical approach utilizing the atomic representation and Hubbard operator formalism \cite{zaitsev-75,izyumov-92,kikoin-01,valkov-04} to treat the regime of strong electron correlations in the wire. According to \cite{izyumov-97} the use of unitary transformation method for the Shubin-Vonsovsky-type model allows to obtain the effective Hamiltonian of the $t-J^{*}-V$-model. In addition to the well-known superexchange \cite{anderson-87} in this work we also derive the effective interactions induced by the Rashba spin-orbit coupling. As a result, the DMRG algorithm is extended to this situation demonstrating higher computation speed and better convergence due to the exclusion of all two-particle states. The DMRG calculations permit to show the transformation of topological-phase diagrams in the strong-interaction regime unveiling both the MM survival and induction of these states by electron-electron correlations. The problem of MM normalization is discussed as well.     

It is important to stress that from a fundamental point of view here we suggest an approach based on the atomic representation to analyze the MM features in the 1D system with substantial influence of Coulomb interactions leading to formation of Hubbard fermions. To the best of our knowledge, it was not done earlier.

One of the possibilities to probe the MMs in the wire is to measure magneto- or electrocaloric effect (MCE or ECE, respectively) \cite{valkov-17}. The features of caloric functions in the 1D system with conventional s-wave SC pairing is related to the oscillations of ground-state fermionic parity which are caused by the hybridization of Majorana wave functions. As a result, the set of quantum transitions emerges as some parameter, e.g. magnetic field or chemical potential, is swept. In turn, one can observe them via extremely strong changes of the MCE and ECE which diverge exactly at the transition points. We show here that in the BDI-class wire in addition to this behavior, that points out to the TSC phase with one MM at each edge, the caloric functions can oscillate with finite amplitude indicating the appearance of TSC phase with two MMs and the preserved ground-state parity. Based on the DMRG data, we argue that these features are able to persist in the strongly correlated regime.

The article has been organized in five sections. The model Hamiltonian and methods used to analyze the TSC phases in the BDI-class system taking into account Coulomb correlations are described in Section 2. The numerical results obtained by the DMRG algorithm and effective model derivation are presented in Section 3. The possibility of MM experimental probe in the BDI-class wire utilizing the caloric effects is discussed in Section 4. Conclusions are given in Section 5.

\section{\label{sec2}Model and methods}

Let us consider a model of one-dimensional quantum wire with the Rashba spin-orbit coupling in external magnetic field. Carriers in the wire experience both on- and inter-site spin-singlet pairing due to the proximity effect with a bulk superconductor. The main goal of the work is to analyze the effects of electron-electron interactions, namely, the on-site Hubbard repulsion and Coulomb interaction within the first coordination sphere.
The tight-binding Hamiltonian of the described system reads
\begin{eqnarray}
\label{Ham_Fermi}
{\mathscr{H}} & = &  \sum_{f \sigma} \xi_{\sigma} a_{f \sigma}^{\dag} a_{f \sigma} - \frac{t}{2} \sum_{f\sigma} \left( a_{f \sigma}^{\dag} a_{f+1 \sigma} + a_{f+1 \sigma}^{\dag} a_{f \sigma} \right) -
\nonumber \\
& - & \frac{\alpha}{2} \sum_{f\sigma} \eta_{\sigma} \left( a_{f \sigma}^{\dag}a_{f+1 \bar{\sigma}} +  a_{f+1 \bar{\sigma}}^{\dag}a_{f \sigma}  \right) +
\nonumber \\
& + & \sum_f \biggl[\Delta a_{f \uparrow} a_{f \downarrow} + \Delta_1 \left( a_{f \uparrow} a_{f+1 \downarrow} + a_{f+1 \uparrow} a_{f \downarrow} \right) + h.c.\biggr]
\nonumber \\
& + &  U \sum_f n_{f \uparrow} n_{f \downarrow} + V \sum_f n_{f} n_{f+1},
\end{eqnarray}
where $\xi_{\sigma} = \xi - \eta_{\sigma}h$, $\xi=\varepsilon_0 - \mu$;
$\varepsilon_0$ is a bare electron energy, $\mu$ is a chemical potential,
and $h$ is the Zeeman splitting; $\eta_{\uparrow\left(\downarrow\right)}=\pm1$; parameters $t$ and $\alpha$ describe hoppings and the Rashba spin-orbit coupling between nearest neighbors, respectively; $\Delta,~\Delta_1$ are parameters of on- and inter-site SC pairing, respectively (which are supposed to be real throughout the article); $U$ is an intensity of on-site Coulomb interaction; $V$ is a parameter characterizing inter-site Coulomb interaction. Henceforth we consider all energy variables in units of $t$ and $t=1$. In general, the $t-U-V$-model \eqref{Ham_Fermi} is the Shubin-Vonsovsky-type one \cite{shubin-34,vonsovsky-79} supplemented by the Rashba spin-orbit couping and s-wave pairings. 

The Hamiltonian \eqref{Ham_Fermi} with $U=V=0$ and $\Delta = 0$
has been studied in \cite{wong-12}. In particular, it was
shown that in the strictly one-dimensional system along with the
electron-hole symmetry the additional time-reversal-like symmetry
takes place leading to the BDI-class of the corresponding Hamiltonian. It implies a richer picture of topological phases in comparison with the popular D-class wire. In particular, aside from the Majorana single modes (MSMs) the formation of two Majorana bound states, Majorana double modes (MDMs), localized at each edge of the open BDI-wire is possible. Note that since the subsequent calculations include $h=0$ and $h\neq0$ cases the MDM term means both the Majorana Kramers pairs \cite{schnyder-08,kitaev-09,qi-09} and modes with the lifted degeneracy, respectively.
\begin{figure}[htbp]
\includegraphics[width=0.5\textwidth]{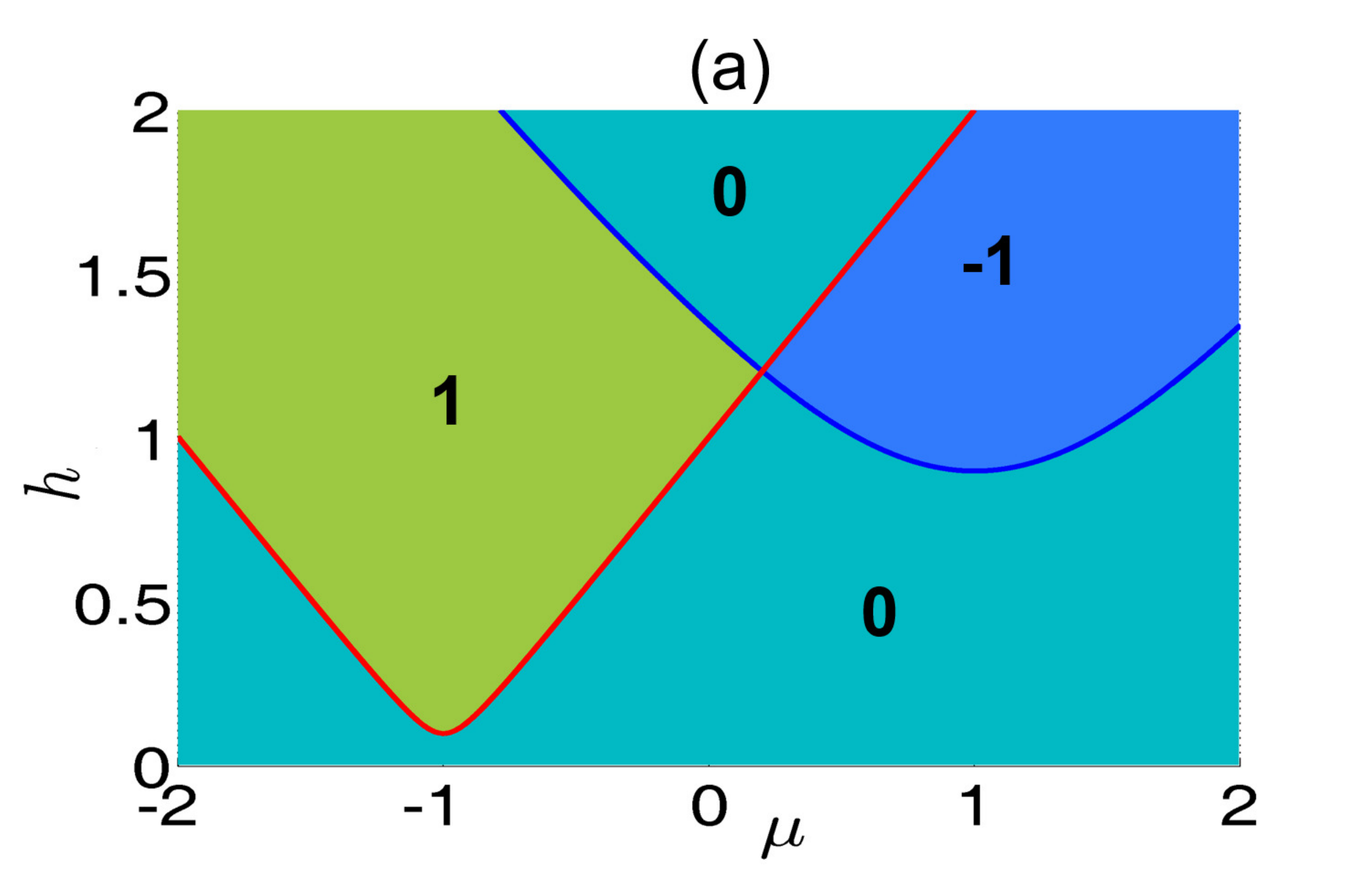}
\includegraphics[width=0.5\textwidth]{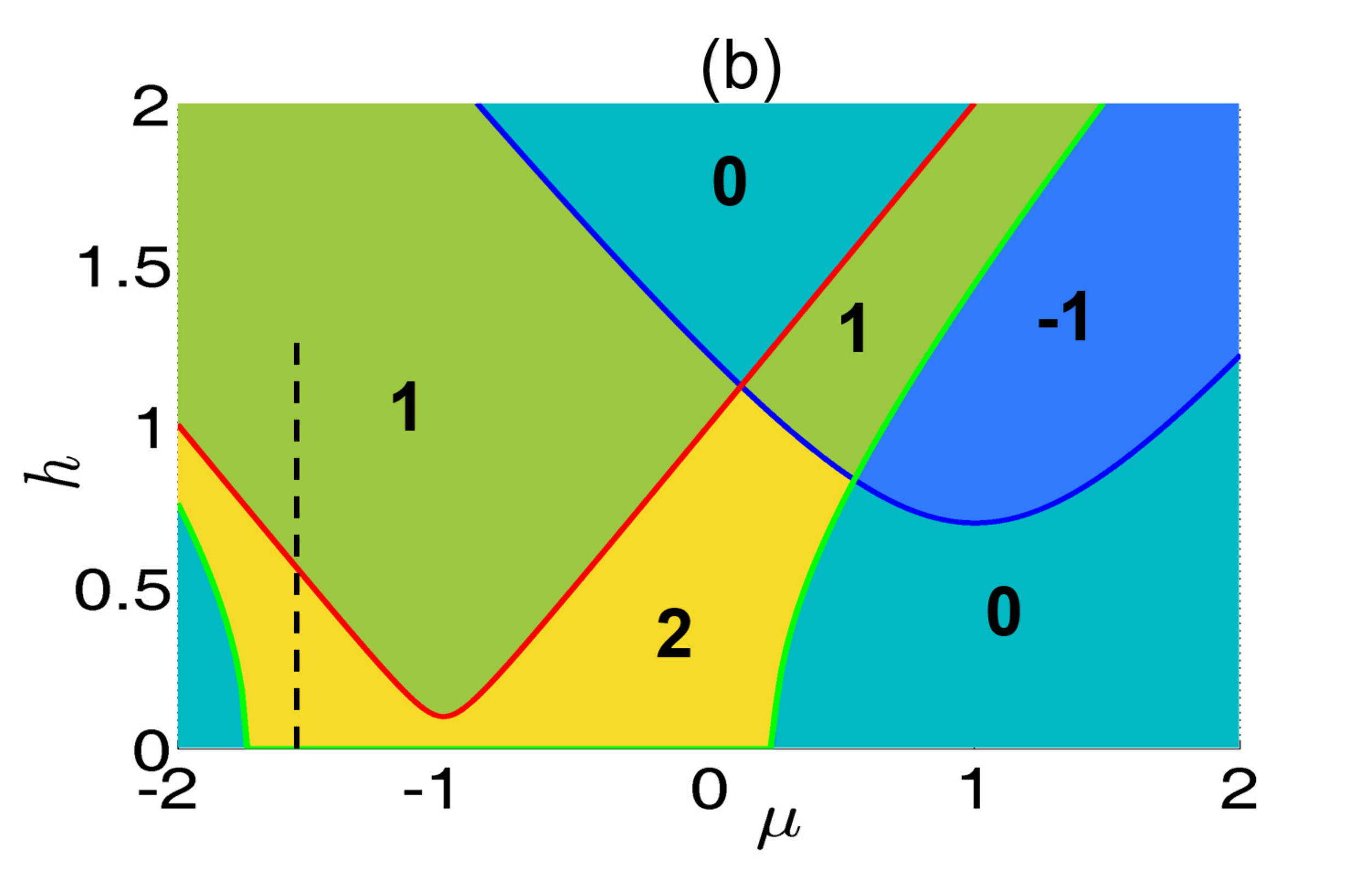}
	\caption{\label{N_BDI_map} Topological phase diagrams of non-interacting BDI-class wire. The invariant $N_{BDI}$ as function of chemical potential and magnetic field at $\Delta=-0.5$ (a) and $\Delta=-0.3$ (b). Other parameters are $\Delta_{1}=0.2$, $\alpha=1.5$. The numbers on the diagrams are values of $N_{BDI}$. The dashed line corresponds to the magnetic-field dependencies on Fig.~\ref{MCE_U0}.}
\end{figure}

These features remain valid in the case of nonzero on-site SC pairing, $\Delta\neq0$, and $ U = V = 0 $. To show it we generalize the analytical results obtained in \cite{wong-12}. Then assuming periodic boundary conditions the Hamiltonian \eqref{Ham_Fermi} in $k$-space has the following Bogoliubov-de-Gennes (BdG) form:
\begin{eqnarray}
\label{Ham_Fermi_BdG}
H(k) = \left( {\begin{array}{*{20}{c}}
	{A\left( k \right)}&{B\left( k \right)}\\
	{{B^ + }( - k)}&{ - {A^T}( - k)}
	\end{array}} \right),
\end{eqnarray}
here $A(k)= \xi_{k}\sigma_{0} + h\sigma_{z} + \alpha_{k}\sigma_{y}$,
$B(k)= i\Delta_{k} \sigma_{y}$, $\xi_{k} = -t\cos k - \mu$,
$\alpha_{k}=\alpha \sin k$, $\Delta_{k} = \Delta + 2\Delta_{1} \cos k$; $\sigma_{0}$ - the unity matrix; $\sigma_{x,y,z}$ - the Pauli matrices acting in spin space.
Under the unitary transformation,
$H(k) \to \tilde{H}(k) = U_{S}H(k)U_{S}^{+}$,
where $U_{S}=\left(\sigma_{0}\otimes\sigma_{x} - i \sigma_{y}\otimes\sigma_{x} \right)/\sqrt{2}$,
the BdG matrix transforms to
\begin{eqnarray}
\label{Ham_Fermi_BdG_Chiral}
\tilde{H}(k) &=& \left( \begin{array}{*{20}{c}}
	{0}&{ Q(k)}\\
		{Q^{T}(-k)}&{0}
		\end{array} \right),\nonumber\\
	Q(k) &=& \xi_{k}\sigma_{0}-h\sigma_{z}-\left(\alpha_{k}+i\Delta_{k}\right)\sigma_{y}.
	\end{eqnarray}
It allows us to introduce a topological (winding) number
	\begin{eqnarray}
	\label{N_BDI}
	N_{BDI}=\frac{-i}{\pi}\int_{k=0}^{k=\pi} \frac{d z(k)}{z(k)};~~~
	z(k) = \frac{\det\left(Q(k)\right)}{|\det\left(Q(k)\right)|}.
	\end{eqnarray}
The topological phase diagrams representing the invariant $N_{BDI}$ as a function of $\mu$ and $h$ are shown in Figure \ref{N_BDI_map}. Each region located between two boundary lines is characterized by an individual
value of the topological index indicated on the diagram, $N_{BDI}=0,~\pm 1,~\pm 2$.  The parametric region where $N_{BDI} = 0$
($N_{BDI} \neq 0$) corresponds to the topologically
trivial (nontrivial) phase. In the case of non-trivial topology the absolute value of $N_{BDI}$ points out the number of Majorana bound states in the open wire. The solid lines in Fig. \ref{N_BDI_map} are obtained from the condition of presence of gapless excitations in the bulk energy spectrum. These lines for the condition $|\Delta|<2|\Delta_{1}|$ are defined as:
\begin{eqnarray}
\label{h_gapless}
h_1^2 & = & (t+\mu)^2 + (\Delta+2\Delta_{1})^{2},
\nonumber \\
h_2^2 & = & (t-\mu)^2 + (\Delta-2\Delta_{1})^{2},
\nonumber \\
h_3^2 & = & \text{Re} \left[\left(\mu - \tilde{\Delta} \right)^2 - \alpha^2\sin^2\phi \right].
\end{eqnarray}
where $\tilde{\Delta} = \Delta/2\Delta_{1}$, $\phi = \arccos(\tilde{\Delta})$.
Then, non-zero values of $N_{BDI}$ occur under the conditions:
\begin{eqnarray}
\label{O_BDI}
N_{BDI} & = & \text{sgn}(\Delta_{1}) \, \, \, \text{if} \, \, \, |h_{2 \, (1)}|, |h_3| < |h| < |h_{1 \, (2)}|;
\nonumber \\
N_{BDI} & = & -\text{sgn}(\Delta_{1}) \, \, \, \text{if} \, \, \, |h_{2 \, (1)}| < |h| < |h_{1 \, (2)}|, |h_3|;
\nonumber \\
N_{BDI} & = & 2\cdot\text{sgn}(\Delta_{1}) \, \, \, \text{if} \, \, \, |h_3| < |h| < |h_1|, |h_2|.
\end{eqnarray}
Thus, there are the MDMs in the open wire if $|\Delta|<2|\Delta_{1}|$ due to the bulk-boundary correspondence (see Fig. \ref{N_BDI_map}b). When $|\Delta| \geq 2 |\Delta_{1}|$ the topological-phase transitions are determined only by the expressions
$h_{1 (2)}^2 = (t\pm\mu)^2 + (\Delta \pm 2\Delta_{1})^{2}$. In this case the topological invariant equals $N_{BDI} = -\text{sgn}(\Delta)$ if $|h_{1}| < |h| < |h_{2}|$ or $N_{BDI} = \text{sgn}(\Delta)$ if $|h_{2}| < |h| < |h_{1}|$. In the variables $\mu$ and $h$ these relations result in two parabola-shaped regions in Fig. \ref{N_BDI_map}a. For simplicity, we will use the notations "left parabola"$~$and "right parabola"$~$with regard to them.
	
It is necessary to stress two features. Firstly, as it follows from the analysis of \cite{wong-12}, the BDI-symmetry class is realized only for strictly one-dimensional system. If several electronic subbands are taken into account the system drops into the D-symmetry class. Secondly, the condition $|\Delta|<2|\Delta_{1}|$ for the MDM realization is equivalent to the presence of nodal points of SC order parameter $\Delta_{k}$. This inequality is violated in the interacting system. In the next section we will show that the similar single-particle excitations emerge even under $|\Delta|>2|\Delta_{1}|$ if $U,~V\neq 0$. 
	
Main approach used here to study the topological phases in 1D system under the presence of electron correlations is the DMRG method. This tool is powerful to investigate the properties of lowest-energy many-body states of 1D- and quasi-1D systems taking into account interactions \cite{peschel-99}. In the present work the DMRG tool is used to study both the initial Hamiltonian \eqref{Ham_Fermi} as well as effective models in strongly correlated regime. It turned out that consideration of the effective Hamiltonians made it possible to increase the speed and accuracy of DMRG calculations that additionally underlines the necessity to develop the atomic-representation description of interacting TSC structures.

For both initial and effective models, the many-body Hilbert space has been divided into sectors with an even and odd number of fermions. In each sector the quantum states $|\Psi^{ev(od)}_{1,2}\rangle$ and corresponding energy levels $E^{ev(od)}_{1,2}$ have been calculated. Since the many-body density matrix is also obtained one can investigate the behavior of different equilibrium averages.

For finite-size systems the $N_{BDI}$ index can only approximately describe the conditions of MM existence. A convenient approach to analyze these states in the wire with finite length is based on the Majorana polarization (MP) quantity introduced earlier in \cite{sedlmayr-15, sedlmayr-16} as a measure of the Majorana spatial distribution of a single-particle wave function. We consider a direct generalization of the MP for many-body states. Then the MP definition is
\begin{eqnarray}
\label{MP_def}
&~&
MP_{j}=\frac{\sum\limits_{f\sigma}{'}\left(w_{jf\sigma}^2 - z_{jf\sigma}^2\right)}
{\sum\limits_{f\sigma}{'}\left(w_{jf\sigma}^2 + z_{jf\sigma}^2\right)},~j=1,2,\\
&~&w_{jf\sigma}=\langle \Psi_{j} | \left( a_{f\sigma} + a^{+}_{f\sigma} \right) | \Psi_{0} \rangle,\nonumber\\
&~&z_{jf\sigma}=\langle \Psi_{j} | \left( a_{f\sigma} - a^{+}_{f\sigma} \right) | \Psi_{0} \rangle,\label{wz}
\end{eqnarray}
where $|\Psi_{0} \rangle=|\Psi^{ev/od}_{1}\rangle$ is a ground even- or odd-parity state; $|\Psi_{j} \rangle=|\Psi^{od/ev}_{j}\rangle$ are first ($j=1$) and second ($j=2$) excited many-body states from the dual-parity sector of the Hilbert space. The apostrophe indicates that the summation over $l$ is carried out for the half of wire sites. It is seen that the value of $MP_{j}$ determines the overlap between Majorana-type coefficients $w_{jf\sigma}$ and $z_{jf\sigma}$.
				
In the absence of Coulomb interactions ($U=V=0$) the definition \eqref{MP_def} coincides with the one introduced in \cite{sedlmayr-15,sedlmayr-16}. As we mentioned above $MP_{j}$ just partly agrees with $N_{BDI}$ in the finite-length wire. However, a clear correspondence is obtained when the number of sites $N \to \infty$: $MP_{j} \to 0$ for a bulk excitation and $MP_{j} \to 1$ in case of the MM. Therefore, it is assumed that if $N \to \infty$ the MSMs appear with $MP = MP_1 + MP_2 = 1$ in the topological phases characterized by $N_{BDI} = \pm 1$. The MDMs are realized having $MP = 2$ in the phase with $N_{BDI} = 2$. Finally, $MP = 0$ is in the trivial phase where $N_{BDI} = 0$. In the wire with finite $N$ the spatial distribution of particular excitation changes continuously from bulk- to edge-like, especially in the vicinity of topological phase boundaries. Hence, $MP$ also varies between $0$ and $1$. For simplicity, we will assume that the edge-like excitation is dominant and MMs are formed if at least $MP_{j} > 0.8$.

For interacting systems the topological classification can be carried out by analyzing the entanglement spectrum of the reduced density matrix, $\mathcal{H}=-\ln\tr_{N/2} \rho$, where $\rho$ - many-body density matrix \cite{turner-11,stoudenmire-11,gergs-16}. For the system under consideration an entanglement spectrum degeneracy, $d=d\left[\mathcal{H}\right]$, can be one-, two- and fourfold. In the limit $N \to \infty$ the correspondence between $MP$ and $d$ is as follows: $MP = 0 \leftrightarrow d = 1$; $MP = 1 \leftrightarrow d = 2$; $MP = 2 \leftrightarrow d = 4$. These relations are relevant also in the strongly correlated regime. As before $MP$ has no topological origin. Nevertheless, it allows to identify the MMs as well as their hybridization, therefore, describing edge effects. The boundaries of topological phases obtained using the invariant $d$ and the wire-length dependence of $MP$ in the presence of Coulomb interactions will be presented in Section \ref{secE}.
	
The second approach utilized in the article is the generalized mean-field description (GMF). Technically, it is based on the Bogoliubov transformation of four-fermion operators with consequent renormalization of the operator terms \cite{kukharenko-75,valkov-91}. In such an approach the equations for the transformation
coefficients become nonlinear since the effective quadratic form of Hamiltonian depends on the transformation parameters. This approach was used to study the D-class wires in \cite{valkov-17,valkov-19a}. Comparison of the GMF with the exact-diagonalization (DMRG) results for the short (long) BDI wires shows qualitative agreement at $U \lesssim 1$, $V \lesssim 0.5$ and considerable deviation in the strongly correlated regime. The GMF details for the BDI system \eqref{Ham_Fermi} are presented in Appendix \ref{apxA}. 
\begin{figure*}[htbp]
	\includegraphics[width=0.45\textwidth]{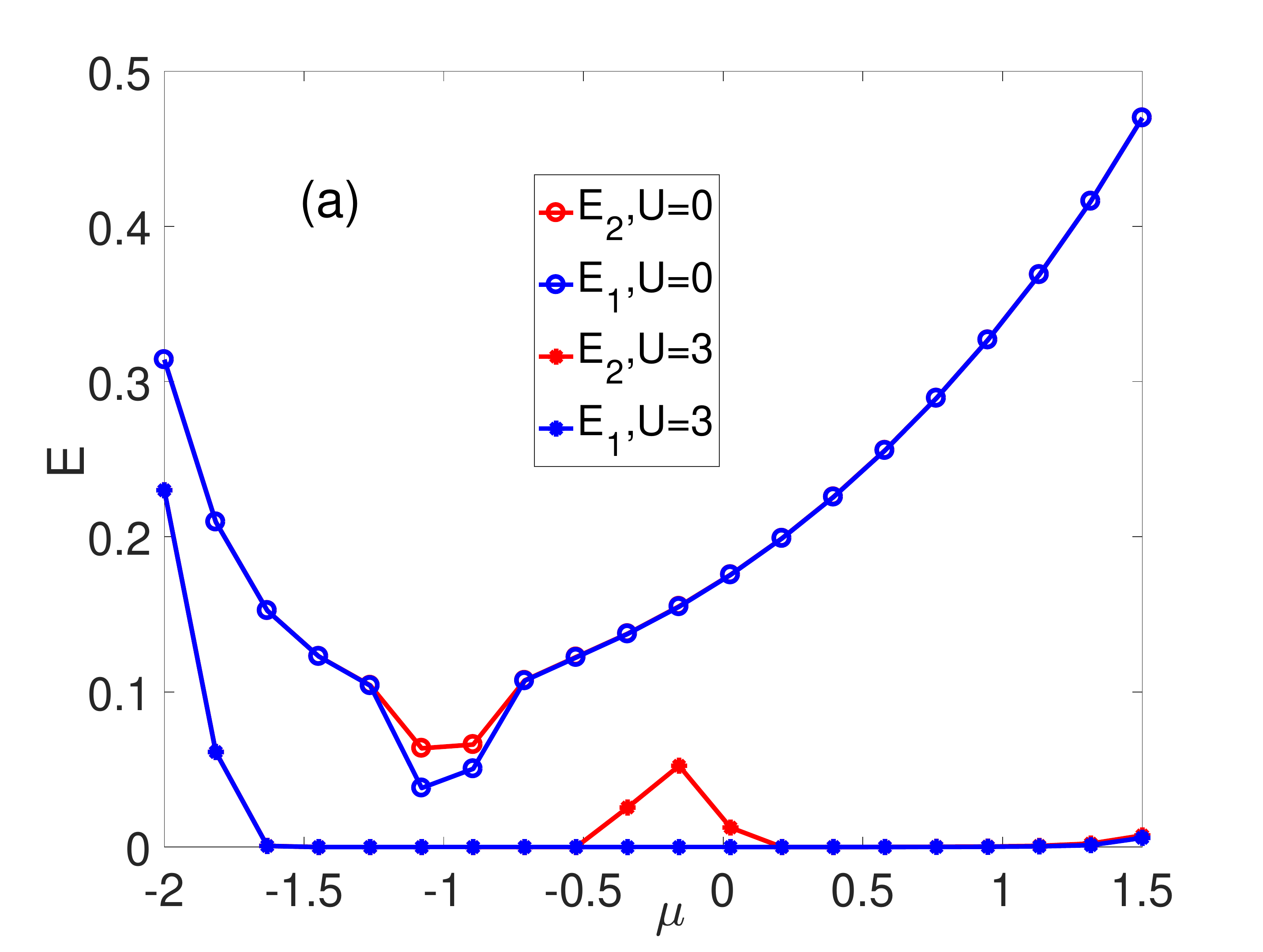}
	\includegraphics[width=0.45\textwidth]{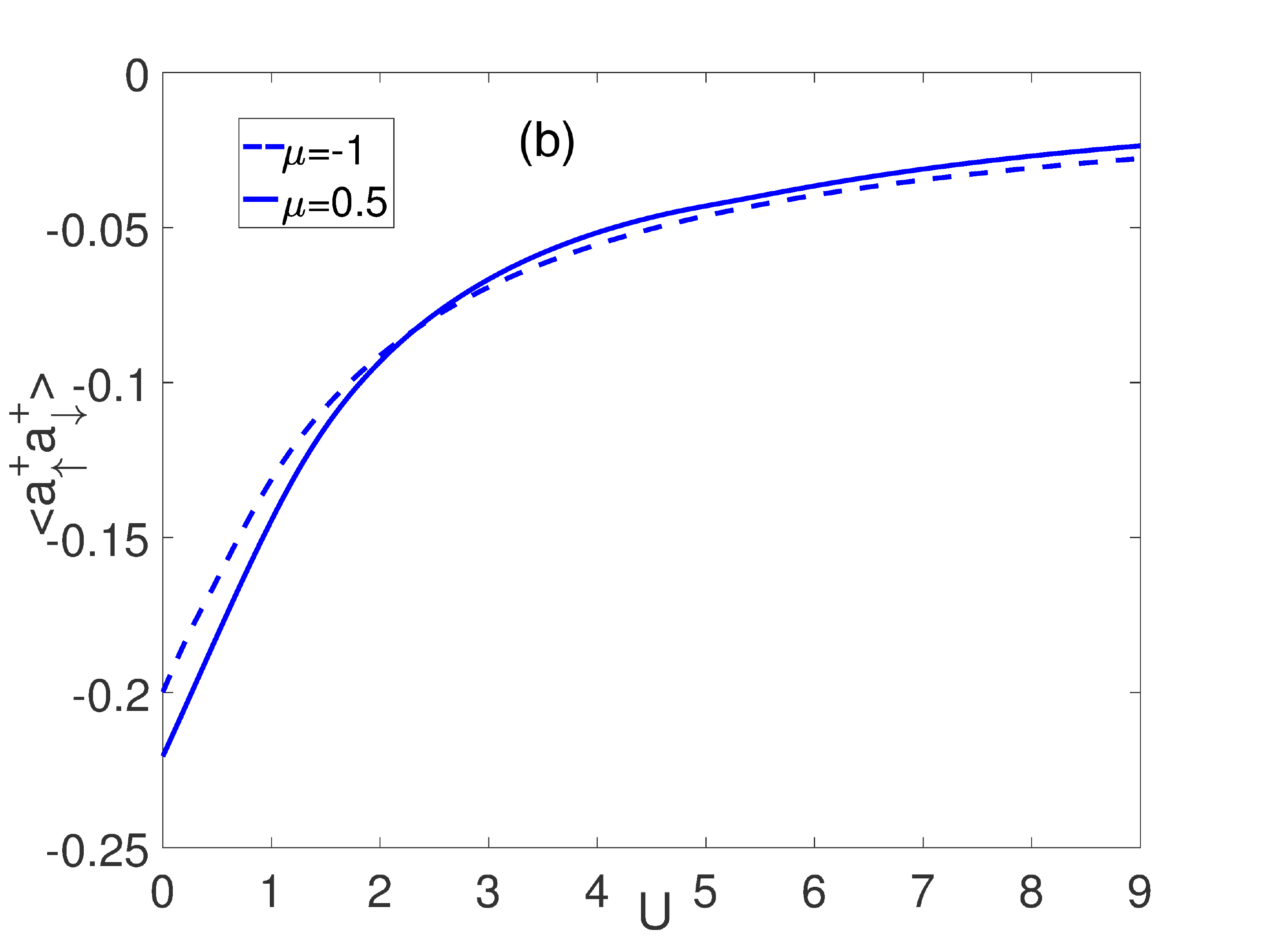}
	\includegraphics[width=0.45\textwidth]{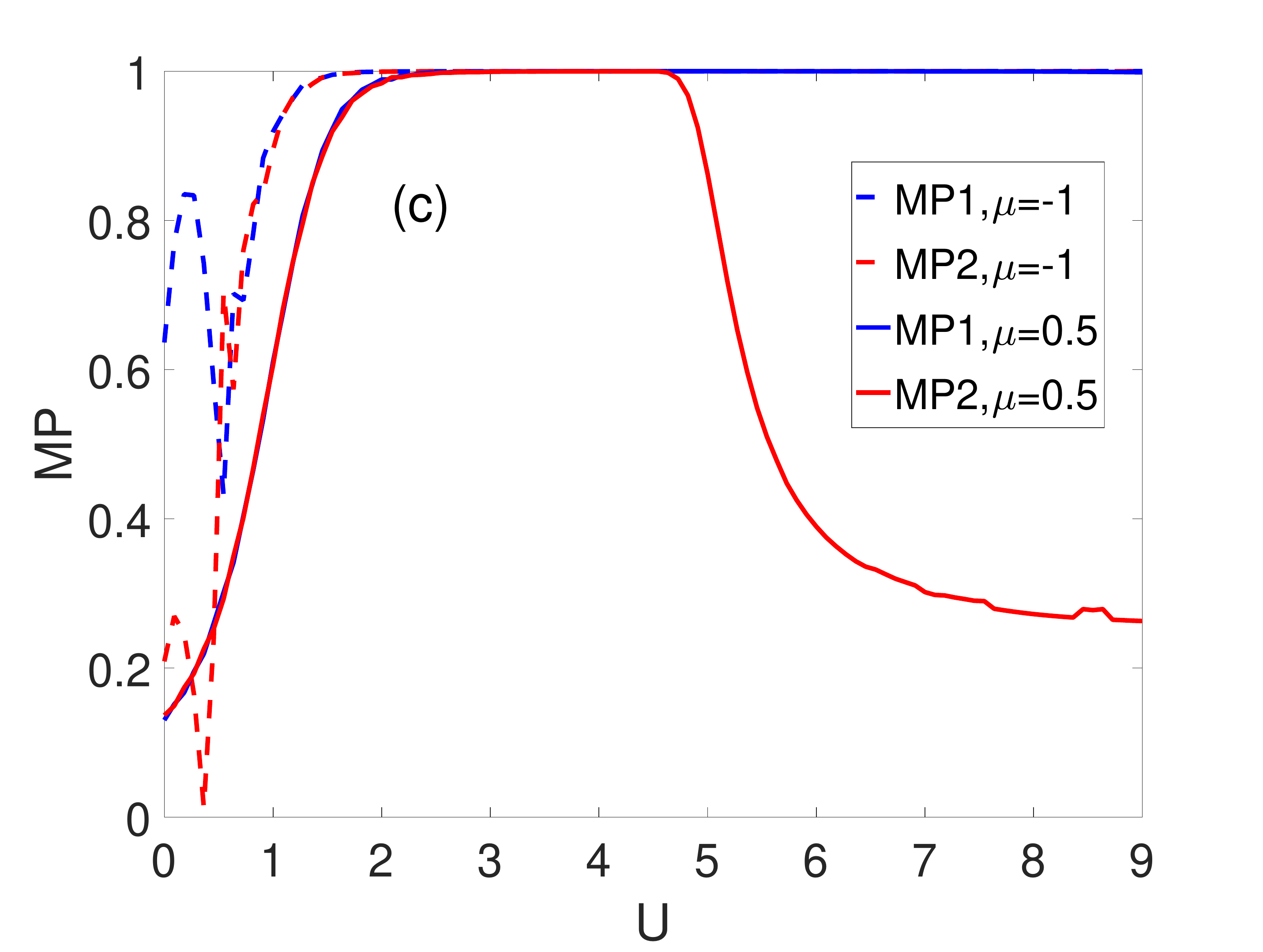}
	\includegraphics[width=0.45\textwidth]{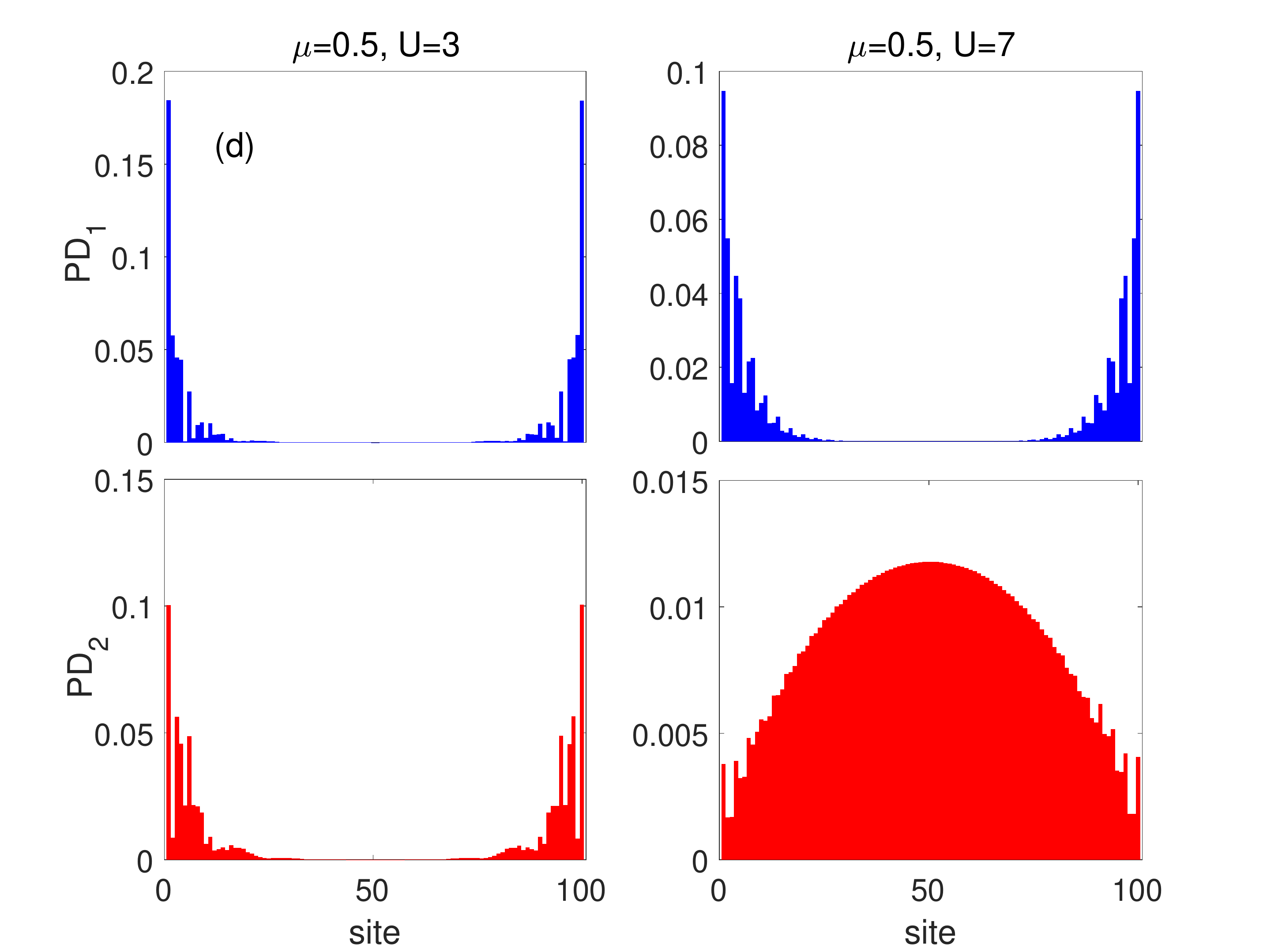}
	\caption{\label{indMDM} Occurrence of the Majorana double modes due to the strong on-site Coulomb correlations. (a) The chemical-potential dependence of the first two excitation energies. (b,c) The dependence of average on-site anomalous correlator and Majorana polarization on $U$, respectively. (d) Two graphs in the left (right) column represent the spatial distributions of probability densities of the first two excitations, $PD_{1}$ and $PD_{2}$, at $\mu=0.5$, $U=3$ ($U=7$). Parameters: $h=0.1$, $V=0$.}
\end{figure*}

\section{\label{sec3} Results and discussion}

For the subsequent numerical simulations we fix some of the BDI-nanowire parameters: $N=100$, $\varepsilon_{0}=0$, $\Delta=-0.5$, $\Delta_{1}=0.2$, $\alpha=1.5$. We will provide a semiquantitative analysis. Therefore, $64$ ($27$) quantum states for the basic (effective) model are kept. The truncation error in both cases did not exceed $10^{-5}$. 

\subsection{\label{secA} General findings}

Let us proceed to the numerical results obtained by the DMRG method to analyze the interaction influence on topological phases in the BDI-class wire \eqref{Ham_Fermi}. DMRG has been already utilized earlier to study this issue in the D-class wire \cite{stoudenmire-11}. In particular, considering the left parabola with the MSMs inside at low $U$ Stoudenmire et al. showed that its right border moves to the right and the minimum shifts right and down while $U$ is rising. The left border remains approximately at the same place because of the low electron concentration. In other words, the nontrivial phase emerges at the lower magnetic fields and higher chemical potentials in comparison with the $U=0$ case. This behavior can be qualitatively explained by effective enhancement of the Zeeman splitting and suppression of the on-site SC pairing. Such features are distinctly manifested already in the GMF description (see the expressions for $\left( A_{\sigma \sigma} \right)_{f,f}$ and $\left( B_{\uparrow \downarrow} \right)_{f,f}$ in \eqref{A_B_int}). Additionally, the trivial-phase gap appears between the MSM areas. Finally, starting from the regime of intermediate electron correlations, $U>2$, the left and right parabolas settle down in the lower and upper Hubbard subbands, respectively.

Similar effects occur in the BDI-class wire. Next, according to the relations \eqref{h_gapless}, \eqref{O_BDI}, if $\Delta,~\Delta_{1}\neq0$ the phase diagram becomes asymmetric (for $U=0$ in the D-class system the bottoms of both parabolas are at $h=\Delta$). Consequently, when $\Delta < 0$, $\Delta_1 > 0$, and $|\Delta|<2|\Delta_{1}|$ the area with the MDMs in the $\mu-h$ parametric space is located around the left parabola. In turn, its width is defined by $\alpha$.

\subsection{\label{secB}Correlation-induced Majorana modes}

One can see from the above-discussed results that the control of relation between $\Delta$ and $\Delta_{1}$ leads to the different topological-phase diagrams. In particular, the MDMs vanish under the $\Delta$ increase. On the contrary, the on-site Coulomb interaction has to suppress the corresponding SC pairing and we expect the MDM phase to recover. This assumption is confirmed by the numerical calculations. In Fig.\ref{indMDM}a such a phenomenon is displayed in a chemical-potential dependence of two lowest-excitation energies, $E_{1,2}$, for the small value of magnetic field, $h=0.1$. When $U=0$ and $\mu\approx-1$ the system is in the parametric region of left-parabola bottom and close to the topological phase transition (see Fig.\ref{N_BDI_map}a). The last is additionally supported by the data in the Fig.\ref{indMDM}a where the energies $E_{1,2}$ are split around $\mu\approx-1$ even though $E_{1}$ is nonzero yet (see blue and red circle-marked curves). When $U$ increases the topological phases with the MSMs can be reached at weaker magnetic fields as it was discussed above. As a result, $E_{1}\approx0$ and $E_{2}\neq0$ at $\mu\approx-0.5$ --- $0.2$ for $U=3$ (see blue and red cross-marked curves). In addition, the MDM phase emerges to the left and right of this area where both $E_{1}$ and $E_{2}$ approximately equal zero in spite of $|\Delta|>2|\Delta_{1}|$.  

The MDM induction at strong $U$ for $|\Delta|>2|\Delta_{1}|$ can be qualitatively accounted to the considerable reduction of effective on-site pairing which is clearly seen via the behavior of corresponding average anomalous correlator, $\left<a^{+}_{\uparrow}a^{+}_{\downarrow}\right>=\sum\limits_{f}\left<a^{+}_{f\uparrow}a^{+}_{f\downarrow}\right>/N$. Its dependence on $U$ is shown in Fig.\ref{indMDM}b displaying about two-time attenuation at $U=2$.

The edge-like character of both excitations in the left and right MDM areas is proved by the MP values which equal 1 for $\mu=-1$ and $\mu=0.5$ (see all the curves at point $U=3$ in Fig.\ref{indMDM}c). Significantly, these states are mainly localized at the wire ends ($MP_{i} \gtrsim 0.9$) only for $U \gtrsim 1$, where the mean-field description becomes invalid \cite{stoudenmire-11,valkov-17}. It is remarkable that the left MDMs survive even at the high intensities of on-site correlations. In opposite, the right MDMs transform into the MSMs as $MP_2$ significantly deviates from 1 for $U>5$ (see red solid curve in Fig.\ref{indMDM}c) that can be explained by the continuing movement of the MSM area to the right on phase diagram. Simultaneously, since the gap between Hubbard subbands develops at half-filling, $\mu\approx1.5$, it gradually shrinks the right MDMs around this point. Note that the many-body-interaction mechanism of MDM formation was also analyzed for quasi-1D DIII-class wire by means DMRG and for the BDI-class one using the Hartree-Fock approximation \cite{haim-14}. 

To display the transition from the MDM- to MSM phase evidently and show direct relation between the MP and probability densities of excitations, $PD_{j}\left(f\right)$ ($j=1,2$), we plot the spatial distributions of the latter in Fig.\ref{indMDM}d. In general, $PD_{j}\left(f\right)$ can be expressed in terms of the Bogoliubov coefficients as follows
\begin{equation} \label{PD}
PD_{j}\left(f\right)=\sum\limits_{\sigma}\left(\mid u_{jf\sigma}\mid^2+\mid v_{jf\sigma}\mid^2\right),
\end{equation}
where $u_{jf\sigma}=\left(w_{jf\sigma}+z_{jf\sigma}\right)/2$, $v_{jf\sigma}=\left(w_{jf\sigma}-z_{jf\sigma}\right)/2$ and the coefficients $w_{jf\sigma}$, $z_{jf\sigma}$ are defined in \eqref{wz}. The left column corresponds to the case of $\mu=0.5,~U=3$. There are two MMs (see top blue and bottom red dependencies). Whereas the right column describes the point $\mu=0.5,~U=7$ where the second state becomes bulk-like (see bottom red distribution). 

\begin{figure}[htbp]
	\includegraphics[width=0.5\textwidth]{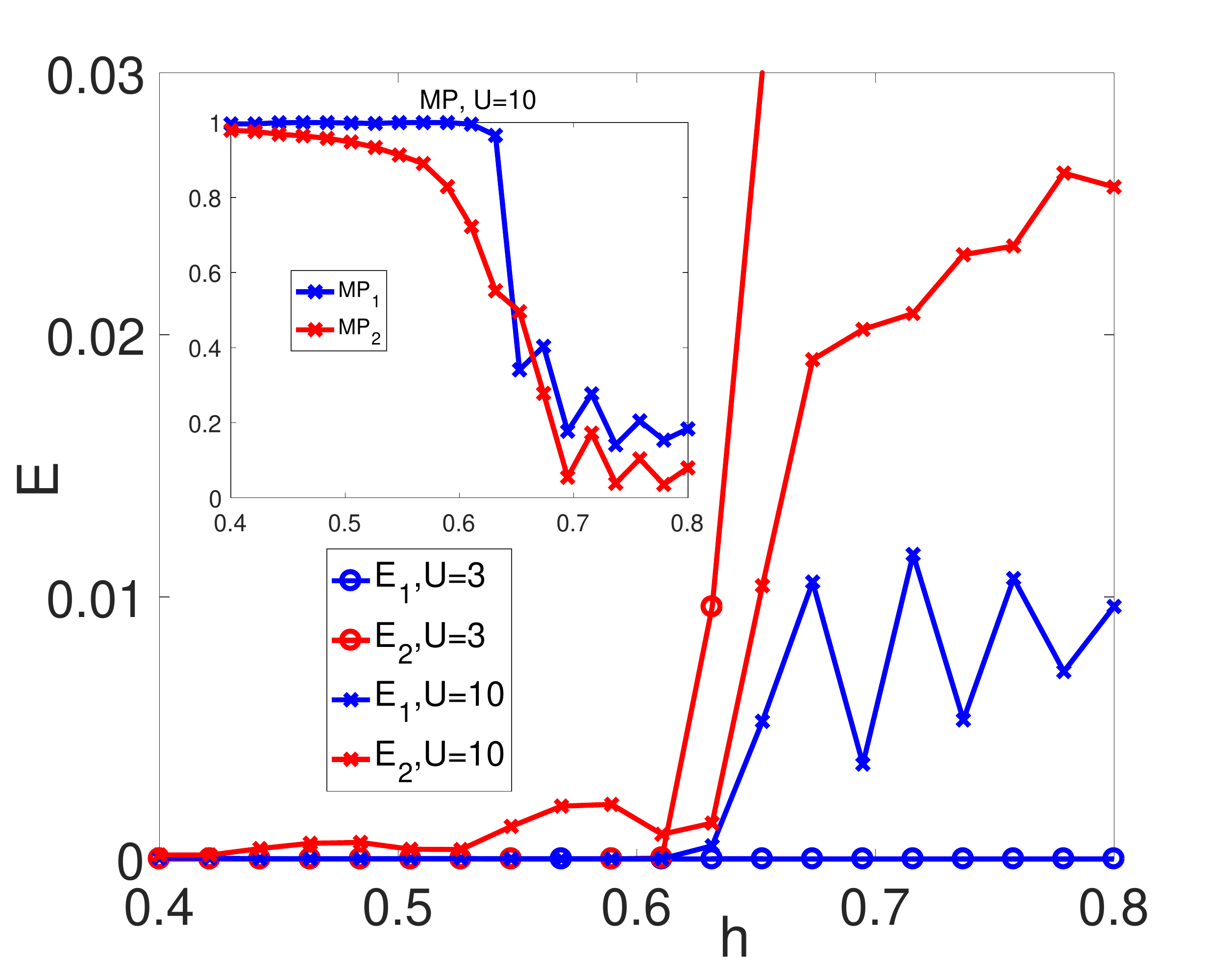}
	\caption{\label{indMSM} Change in the number of Majorana modes for different on-site interaction strengths. The magnetic-field dependence of the first two excitation energies at $U=3$ (circles) and $U=10$ (crosses). Inset: $MP_{1,2}$ versus $h$ at $U=10$. Parameters: $\mu=-1.4$, $V=0$.}
\end{figure}
Since the Kramers degeneracy is absent at $h\neq0$ that the second state in the MDM is earlier affected by the bulk-gap closing while $U$ increases. Hence, it is expected that at sufficiently high magnetic fields the MDMs can be transformed into the MSMs. The described picture is shown in Figure \ref{indMSM}. For $U=3$ the blue and red circle-marked curves correspond to $E_{1}\left(h\right)$ and $E_{2}\left(h\right)$, respectively. At $h \approx 0.6$ the topological phase transition emerges. The phase with two (one) MMs is realized to the left (right) of this point. If the intensity of on-site repulsion grows to $U=10$ that the MSM phase at $h\gtrsim0.6$ is fully suppressed (see blue and red cross-marked curves). However, a part of the MDM area is turned into the MSM one at $0.55 \lesssim h \lesssim0.6$ which is also confirmed by the MP calculations (see the inset of Fig. \ref{indMSM}). 

The MP definition \eqref{MP_def} provides data about overlapping of the wave functions of Majorana single-particle excitations and does not give information
about partial contribution of such excitations to the general structure of many-body quantum transitions. The last can be estimated via the norm of $j$th excitation, $norm_{j}$. If it tends to 1 the MMs form a well-defined quasiparticle excitation that is suitable for quantum-computation purposes. Otherwise,
when $norm_{j} < 1$ the role of quasiparticle dissipation grows. This issue is analyzed in the following section.

\subsection{\label{secC}Dependence of Majorana-mode norm on the on-site Coulomb interaction and electron concentration}

\begin{figure}[htbp]
	\includegraphics[width=0.5\textwidth]{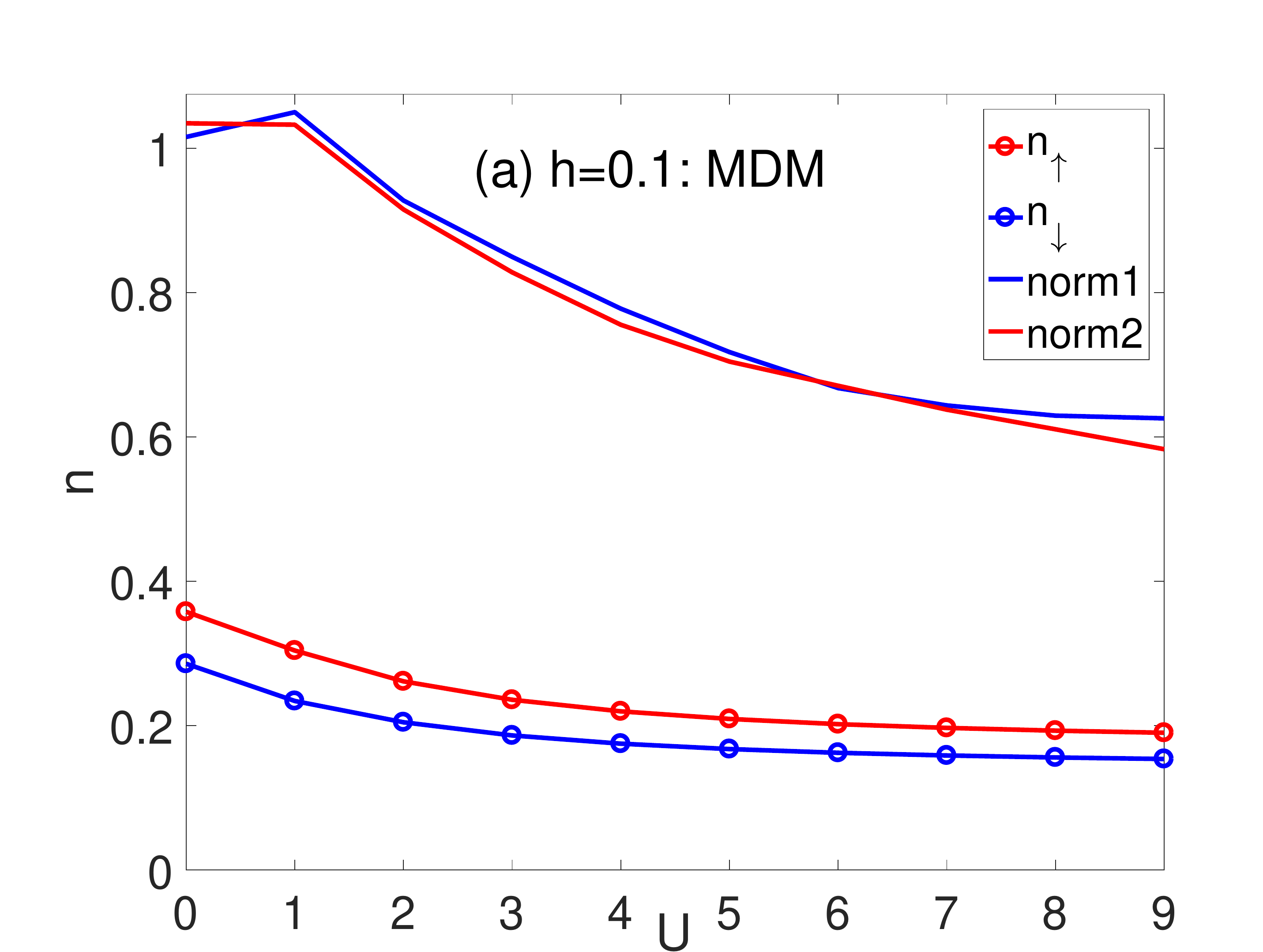}
	\includegraphics[width=0.5\textwidth]{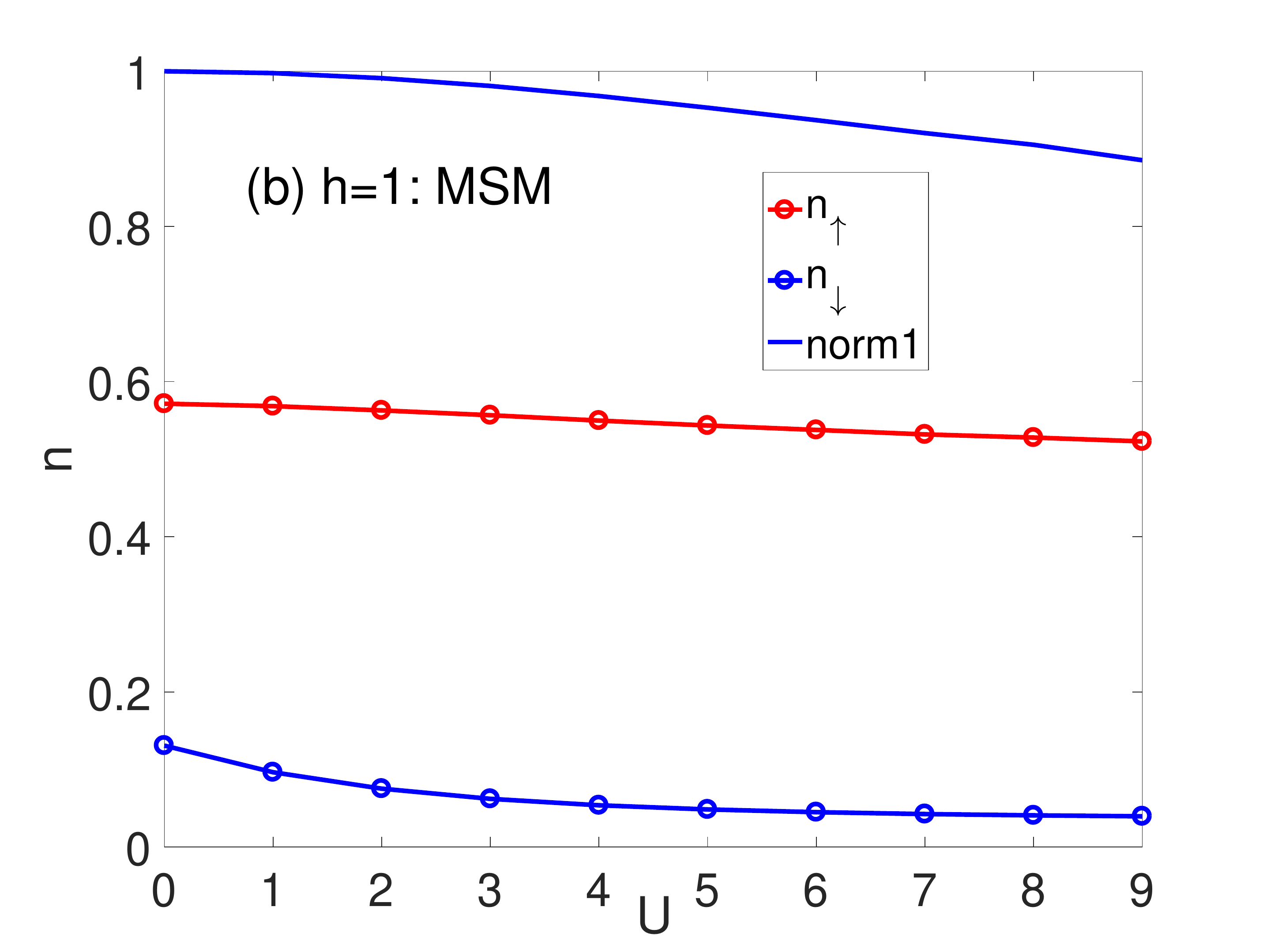}
	\caption{\label{nudnorm_U} The impact of on-site interaction and average spin-dependent concentrations on the norm of Majorana single-particle excitations at the low (a) and high (b) magnetic fields, $h=0.1$ and $h=1$, respectively; $\mu=-1$. The other parameters are the same as in Fig.\ref{indMDM}.}
\end{figure}

To clarify deeper the influence of strong on-site Coulomb interaction on the Majorana-type excitations let us consider the behavior of their norms, $norm_{j}=\sum\limits_{f}PD_{j}\left(f\right)$. The cases of low and high magnetic fields are depicted in Figures \ref{nudnorm_U}a and \ref{nudnorm_U}b, respectively. As it was mentioned above the MDMs occur at $h=0.1$, $\mu=-1$ and $U\geq1$. In turn, their norms dramatically reduce while $U$ increases (see red and blue solid curves in Fig.\ref{nudnorm_U}a) signalizing that the three- and more-fermion terms have to be taken into account to properly characterize the excitations. There is also necessary to remark that the slight exceeding 1 by both norms is related to not enough basis states kept. Obviously, the essential factor influencing on the norms at strong $U$ regime is spin-dependent on-site concentrations. One can see it in Fig.\ref{nudnorm_U}a where average spin-up and -down occupation numbers, $n_{\sigma}=\sum\limits_{f}\left<a^{+}_{f\sigma}a_{f\sigma}\right>/N$, are shown as well (see red and blue circle-marked curves). The norms considerably decrease since $n_{\uparrow}$ and $n_{\downarrow}$ are commensurable quantities. 

The effect of on-site correlations on the Majorana excitation becomes much weaker at high magnetic fields. The corresponding case is depicted in Fig.\ref{nudnorm_U}b. For used parameters, $h=1$ and $\mu=-1$, the MSMs realize. In this situation the wire is in spin-polarized regime as the difference between $n_{\uparrow}$ and $n_{\downarrow}$ becomes significant (see red and blue circle-marked curves). While $U$ enhances the minority-spin occupation, $n_{\downarrow}$, decreases faster than the majority-spin one, $n_{\uparrow}$, leading to $n_{\uparrow}/n_{\downarrow}\approx10$. As a result, even for the large values of on-site Coulomb interaction the norm deviation is still about $10\%$ (see blue solid curve). The established correspondence between $norm_{j}$ and $n_{\sigma}$ might be useful for the
experimental analysis of the MM properties in interacting quantum wires.

\subsection{\label{secD}Effective Hamiltonian in the limit $U \gg \alpha, t, h, \Delta, \Delta_1$}

When performing DMRG calculations the diagonalization of large sparse matrices is carried out by the Lanczos algorithm. Its convergence significantly decreases in the strongly correlated regime, $U > 3$, that becomes especially prominent at $U\sim10$, $V\sim1$. The reason is an appearance of large number of matrix elements which values substantially deviate from zero. Such an effect is qualitatively observed even in the GMF description (see the elements $\left( A_{\uparrow \downarrow} \right)_{f,f}$ and $\left( B_{\sigma \sigma} \right)_{f+1,f}$ in \eqref{A_B_int}) \cite{valkov-17,valkov-19a}. In order to overcome this obstacle we derive here the effective model employing the atomic representation. 

The second advantage of the DMRG extension to the case of effective strong-interaction Hamiltonian is considerable enhancing of numerics speed. It becomes possible since all the states with two electrons on one site are integrated out theoretically due to projection-operator technique. The DMRG algorithm for the final model operates faster in comparison with the one for the model \eqref{Ham_Fermi} since the number of used eigenstates decreases from $4^{N_0}$ to $3^{N_0}$, where $N_{0}$ - a number of sites in the cluster. Additionally, the biggest size of matrices for the corresponding calculations reduces from $4^{2\left(N_0+1\right)}$ to $3^{2\left(N_0+1\right)}$.

Let us introduce the Hubbard operators as
\begin{equation}
\label{Hub_oper}
X_{f}^{pq}=| f,p \rangle \langle f,q |,
\end{equation}
where $p, q = 0; \sigma; 2$ describe quantum states on
the $f$th site \cite{valkov-04}. Then using the connection $a_{f \sigma} = X_f^{0 \sigma} + \eta_{\sigma} X_f^{\bar{\sigma} 2}$ the Hamiltonian \eqref{Ham_Fermi} in the atomic representation acquires the form:
\begin{equation}
\label{Ham_Hub}
{\textit{H}} = {\textit{H}}_0 + {\textit{H}}_1 + {\textit{H}}_2 + {\textit{H}}' + {\textit{H}}_V,
\end{equation}
where the on-site Hamiltonian is
\begin{eqnarray}
\label{Ham_Hub0}
{\textit{H}}_0 = \sum_{f \sigma} \xi_{\sigma} X_f^{\sigma \sigma} + \sum_{f} \left( 2\xi + U \right) X_f^{2 2}.
\end{eqnarray}
The Hamiltonians ${\textit{H}}_1$, ${\textit{H}}_2$ describe processes at the lower and upper Hubbard subbands, respectively,
\begin{eqnarray}
\label{Ham_Hub1}
{\textit{H}}_1 & = & - \frac{t}{2} \sum_{f\sigma} \left( X_f^{\sigma 0} X_{f+1}^{0 \sigma} + X_{f+1}^{\sigma 0} X_{f}^{0 \sigma}  \right) -
\nonumber \\
& - & \frac{\alpha}{2} \sum_{f\sigma} \eta_{\sigma} \left( X_f^{\sigma 0} X_{f+1}^{0 \bar{\sigma}} + X_{f+1}^{\bar{\sigma} 0} X_{f}^{0 \sigma} \right) +
\nonumber \\
& + & \Delta_1 \sum_{f} \left( X_f^{ 0 \uparrow} X_{f+1}^{0 \downarrow} + X_{f+1}^{0 \uparrow} X_{f}^{0 \downarrow}  \right) + h.c.,
\end{eqnarray}
\begin{eqnarray}
\label{Ham_Hub2}
{\textit{H}}_2 & = & - \frac{t}{2} \sum_{f\sigma} \left( X_f^{2 \bar{\sigma}} X_{f+1}^{\bar{\sigma} 2} + X_{f+1}^{2 \bar{\sigma}} X_{f}^{\bar{\sigma} 2}  \right) -
\nonumber \\
& - & \frac{\alpha}{2} \sum_{f\sigma} \eta_{\bar{\sigma}} \left( X_f^{2 \bar{\sigma}} X_{f+1}^{\sigma 2} + X_{f+1}^{2 \sigma} X_{f}^{\bar{\sigma} 2} \right) -
\nonumber \\
& - & \Delta_1 \sum_{f} \left( X_f^{ \downarrow 2} X_{f+1}^{\uparrow 2} + X_{f+1}^{\downarrow 2} X_{f}^{\uparrow 2} \right) + h.c.
\end{eqnarray}
The interaction between the subbands is characterized by the Hamiltonian ${\textit{H}}'$,
\begin{eqnarray}
\label{Ham_Hub_int}
&& {\textit{H}}' = - \frac{t}{2} \sum_{f\sigma} \eta_{\sigma} \left( X_f^{\sigma 0} X_{f+1}^{\bar{\sigma} 2} + X_{f}^{2 \bar{\sigma}} X_{f+1}^{0 \sigma} + h.c.  \right) -
\nonumber \\
& - & \frac{\alpha}{2} \sum_{f\sigma} \left( -X_f^{\sigma 0} X_{f+1}^{\sigma 2} + X_{f}^{2 \sigma} X_{f+1}^{0 \sigma} + h.c. \right) +
\nonumber \\
& + &  \left\{ -\Delta \sum_f X_f^{0 2} +  \Delta_1 \sum_{f} \left( -X_f^{ 0 \uparrow} X_{f+1}^{\uparrow 2} - X_{f+1}^{0 \uparrow} X_{f}^{\uparrow 2} \right. \right. +
\nonumber \\
& + & \left. \left. X_f^{ \downarrow 2} X_{f+1}^{0 \downarrow} + X_{f+1}^{\downarrow 2} X_{f}^{0 \downarrow} \right) \right\} + \left\{ h.c. \right\}
\end{eqnarray}
The term characterizing inter-site Coulomb interaction is
\begin{equation}
\label{Ham_V}
{\textit{H}}_V = V \sum_f n_{f} n_{f+1},
\end{equation}
here $n_f = X_f^{\uparrow\uparrow} + X_f^{\downarrow\downarrow} + 2 X_f^{22}$.

To derive the analogue of $t-J^*-V$-model from the $t-U-V$-model \eqref{Ham_Hub} in the limit of strong electron correlations taking into account spin-orbit coupling the unitary transformation is applied, 
\begin{equation}
\label{Ham_S}
H \to \tilde{H} = e^S H e^{S^{\dag}}, \, \, \, S^{\dag} = - S,
\end{equation}
where the operator $S$ has to satisfy ${\textit{H}}' - \left[ {\textit{H}}_0, S \right]_{-} = 0$. As a result, $S$ is given by
\begin{eqnarray}
\label{S}
&& S =  \frac{t/2}{U} \sum_{f\sigma} \eta_{\sigma} \left( X_f^{\sigma 0} X_{f+1}^{\bar{\sigma} 2} - X_{f}^{2 \bar{\sigma}} X_{f+1}^{0 \sigma} - h.c.  \right) -
\nonumber \\
& - &  \sum_{f\sigma} \frac{\alpha/2}{U-2\eta_{\sigma}h}  \left( X_f^{\bar{\sigma} 0} X_{f+1}^{\bar{\sigma} 2} + X_{f}^{2 \bar{\sigma}} X_{f+1}^{0 \bar{\sigma}} - h.c. \right) +
\nonumber \\
& + & \frac{\Delta}{2\xi + U} \sum_f \left(  X_f^{0 2} - X_f^{2 0}  \right).
\end{eqnarray}

\begin{figure*}[htbp]
	\includegraphics[width=0.48\textwidth]{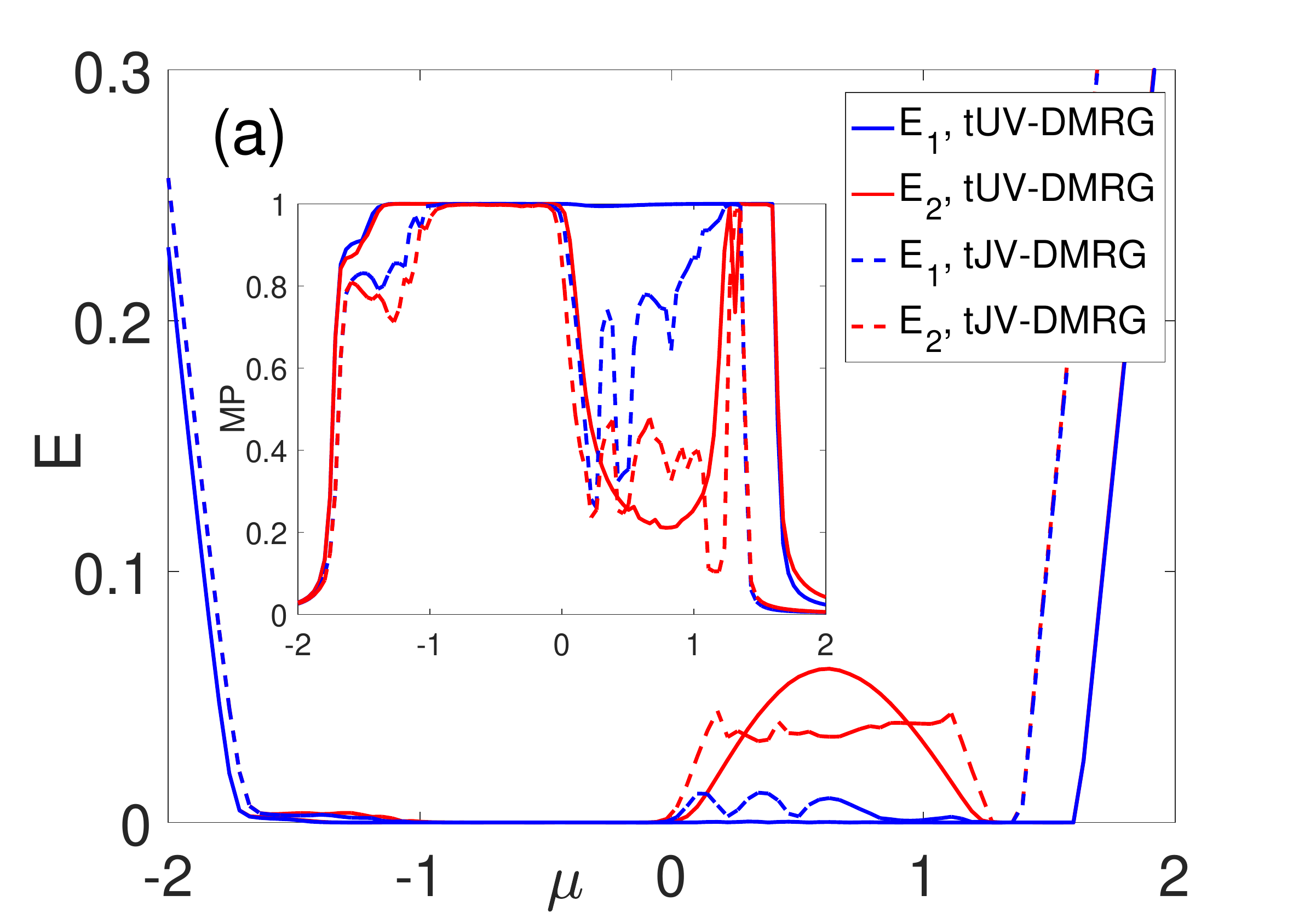}
	\includegraphics[width=0.48\textwidth]{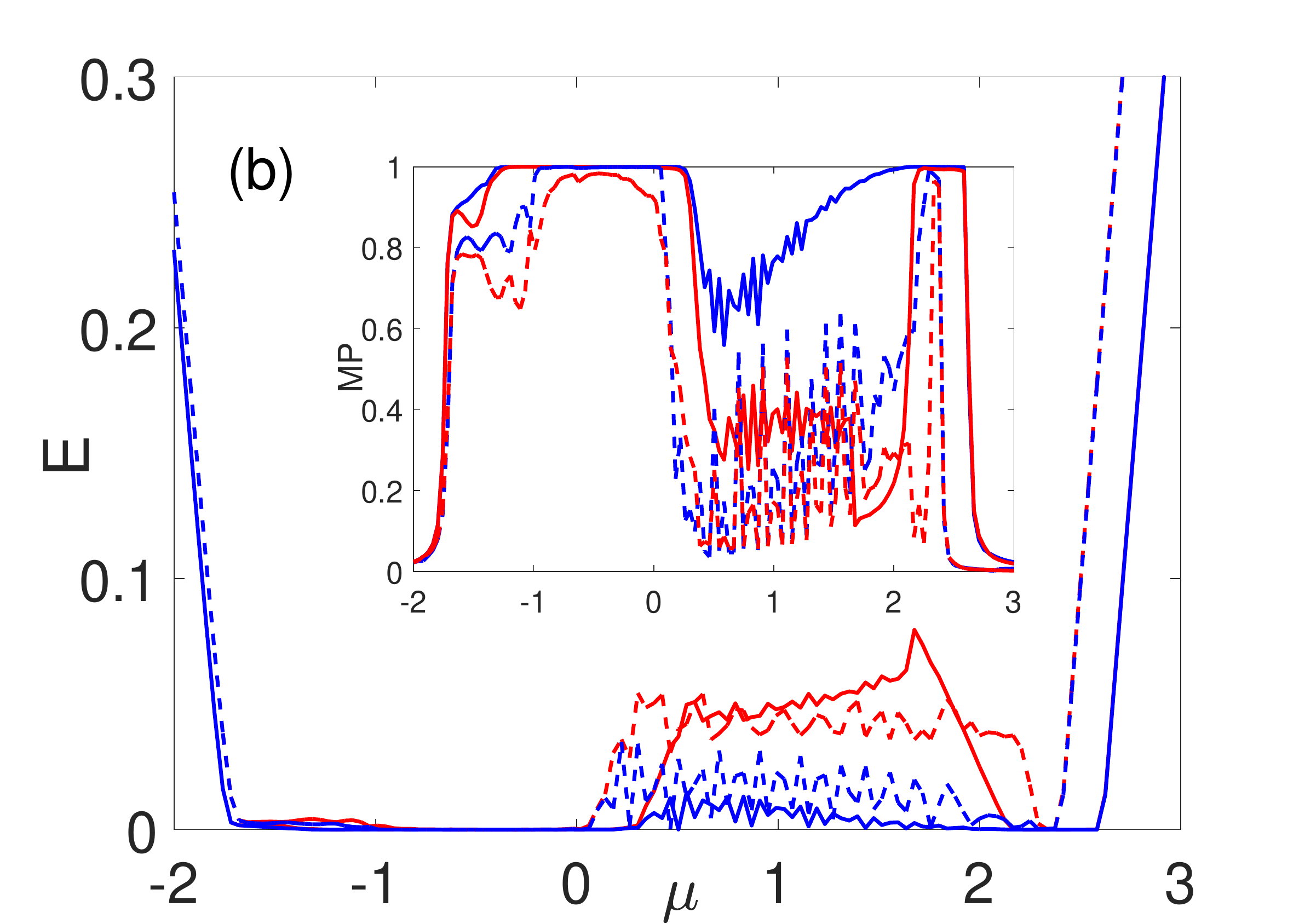}
	\caption{\label{tUV_vs_tJV} The DMRG data for initial ($t-U-V$) and effective ($t-J-V$) models plotted by solid and dashed curves, respectively. The chemical-potential dependencies of excitation energies at $U=10$, $V=0$ (a) and $U=10$, $V=0.5$ (b). Insets: the $\mu$ dependencies of the Majorana polarizations $MP_{1}$ and $MP_{2}$ depicted by blue and red curves, respectively. Parameters: $h=0.1$.}
\end{figure*}
By projecting the states on the lower Hubbard subband in the limit $U \gg \alpha, t, h,\Delta, \Delta_1$ we obtain the effective Hamiltonian,
\begin{equation}
\label{Ham_eff}
{\textit{H}}_{t-J^*-V} = \sum_{f \sigma} \xi_{\sigma} X_f^{\sigma \sigma} - \frac{\Delta^2}{2\xi+U} \sum_f X_f^{00} + {\textit{H}}_1 + {\textit{H}}_{int} + {\textit{H}}_{3} + {\textit{H}}_V,
\end{equation}
where the interaction term is described by
\begin{eqnarray}
\label{Ham_int}
&& {\textit{H}}_{int} = \frac{t^2}{U} \sum_{f} \left( \textbf{S}_f \textbf{S}_{f+1} - \frac{1}{4} n_f  n_{f+1} \right)
- \nonumber \\
& - & \frac{t\alpha}{U} \sum_{f} \left\{ \left( X_f^{\uparrow\downarrow} + X_f^{\downarrow \uparrow} \right) S_{f+1}^{z} - S_{f}^{z} \left( X_{f+1}^{\uparrow\downarrow} + X_{f+1}^{\downarrow \uparrow} \right) \right\}
- \nonumber \\
& - & \frac{\alpha^2}{U} \sum_{f} \left\{ \frac{1}{2}\left( X_f^{\uparrow\downarrow} X_{f+1}^{\uparrow \downarrow} + X_f^{\downarrow \uparrow} X_{f+1}^{\downarrow \uparrow} \right) + S_f^z S_{f+1}^z + \right.
\\
& + & \left. \frac{1}{4} n_f  n_{f+1} \right\}  - \frac{\alpha \Delta (\xi + U)}{U(2\xi + U)} \sum_{f \sigma} \left( X_f^{0 \sigma} X_{f+1}^{0 \sigma} + h.c.  \right) -
\nonumber \\
& - & \frac{t \Delta (\xi + U)}{U(2\xi + U)} \sum_{f} \left( X_f^{0 \downarrow} X_{f+1}^{0 \uparrow} - X_f^{0 \uparrow} X_{f+1}^{0 \downarrow} + h.c. \right),
\nonumber
\end{eqnarray}
and the three-center term is
\begin{eqnarray}
\label{Ham_3}
&& {\textit{H}}_{3} =
\nonumber \\ &-& \frac{t^2}{4U} \sum_{f\sigma} \left( X_{f-1}^{\sigma 0} X_{f}^{\bar{\sigma}\bar{\sigma}} X_{f+1}^{0 \sigma} - X_{f-1}^{\bar{\sigma} 0} X_{f}^{\sigma \bar{\sigma}} X_{f+1}^{0 \sigma} + h.c. \right)
+ \nonumber \\
& + & \frac{t\alpha}{4U} \sum_{f \sigma} \eta_{\sigma} \left\{ X_{f-1}^{\bar{\sigma} 0} \left( X_f^{\sigma\sigma} + X_f^{\bar{\sigma}\bar{\sigma}} \right) X_{f+1}^{0 \sigma} \right.
+ \\
&+& \left. X_{f-1}^{\sigma 0} \left( X_f^{\sigma\bar{\sigma}} - X_f^{\bar{\sigma}\sigma} \right) X_{f+1}^{0 \sigma} + h.c. \right\}
+ \nonumber \\
& + & \frac{\alpha^2}{4U} \sum_{f \sigma}  \left( X_{f-1}^{\sigma 0} X_{f}^{\sigma \sigma} X_{f+1}^{0 \sigma} + X_{f-1}^{\bar{\sigma} 0} X_{f}^{\bar{\sigma} \sigma} X_{f+1}^{0 \sigma} + h.c. \right).
\nonumber
\end{eqnarray}
Now $n_f$ is determined on the reduced Hilbert space, i.e. $n_f = X_f^{\uparrow\uparrow} + X_f^{\downarrow\downarrow}$, which is used in the definition of inter-site Coulomb interaction, ${\textit{H}}_V$, in \eqref{Ham_eff}. It is necessary to notice that  we neglect the contributions from the SC pairings between nearest sites to the effective and three-center interactions assuming $\Delta_1 \ll \alpha, t$. Nevertheless, the pairings in the lower Hubbard subband proportional to $\Delta_1$ are taken into account.

It is seen from \eqref{Ham_int} that the spin-orbit coupling induces the anomalous terms like $S_f^+S_{f+1}^+$ in addition to the super-exchange interaction with parameter $t^2/U$. The local character of on-site SC pairing leads to the appearance of two-site terms only. At the same time, the combination of this pairing and electron hoppings induces the spin-singlet pairings between nearest sites. On the other hand, the interplay of spin-orbit coupling and on-site pairing results in the spin-triplet pairings on the nearest neighbours.

Earlier the effective Hamiltonian for two-band Hubbard model with spin-orbit interaction in the strongly correlated limit was obtained \cite{koshibae-93}. However, to the best of our knowledge the effective interactions \eqref{Ham_eff} induced by the Rashba spin-orbit coupling have not been derived previously and are of fundamental interest themselves. It is useful to notice that the model \eqref{Ham_eff} can be easily brought to the D-class case by setting $\Delta_{1}$ to zero.

In the limit $U \to \infty$ the wire Hamiltonian \eqref{Ham_eff} is reduced to the $t$-model,
\begin{equation}
\label{Ham_t}
{\textit{H}}_{t} = \sum_{f \sigma} \xi_{\sigma} X_f^{\sigma \sigma} + {\textit{H}}_1 + {\textit{H}}_V.
\end{equation}
It is obvious
that in this limit the proximity induced on-site SC pairing is fully
suppressed as $\Delta/U \to 0$. Therefore, the D-class-like situation can not be implemented here. Next, the $t-J-V$- ($H_{3}$ term is neglected) and $t$-models allow to modify the numerical approach for more fruitful treatment of the strong-correlation regime by means of the $tJV$- and $t$-DMRG algorithms, respectively.   

\subsection{\label{secE}DMRG in the atomic representation}

Let us turn to the numerical results obtained by the DMRG method for the effective model. In this case the coefficients $\omega_{jf \sigma}$ and $z_{jf \sigma}$ are defined similarly to \eqref{wz} using the Hubbard operators,
\begin{eqnarray}
\label{wz2}
w_{jf\sigma}=\langle \Psi_{j} | \left( X_{f}^{0\sigma} + X_{f}^{\sigma 0} \right) | \Psi_{0} \rangle,\nonumber\\
z_{jf\sigma}=\langle \Psi_{j} | \left( X_{f}^{0\sigma} - X_{f}^{\sigma 0} \right) | \Psi_{0} \rangle.
\end{eqnarray}
To show correspondence between the results provided by the DMRG for the initial $t-U-V$-model ($tUV$-DMRG) and $tJV$-DMRG the chemical-potential dependencies of $E_{1,2}$ are plotted in Figure \ref{tUV_vs_tJV}. Note that the on-site Coulomb parameter is chosen to be quite high, $U=10$, in order to exclude the higher orders of perturbation theory in the effective model \eqref{Ham_eff}.

\begin{figure*}[htb!]
	\includegraphics[width=0.425\textwidth]{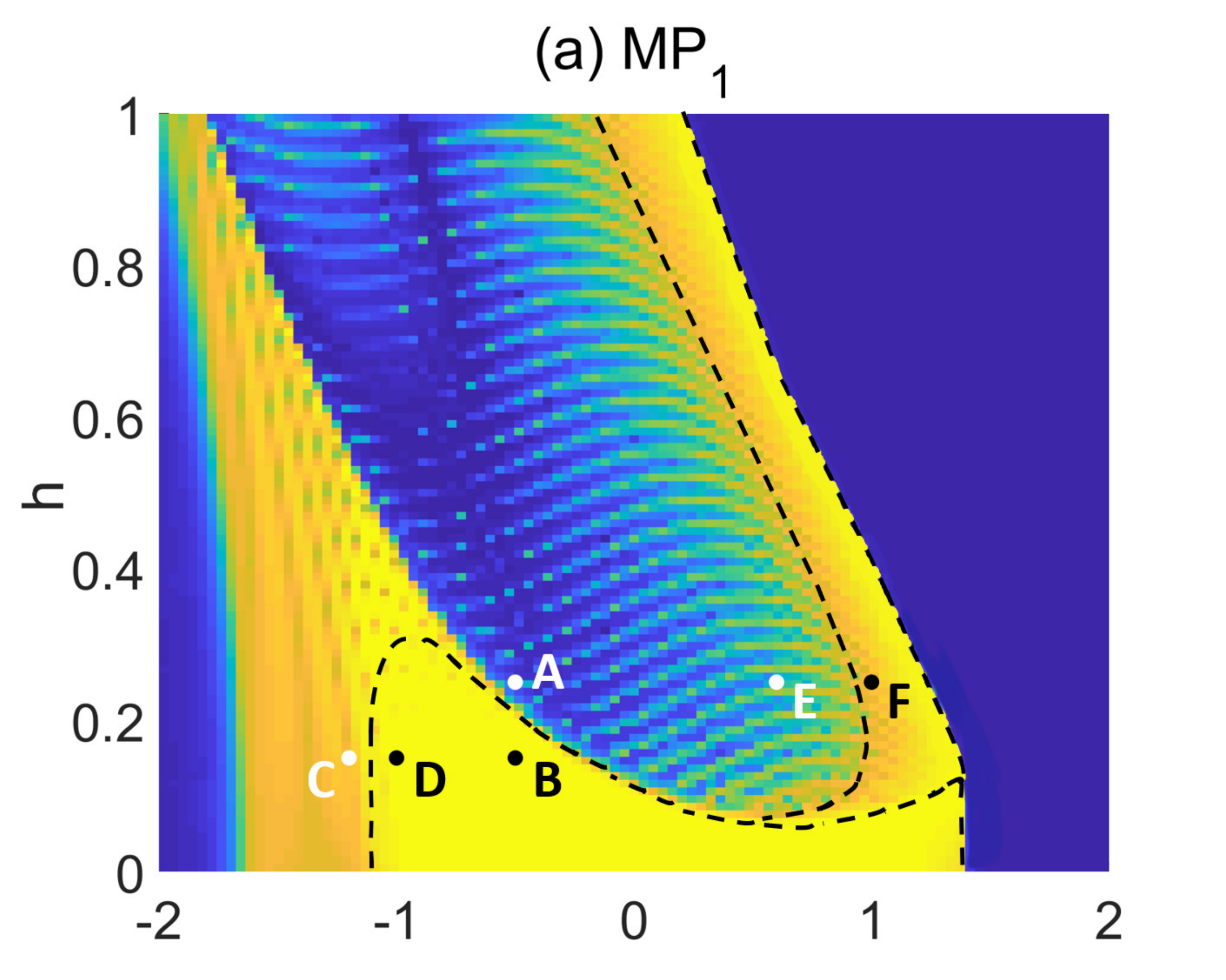}
	\includegraphics[width=0.45\textwidth]{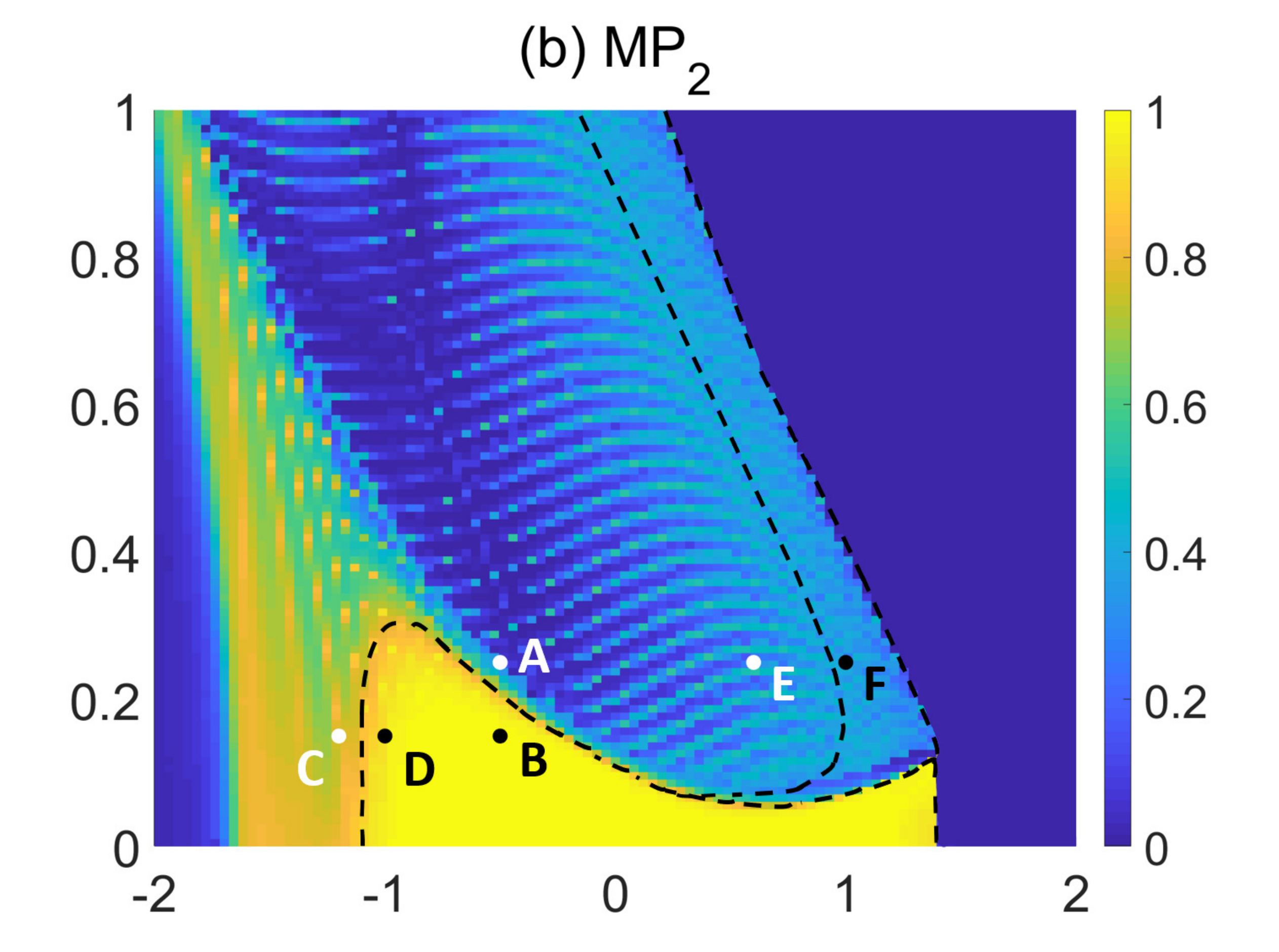}
	\includegraphics[width=0.425\textwidth]{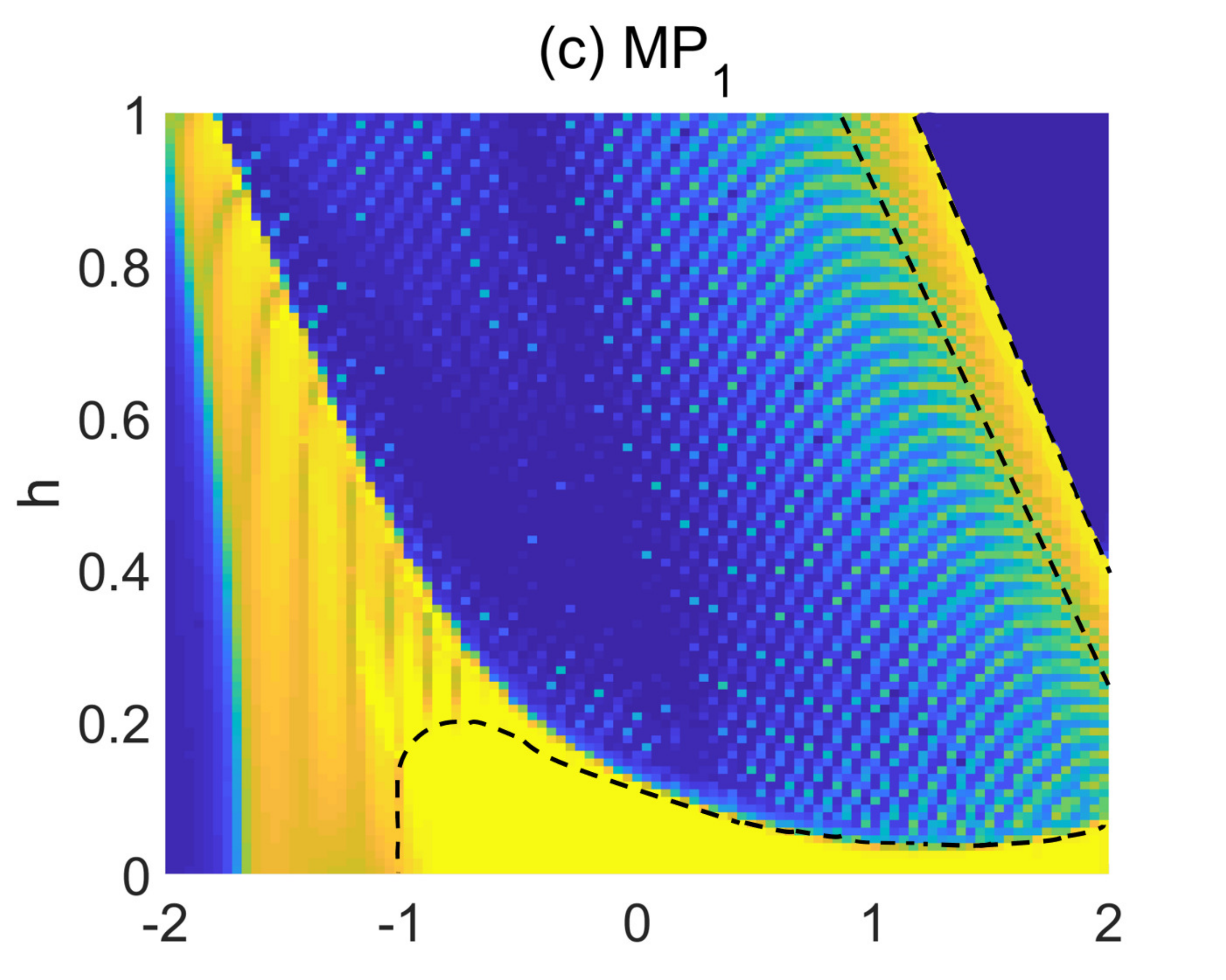}
	\includegraphics[width=0.45\textwidth]{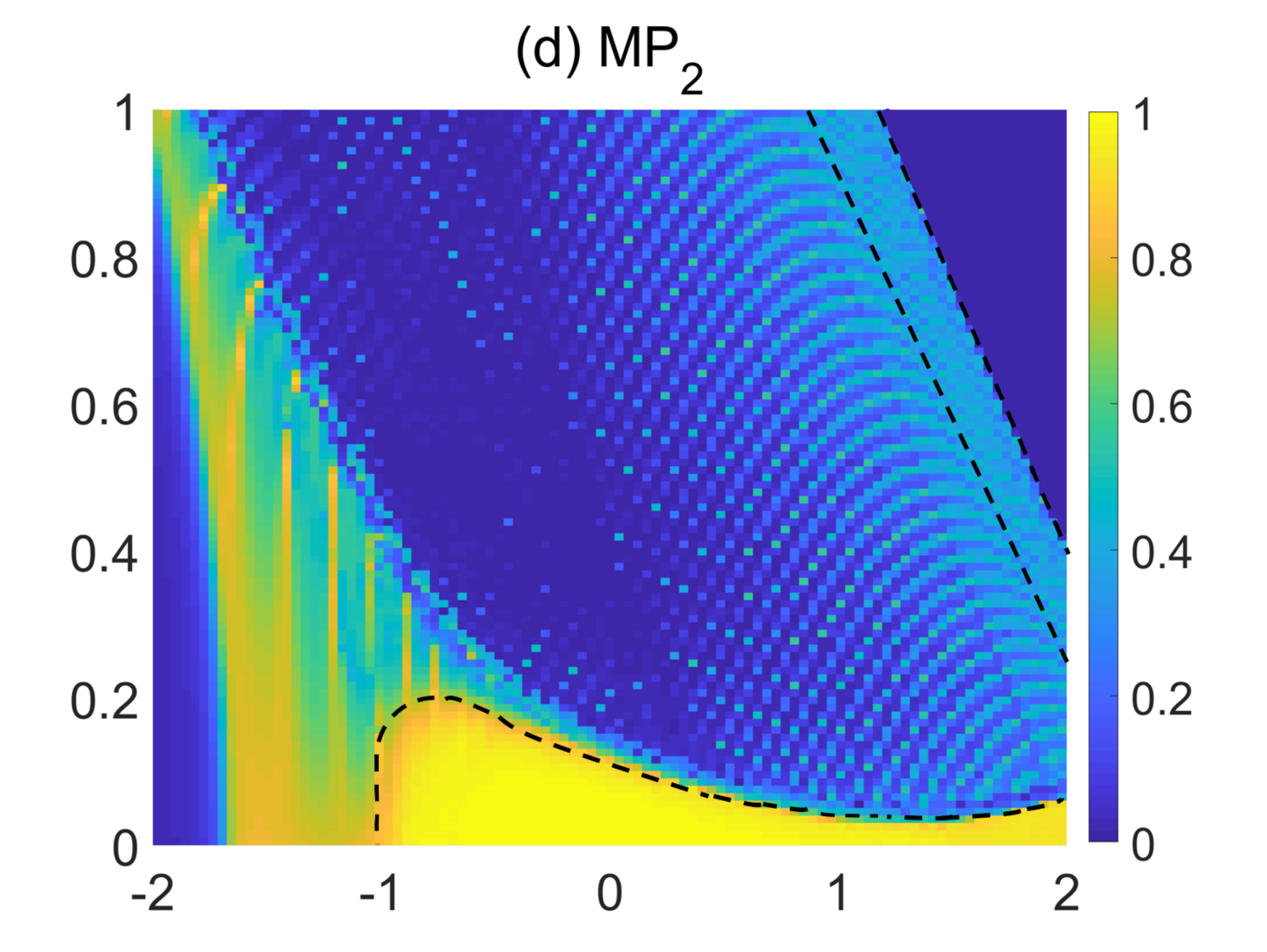}
	\includegraphics[width=0.425\textwidth]{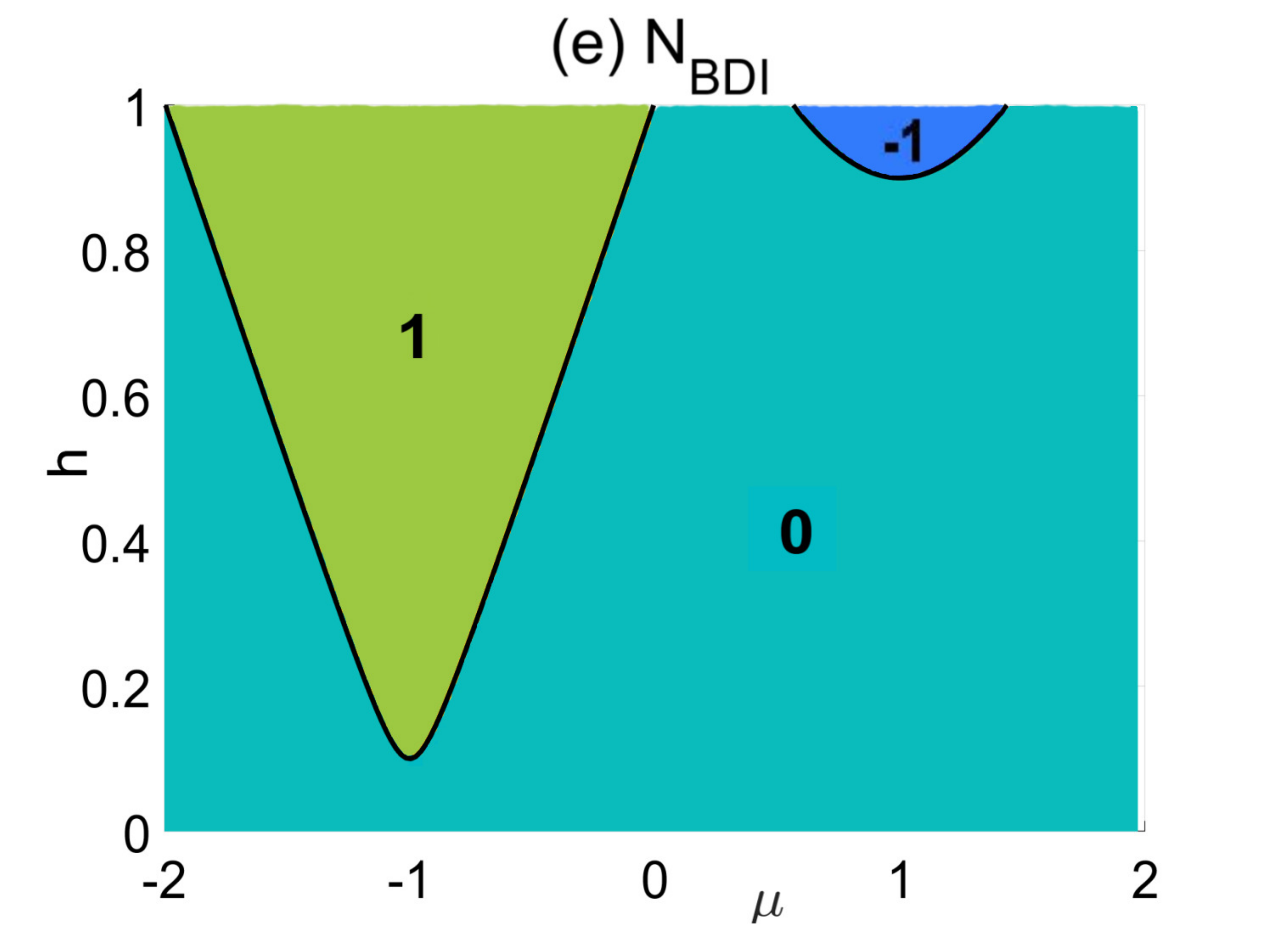}
	\includegraphics[width=0.425\textwidth]{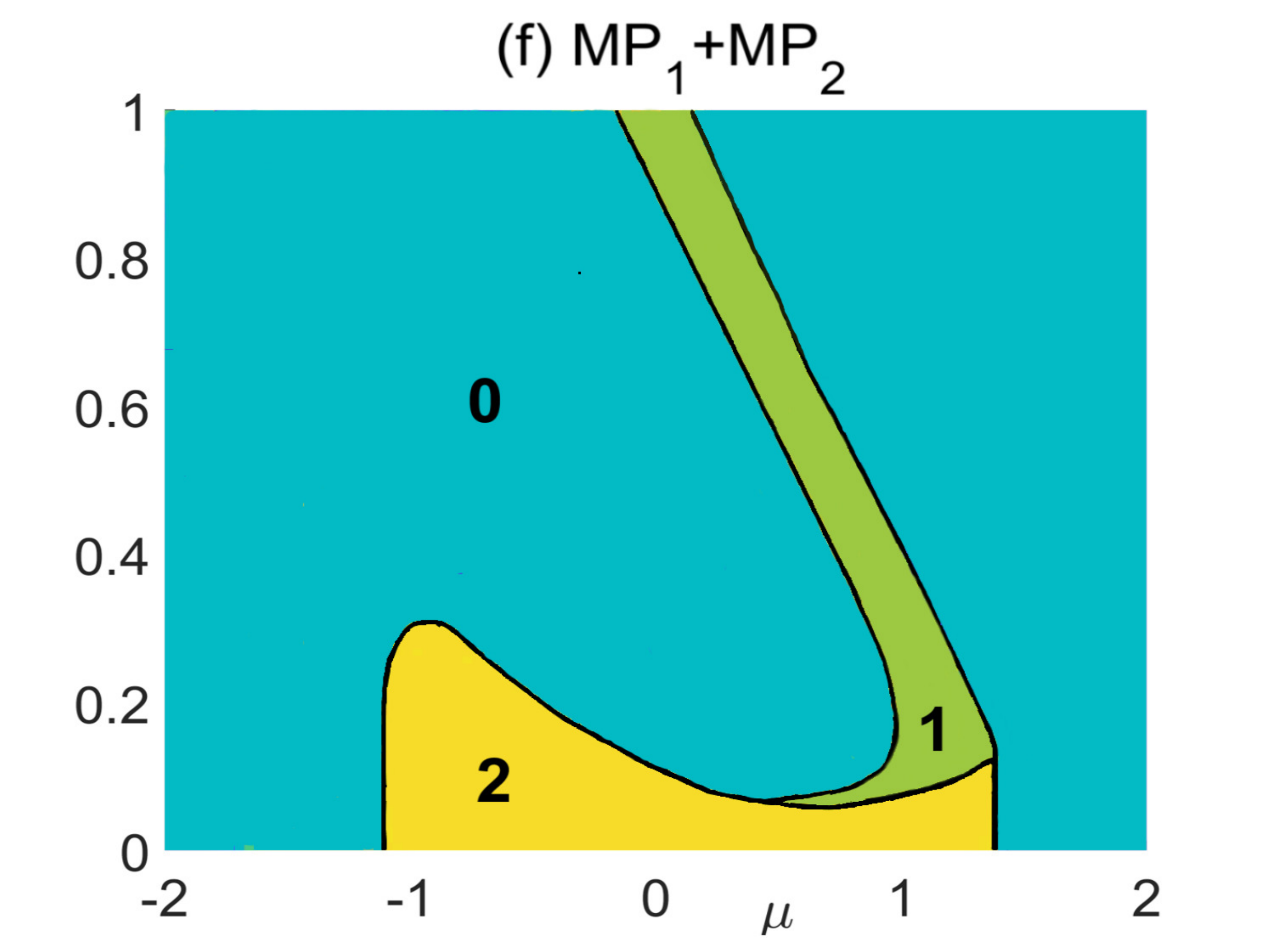}
	\caption{\label{MP_tJV} The effect of Coulomb interactions on the topological phases of the BDI-class wire. $MP_{1,2}$ as functions of chemical potential and magnetic field at $U=10$, $V=0$ (a,b) and $U=10$, $V=0.5$ (c,d) calculated by the DMRG approach for $t-J-V$-model. The dashed curves display the boundaries between different topological phases. The 'A' - 'F' points denote the parameters taken to plot Figures \ref{MP_N}a-\ref{MP_N}c. (e,f) The non-interacting and strongly interacting topological phase diagrams, respectively. The last is obtained at $U=10$, $V=0$ based on the wire length dependencies of Majorana polarizations.}
\end{figure*}
Without the inter-site Coulomb correlations the excitation energies obtained by the $tUV$-DMRG at $U=10$ in Fig. \ref{tUV_vs_tJV}a (see blue and red solid curves) retains the features found at $U=2$ in Fig. \ref{indMDM}a. In particular, there are the left and right MDMs (where $E_{1,2} \approx 0$) around the MSMs (where $E_{1} \approx 0$, $E_{2}\neq0$). The last are realized at $\mu\approx0$ --- $1.2$. Both energies sharply increase at $\mu\approx1.5$ due to the gap between the Hubbard subbands. The appearance of the MDMs and MSMs is proved by the $\mu$ dependence of $MP_{1,2}$ (see blue and red solid curves in the inset of Fig. \ref{tUV_vs_tJV}a). The $tJV$-DMRG data demonstrate qualitatively similar behavior at low $\mu$, $\mu<0$ (see blue and red dashed curves in Fig. \ref{tUV_vs_tJV}a). The differences become stronger at the higher concentrations. Here the splitting of $E_{1}$ and $E_{2}$ is reduced and accompanied by the oscillations of excitation energies. Hence, the MSM region is shorter in comparison with the $tUV$-DMRG results that is confirmed by $MP_{1}<0.9$ at $\mu\approx0$ --- $1$ (see blue dashed curve in the inset of Fig. \ref{tUV_vs_tJV}a). Finally, the area near $\mu = 1.5$ with the MDMs is reduced as well.

It is clearly seen from Figure \ref{tUV_vs_tJV}b that the mentioned partial agreement between the $tUV$- and $tJV$-DMRG data is kept when the inter-site correlations, $V$, are taken into account. In turn, there are two effects the nonzero $V$ leads to. First, it additionally stretches the lower Hubbard subband to the right. In other words, the inter-site interactions effectively increase the on-site energy (see the expression for $\left( A_{\sigma \sigma} \right)_{f,f}$ in \eqref{A_B_int}). Consequently, one has to raise $\mu$ to reach the same concentration level in comparison with the situation of $V=0$. Second, the nonzero $V$ decreases the excitation-spectrum gap much stronger than $U$ even though $V \ll U$ \cite{wieckowski-19}. As a result, the $E_{1,2}$ oscillations occur in the $tUV$-DMRG solution leading to the MSM suppression that is corroborated by the $MP_{1}$ behavior at $\mu\approx0.25$ --- $1.5$ (see blue solid curve in the inset of Fig. \ref{tUV_vs_tJV}b). Nevertheless, the MSMs survive in the area of $\mu\approx1.5$ --- $2.25$. The regions with the of left and right MDMs roughly conserve their widths. Then the $tJV$-DMRG scheme gives shorter areas of the MSMs and right MDMs. Thus, as it was already noticed above the differences between $tUV$- and $tJV$-DMRG results at $V\neq0$ also strengthen when the electron concentration grows. The observed deviations of two DMRG schemes are attributed to the absence of three-center terms \eqref{Ham_3} in the $tJV$-algorithm which is more powerful at higher electron densities. 

\begin{figure*}[htb!]
	\includegraphics[width=0.45\textwidth]{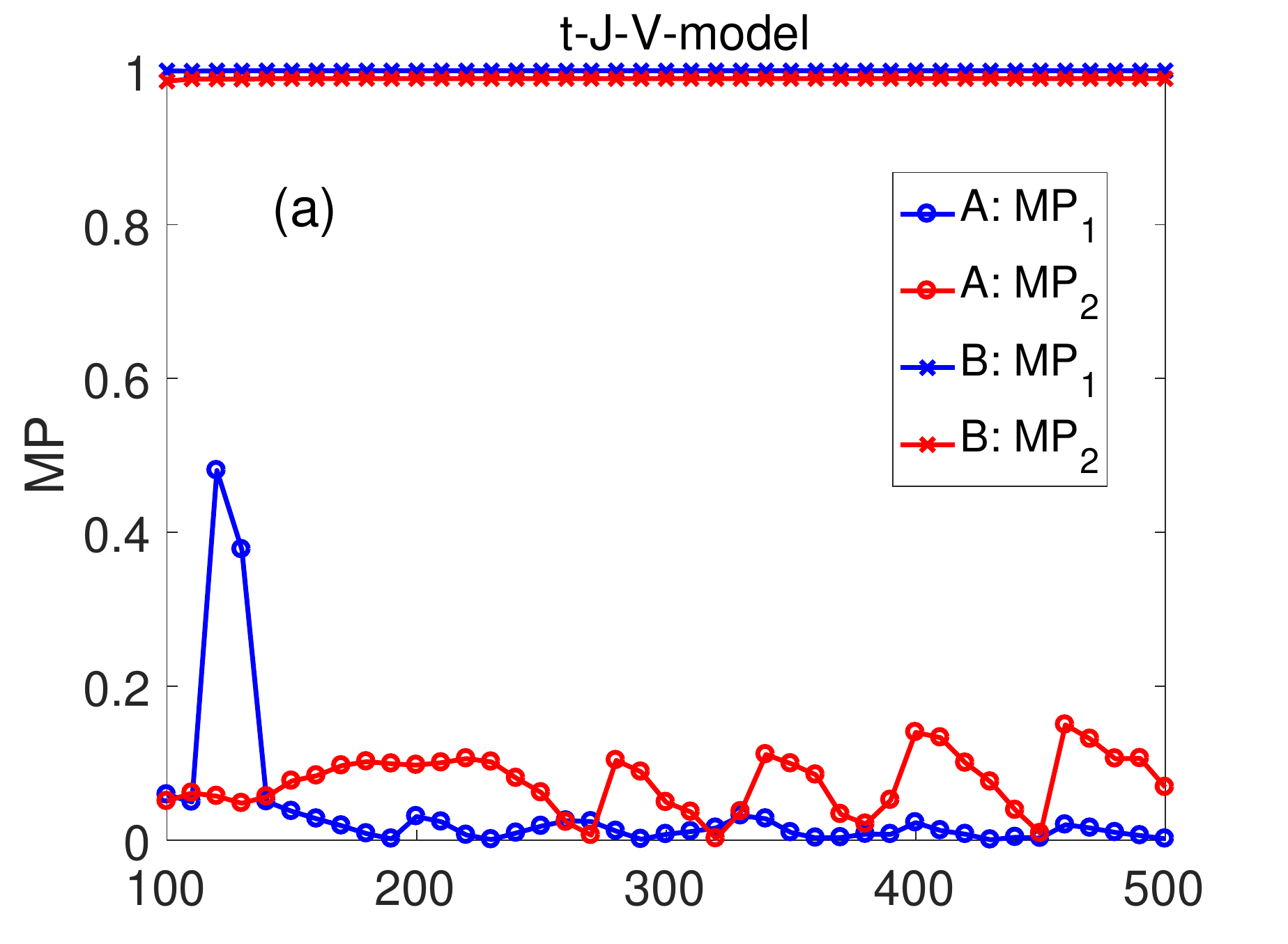}
	\includegraphics[width=0.45\textwidth]{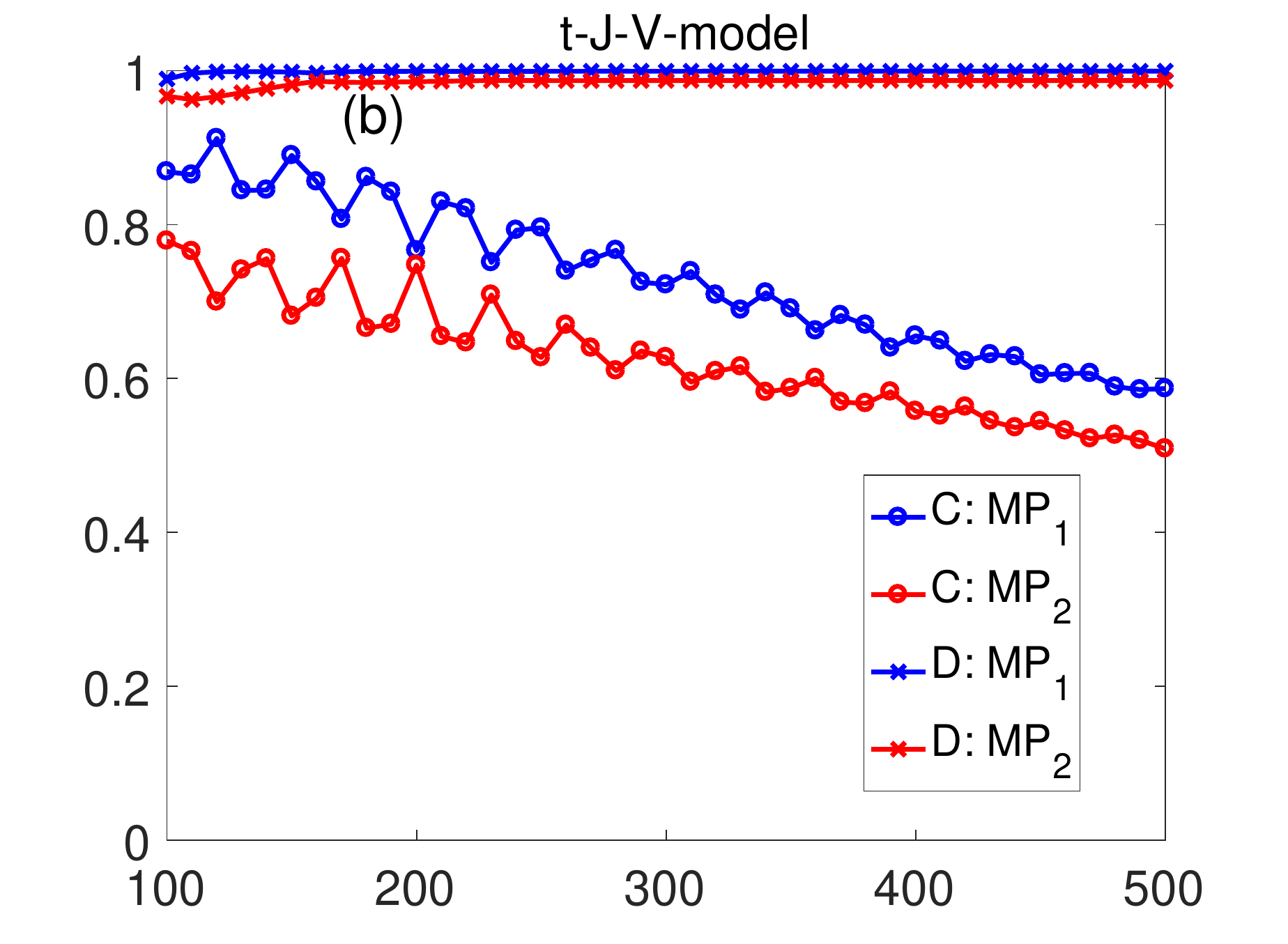}
	\includegraphics[width=0.45\textwidth]{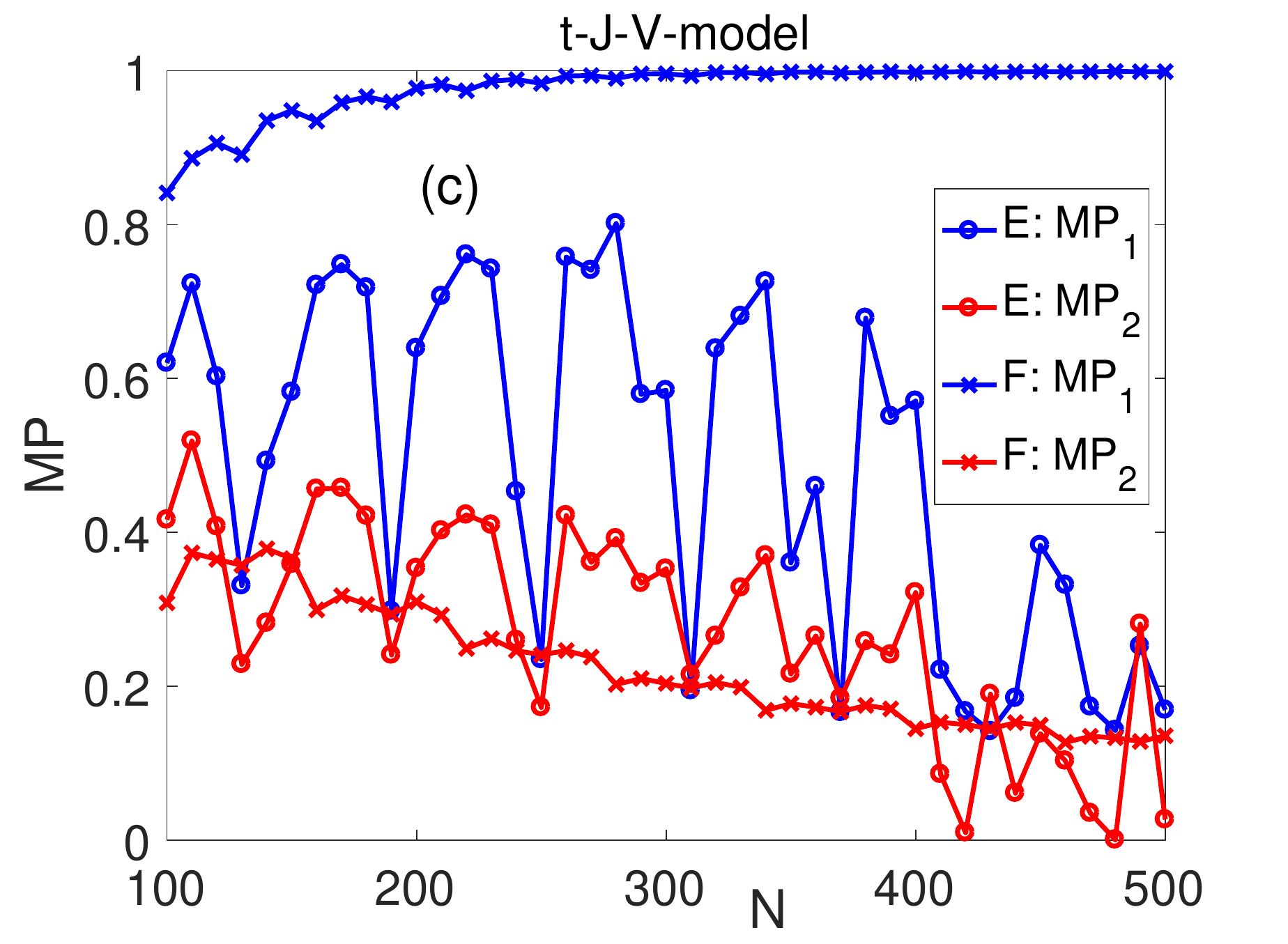}
	\includegraphics[width=0.45\textwidth]{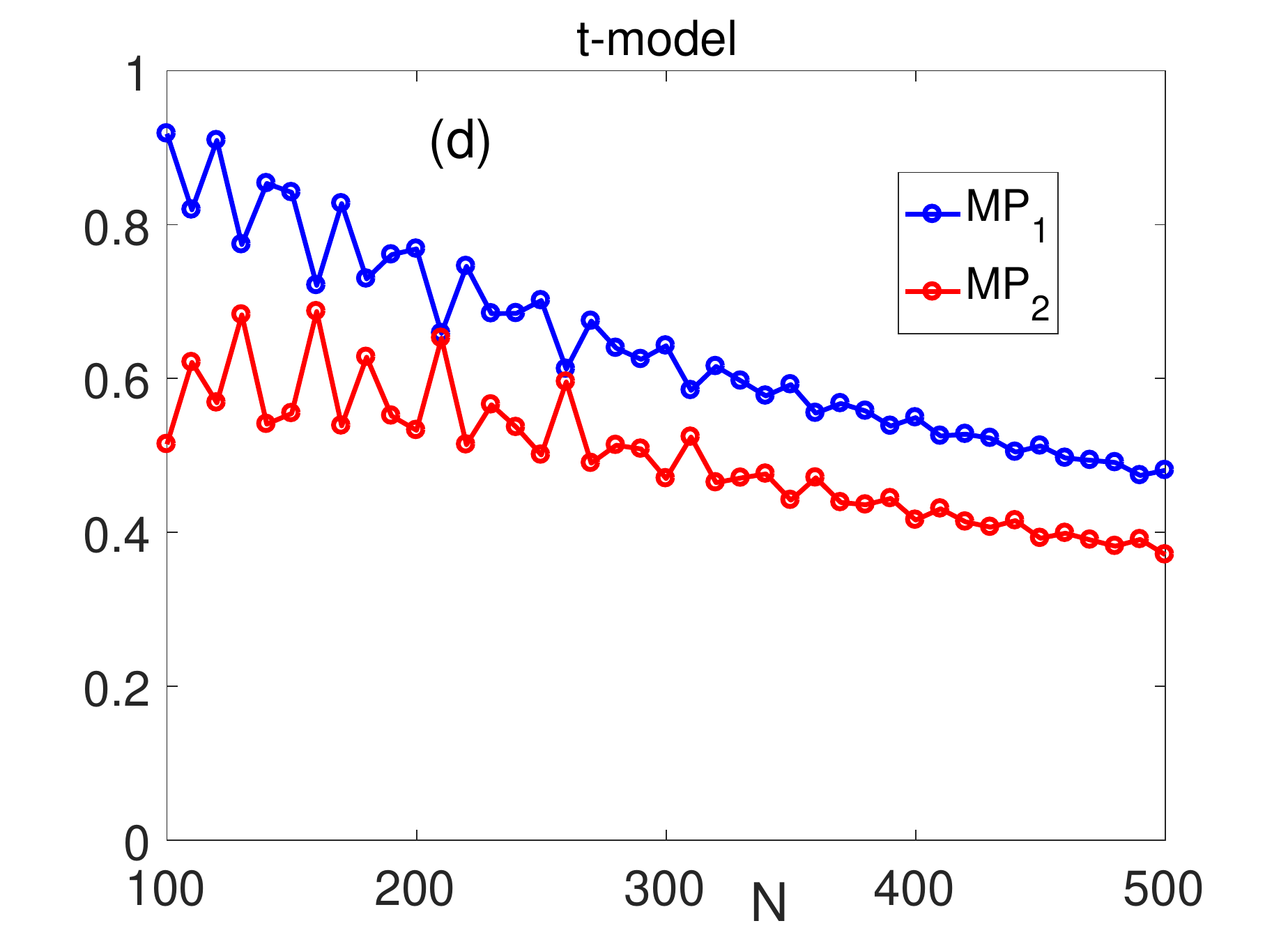}
	\caption{\label{MP_N} The wire length dependence of Majorana polarizations. (a) $MP_{1,2}\left(N\right)$ correspond to 'A' (circles) and 'B' (crosses) points in Fig. \ref{MP_tJV}a; (b) $MP_{1,2}\left(N\right)$ correspond to 'C' (circles) and 'D' (crosses) points in Fig. \ref{MP_tJV}a; (c) $MP_{1,2}\left(N\right)$ correspond to 'E' (circles) and 'F' (crosses) points in Fig. \ref{MP_tJV}a; (d) $MP_{1,2}\left(N\right)$ correspond to the white point in Fig. \ref{MP_t}.}
\end{figure*}
Since the $tJV$-DMRG algorithm also yields both MSMs and MDMs now we turn to the numerical simulations based on the $t-J-V$-model. In Figure \ref{MP_tJV} color plots of $MP_{1,2}$ versus $\mu$ and $h$ are presented. Starting from the $N_{BDI}$ map found in the no-interaction limit and shown in Fig. \ref{N_BDI_map}a, one can trace the evolution of topological phases induced by strong electron correlations in the lower Hubbard subband (related to the left parabola). If the inter-site Coulomb correlations are omitted the left parabola is cut off at $\mu\approx1.5$ by the Mott-Hubbard gap (see Figs. \ref{MP_tJV}a, \ref{MP_tJV}b). The MSMs are largely suppressed inside the parabola persisting only at its right edge in the strip-shaped region where their norm close to 1. Note that taking into account the above-described comparison between the $tUV$- and $tJV$-results we expect this area being much wider in the former DMRG approach. Below the parabola the MDMs emerge as $MP_{1,2}\to1$. Note that such a behavior is similar to the situation shown in Fig. \ref{N_BDI_map}b where, in opposite, $|\Delta|<2|\Delta_{1}|$. The MDM norms vary roughly from 0.5 to 0.8.

When the inter-site electron-electron interactions are turned on and $V \ll U$ the mentioned effects still exist (see Figs. \ref{MP_tJV}c, \ref{MP_tJV}d). Meanwhile, the parabola is additionally stretched out to the right (the right edge is not shown entirely) and its bottom is shifted down. The MSM strip becomes narrower and the MDM-region size is decreased. 

To obtain the boundaries of different topological phases showed by dashed curves in Figs. \ref{MP_tJV} we employ two criteria: the entanglement spectrum degeneracy, $d$, and the length dependencies of MPs, $MP_{1,2}\left(N\right)$. They demonstrate good agreement with each other. The dashed curves divide the areas with different $d$ (to calculate $d$ we took $N=1400$). Additionally, the $MP_{1,2}\left(N\right)$ behave differently in these regions. To show it three pairs of points in the $\left(\mu,h\right)$-space are considered (see Fig. \ref{MP_tJV}a). The MPs as functions of the wire length for each pair are displayed in Figures \ref{MP_N}a-\ref{MP_N}c, respectively. It is seen that the points 'A', 'C', and 'E' are located in the trivial phase with $d = 1$. In turn, the 'B' and 'D' points are in the topological phase with $d = 4$ where the MDMs appear. Finally, The 'F' point is in other topological phase with $d = 2$ corresponding to the phase with the MSMs. 

The wire length dependencies of MPs can be used to approximately receive the phase diagram in case of the infinitely long structure. As it was discussed in Sec. \ref{sec2} if $N \to \infty$, $U=V=0$ there is a clear correspondence between $N_{BDI}$ and $MP_{1}+MP_{2}$ allowing to directly compare the non-interacting and strongly interacting phase diagrams. They are shown in Figs. \ref{MP_tJV}e and \ref{MP_tJV}f, respectively. Here one can explicitly see the significant transformation of the left parabola with the MSMs inside and the induction of MDM region by the on-site Coulomb repulsion.     

\begin{figure*}[htb!]
	\includegraphics[width=0.4\textwidth]{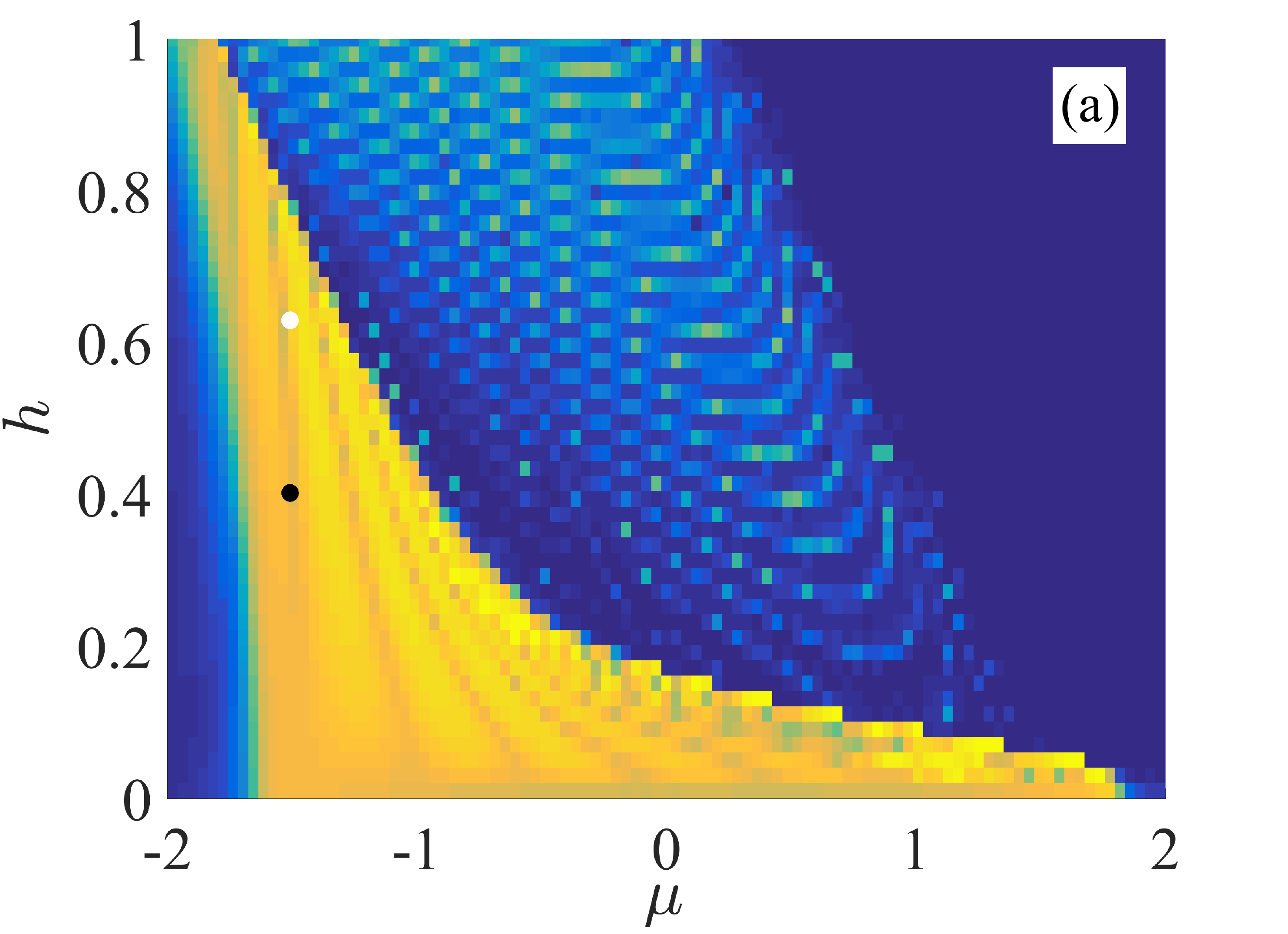}
	\includegraphics[width=0.4\textwidth]{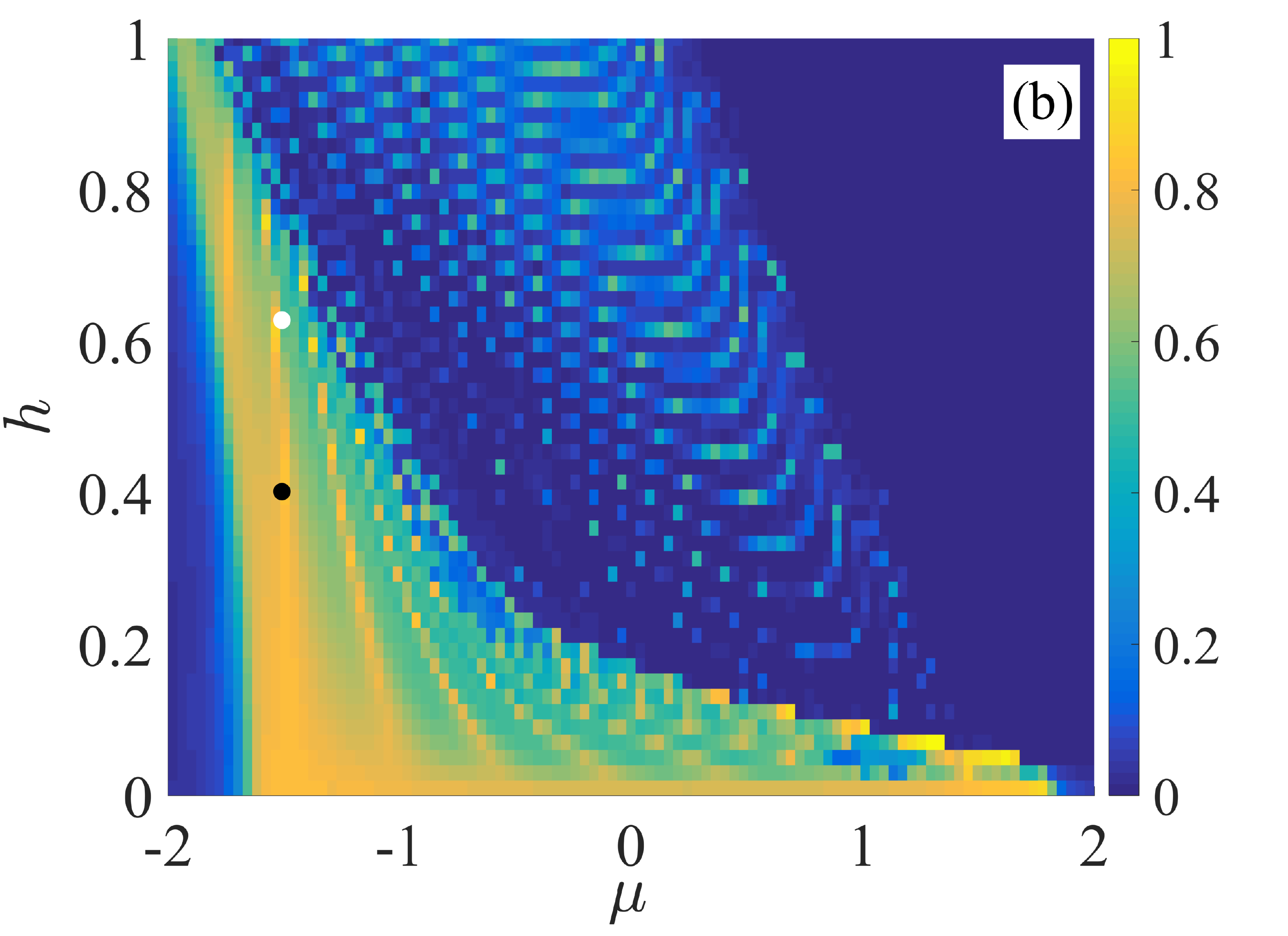}
	\caption{\label{MP_t} The influence of infinite on-site repulsion on edge states of the BDI-class wire. The chemical potential-magnetic field diagrams of the $MP_{1}$ (a) and $MP_{2}$ (b) obtained by the DMRG scheme for the $t$-model at $V=0$. In this limit there are no boundaries of the topological phases on the maps since only the trivial phase with $d=1$ is possible. The black and white points denote the parameters for which the Majorana-type spatial distributions of the first two modes are plotted in Figs. \ref{wz_t}a-\ref{wz_t}d.}
\end{figure*}
Next, it is useful to consider the limit of $U \to \infty$. The $MP_{1,2}$ maps in the variables $\mu$ and $h$ calculated by the $t$-DMRG algorithm are shown in Figures \ref{MP_t}a and \ref{MP_t}b, respectively. As it was mentioned before the results do not depend on $\Delta$ as $\Delta/U \to 0$. One can see that the MSMs, surviving at high and finite $U$ close to the right boundary of the left-parabola region characterized by strong spin polarization, are destroyed if $U \to \infty$. The MDMs in the parametric area under the left parabola are suppressed as the $MP_{2}$ is far from 1. Nevertheless, the single edge modes persist in this region as it is displayed in Fig. \ref{MP_t}a. The norms of such states exceed 0.7 in the wide range of parameters.

\begin{figure}[htbp]
	\includegraphics[width=0.22\textwidth]{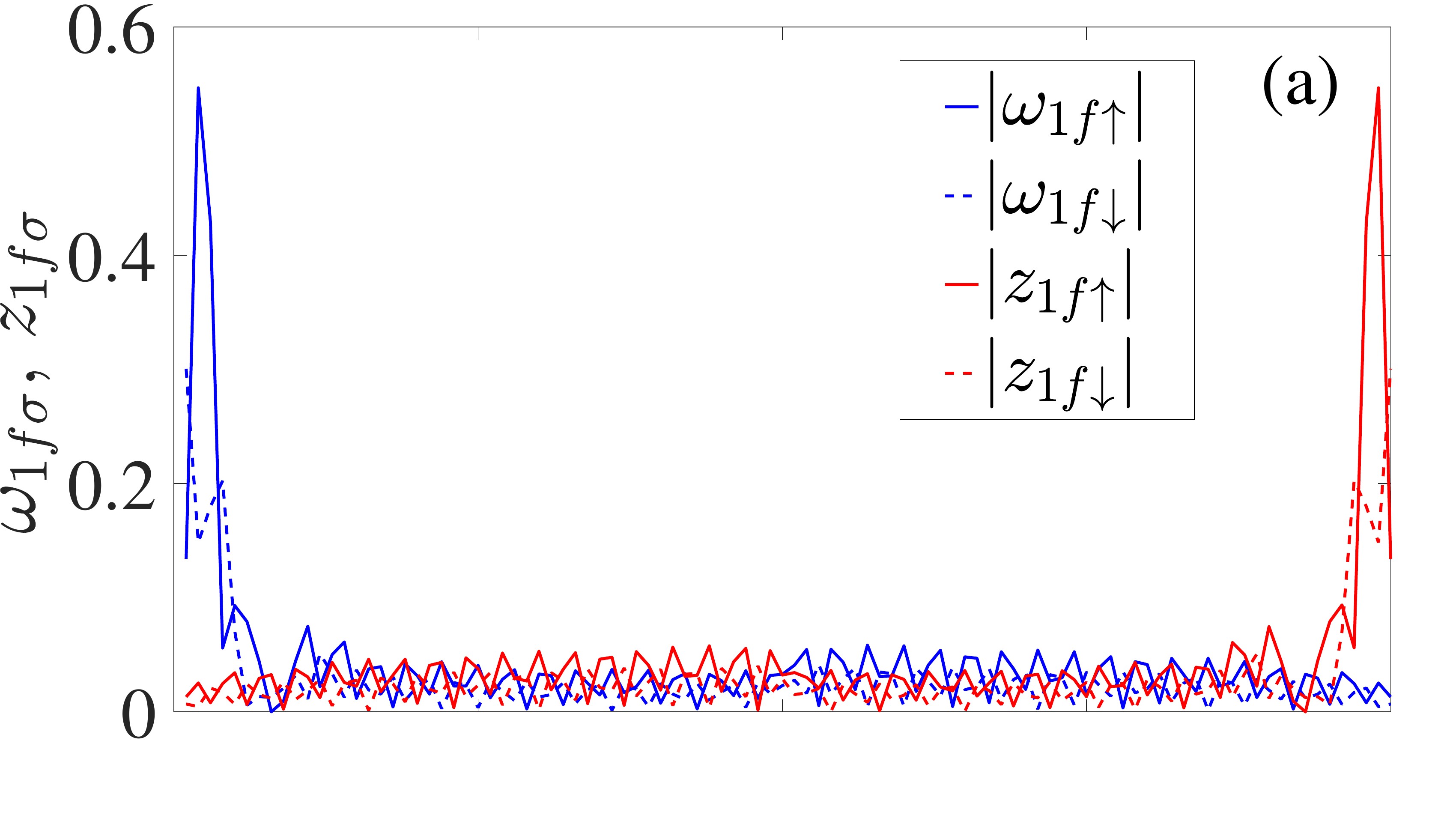}
	\includegraphics[width=0.22\textwidth]{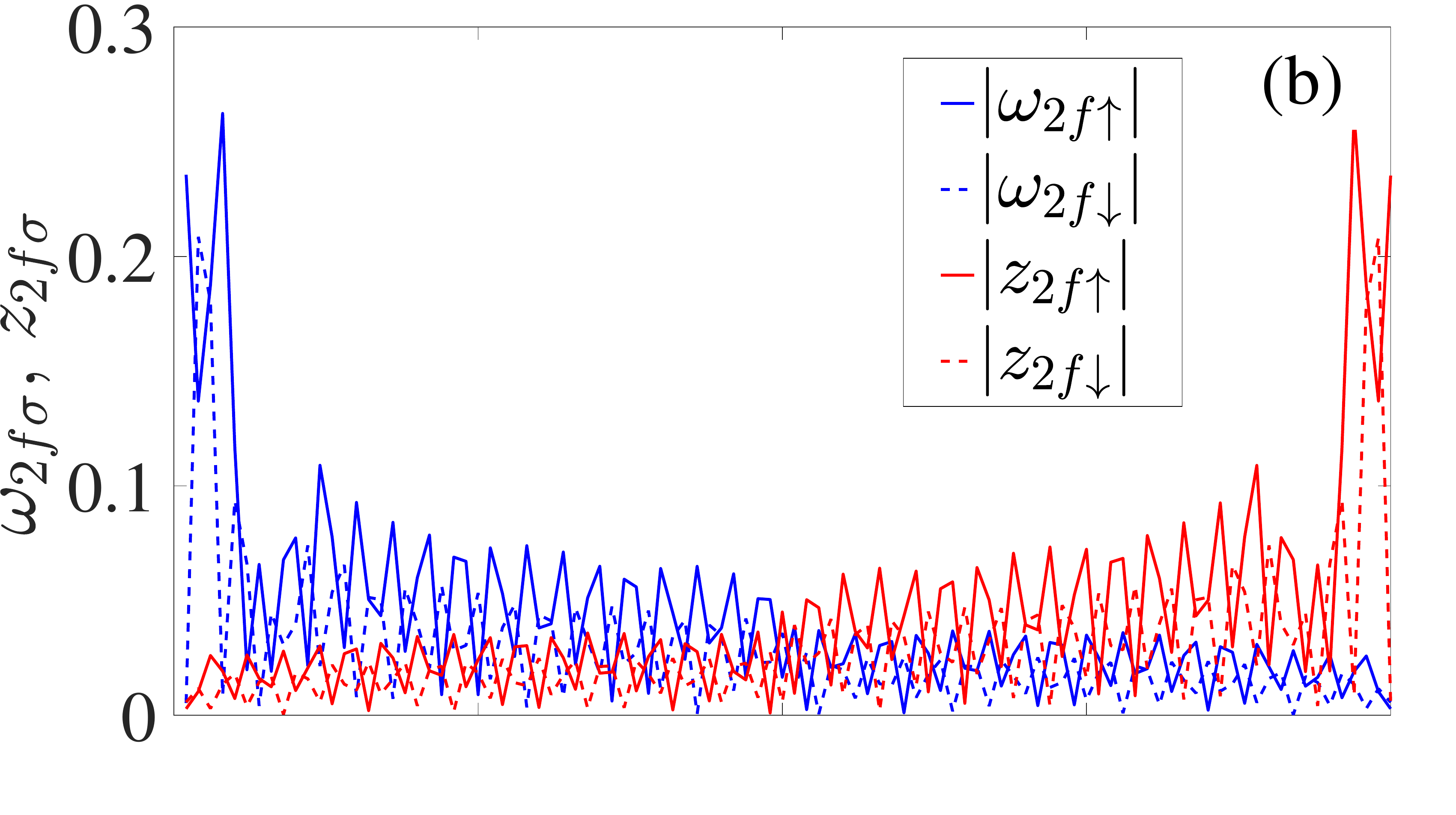}
	\\
	\includegraphics[width=0.22\textwidth]{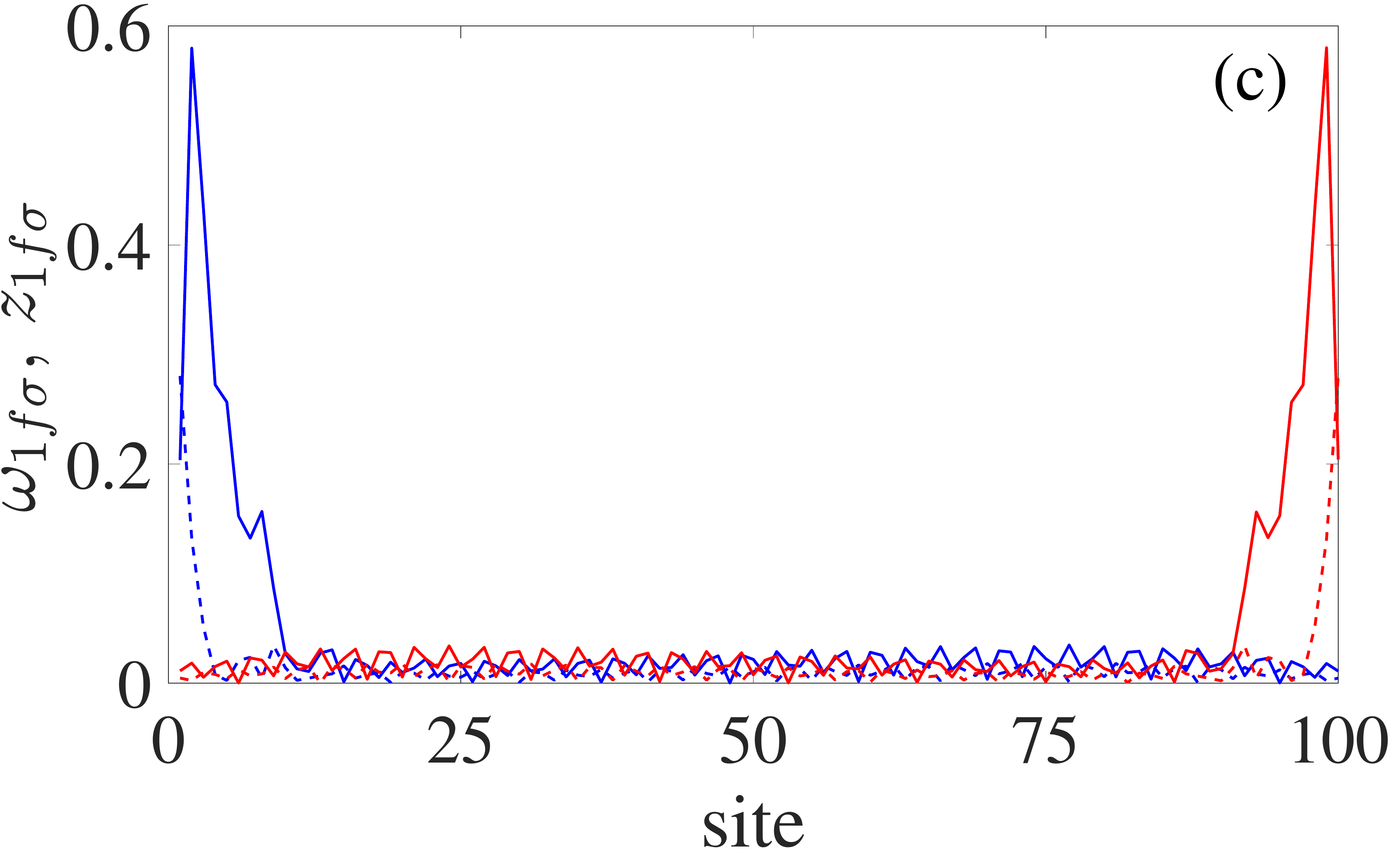}
	\includegraphics[width=0.22\textwidth]{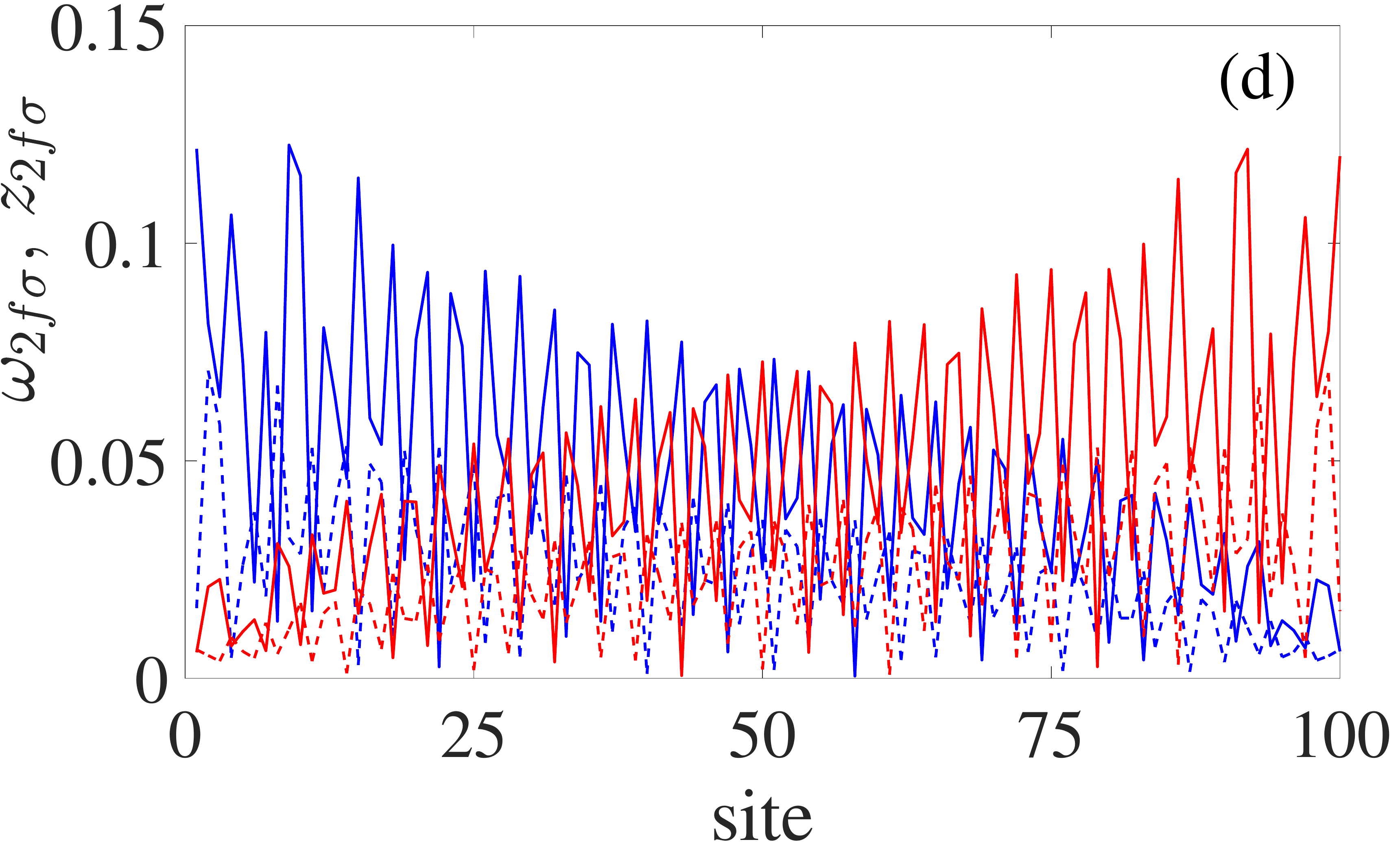}
	\caption{\label{wz_t} The spatial distributions of the first (a), (c) and  second (b), (d) excitations in the $t$-model. The top (bottom) plots correspond to the black (white) point on Figs. \ref{MP_t}.}.
\end{figure}
We denote the parameters $h=0.4$, $\mu=-1.5$ corresponding to the high value of $MP_2$ by the black point in Figs. \ref{MP_t}. For these parameters the spatial distributions of first two excitations, $\omega_{jf \sigma}$ and $z_{jf \sigma}$ ($j=1,2$), are shown in Figures \ref{wz_t}a and \ref{wz_t}b. The MP and norm values are $MP_1 \approx 0.82$, $MP_2 \approx 0.8$ and $norm_1 \approx 0.87$, $norm_2 \approx 0.52$, respectively. For comparison the same spatial distributions at $h=0.63$, $\mu=-1.5$ (see the white point in Figs. \ref{MP_t}) are provided in Figures \ref{wz_t}c , \ref{wz_t}d. In this situation the MP and norm values are $MP_1 \approx 0.94$, $MP_2 \approx 0.54$ and $norm_1 \approx 0.92$, $norm_2 \approx 0.47$, respectively. 

It is seen from Figs. \ref{wz_t} that, while the second excitation demonstrates the pronounced overlapping behavior typical for bulk state (see Figs. \ref{wz_t}b and \ref{wz_t}d), the first one possesses the features typical for the MSM even in the extreme case of $U \to \infty$ as the well-defined maxima of the distributions near both edges occur (see Figs. \ref{wz_t}a and \ref{wz_t}c). However, the length dependencies of MPs reveal the instability of observed edge states if the strength of on-site correlations is infinite. Figure \ref{MP_N}d shows that both $MP_{1}$ and $MP_{2}$ reduce for longer wires. These data are corroborated by the calculations of entanglement spectrum degeneracy which yield $d=1$ for all values of $\mu$ and $h$. 

Thus, one can observe the following modification of the phase diagram for $|\Delta|>2|\Delta_{1}|$: 1) in the no-interaction case there are the MSM and trivial phases (see Fig. \ref{N_BDI_map}a); 2) in the strongly correlated regime the MSM, MDM and trivial phases can be realized (see Figs. \ref{MP_tJV}); 3) in the limit of infinite on-site repulsion the wire is in the trivial phase.  

It is essential to emphasize that the above-discussed DMRG results inherently involve the contribution from zoo of different many-body processes. To show their role more prominently one can analytically consider the $t$-model \eqref{Ham_t} in the simplest Hubbard-I approximation. The corresponding details are given in Appendix \ref{apxB}.  
By solving the system of equations for the Zubarev's Green functions (see Eq. \eqref{Sys_GF}) the quasiparticle operator in the strongly correlated limit is given by
\begin{eqnarray}
\label{DEFUV2} \tilde{\alpha}_{j} & = & \frac{1}{2} \sum_{f=1}^{N} \sum_{\sigma} \left(\tilde{w}_{jf \sigma}
\tilde{\gamma}_{A f \sigma} + i\tilde{z}_{jf \sigma} \tilde{\gamma}_{B f \sigma}\right),
\end{eqnarray}
where the Majorana operators in the atomic representation are expressed as
\begin{eqnarray}
\label{gammaX}
\tilde{\gamma}_{A f \sigma} = X_f^{0 \sigma} + X_f^{\sigma 0},~~~
\tilde{\gamma}_{B f \sigma} = i\left(X_f^{\sigma 0} -  X_f^{0 \sigma}\right).
\end{eqnarray}
The MP in the Hubbard-I approximation is defined similarly to \eqref{MP_def} substituting the $\tilde{w}_{jf \sigma}$, $\tilde{z}_{jf \sigma}$ coefficients.
Initially, such an approach was used to analyze the coexistence phase of superconductivity and noncollinear magnetic ordering in the strongly correlated limit for the quasi-1D system \cite{valkov-19b}.

\begin{figure}[htbp]
	\includegraphics[width=0.4\textwidth]{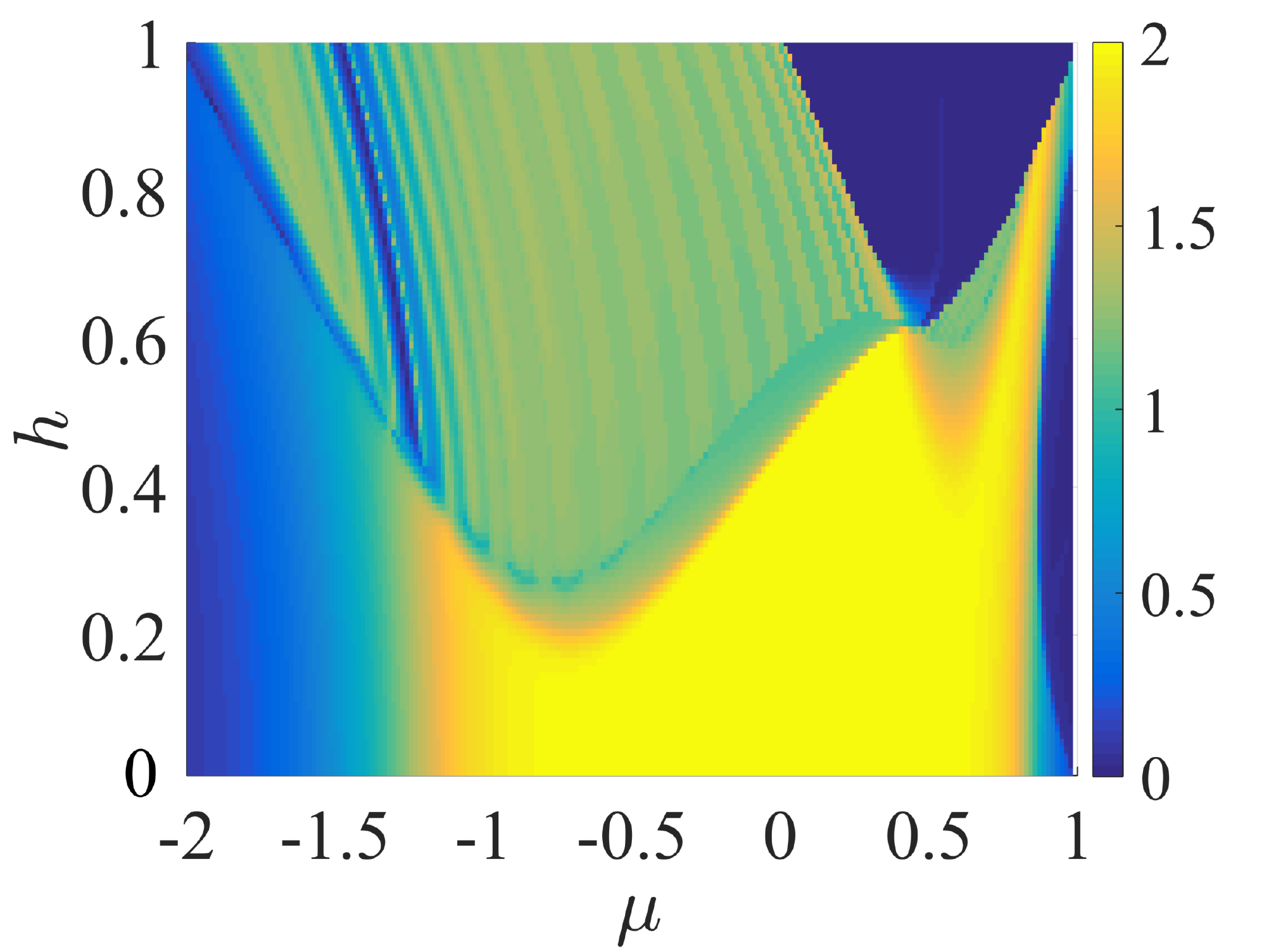}
	\caption{\label{MP_t_HubI} The total Majorana polarization of two excitations, $MP_{1}+MP_{2}$, as a function of $\mu$ and $h$ calculated by means of the Hubbard-I approximation in the $t$-model at $V=0$.}
\end{figure}
The color plot of total MP, $MP_{1}+MP_{2}$, versus $\mu$ and $h$ is displayed in Figure \ref{MP_t_HubI}. It is seen that in the wide range of parameters the total MP is equal to 2 indicating the MDM emergence. In the left-parabola region the inequalities $1<MP_{1}+MP_{2}\lesssim 1.5$ mainly hold. Here the MSMs are well defined as $MP_{1}\approx1$, $MP_{2}\lesssim 0.5$. Apparently, the Hubbard-I approximation leads to the results quantitatively different from the DMRG simulations (compare Figs. \ref{MP_t_HubI} and Fig. \ref{MP_t}). In particular, the former gives rise to the shift in the chemical potential in comparison with the DMRG data due to the differences in the energy spectrum and Fermi momentum. In other words, the same value of $\mu$ in the Hubbard-I approximation and DMRG approach corresponds to different electron densities, e.g. in the absence of magnetic field the filling $\left\langle n_f \right\rangle = 1$ is achieved at $\mu \approx 1\left(2\right)$ in the Hubbard-I (DMRG) approach. Moreover, the MSMs and MDMs persist in the Hubbard-I approximation. Whereas the DMRG yields to the complete suppression of the MSMs and MDMs. 

The reason of highlighted discrepancies is rather obvious: the simplest Hubbard-I approximation does not take into account contributions from spin and charge fluctuations. It means that the quasiparticles related to the Hubbard fermions are well defined. That is why the quasiparticle norm is always equal to 1. On the other hand, spin and charge fluctuations (for example, in the one-loop approximation) will modify the real part of energy spectrum, but also might lead to the damping effect (since the imaginary part of energy spectrum becomes nonzero). We suppose that such corrections will improve the agreement with the $t$-DMRG results and the nonzero quasiparticle damping will give rise to the decrease of its norm which resembles the effect occurring in the DMRG method.

Nevertheless, the Hubbard-I approximation is meaningful from fundamental point of view since that is the first step allowing to define the Majorana fermions in the strongly correlated limit (see \eqref{DEFUV2}). Thus, it opens a route to analytically describe the influence of spin and charge fluctuations on the MMs in this regime. But consideration of similar effects is beyond the scope of current work and will be analyzed in further studies.

\section{\label{sec4} Caloric functions}

Finally, we would like to discuss the possibility of experimental detection of topological phases in the strongly correlated system \eqref{Ham_Fermi} employing caloric functions. There are a few reasons to use this tool here. Firstly, these effects as a way to identify topological phases are studied insufficiently in comparison with the transport properties of the SC wires. Secondly, a series of caloric anomalies indicating the nontrivial-phase formation in the D-class wire persists in weak Coulomb interactions when the GMF approach is valid \cite{valkov-17}. The caloric anomalies in these structures are related to the quantum phase transitions \cite{zhu-03,garst-05} which, in turn, are caused by the hybridization of MMs localized at the opposite edges. It is clearly seen from the above numerical data that the strong electron correlations in the BDI-class wire enhance this effect due to the decrease of bulk gap. Thus, one can expect at least to observe similar features in our system.

The MCE and ECE are defined by the change of system temperature, $T$, under the adiabatic change of magnetic field or chemical potential, respectively,
\begin{eqnarray}
\label{MCE_ECE}
&~&-\frac{1}{T}{\left( {\frac{{\partial T}}{{\partial h}}} \right)_{S,\mu }} = \left(\frac{\partial\langle \hat{M} \rangle /\partial T}{C(T)}\right)_{\mu,h};~~~\hat{M} = \sum_{f=1;\sigma}^{N} \sigma a^{+}_{f\sigma}a_{f\sigma};
\nonumber\\
&~&-\frac{1}{T}{\left( {\frac{{\partial T}}{{\partial \mu}}} \right)_{S,h }} = \left(\frac{\partial\langle \hat{N} \rangle /\partial T}{C(T)}\right)_{\mu,h};~~~\hat{N} = \sum_{f=1;\sigma}^{N} a^{+}_{f\sigma}a_{f\sigma};\nonumber\\
\end{eqnarray}
where $C\left(T\right)$ - a specific heat of the system. Using the scaling theory it was shown that the derivatives \eqref{MCE_ECE} have to diverge in quantum critical points at low temperatures  \cite{zhu-03,garst-05}. In the vicinity of quantum critical points these quantities have different sign. The last follows from the definition \eqref{MCE_ECE} since $\partial\langle \hat{M} \rangle /\partial T$ and $\partial\langle \hat{N} \rangle /\partial T$ must have opposite signs in the left and right neighborhood of quantum critical point. It is demonstrated below that the described behavior should take place if either ground or excited state is changed. 

\begin{figure}[htbp]
	\includegraphics[width=0.5\textwidth]{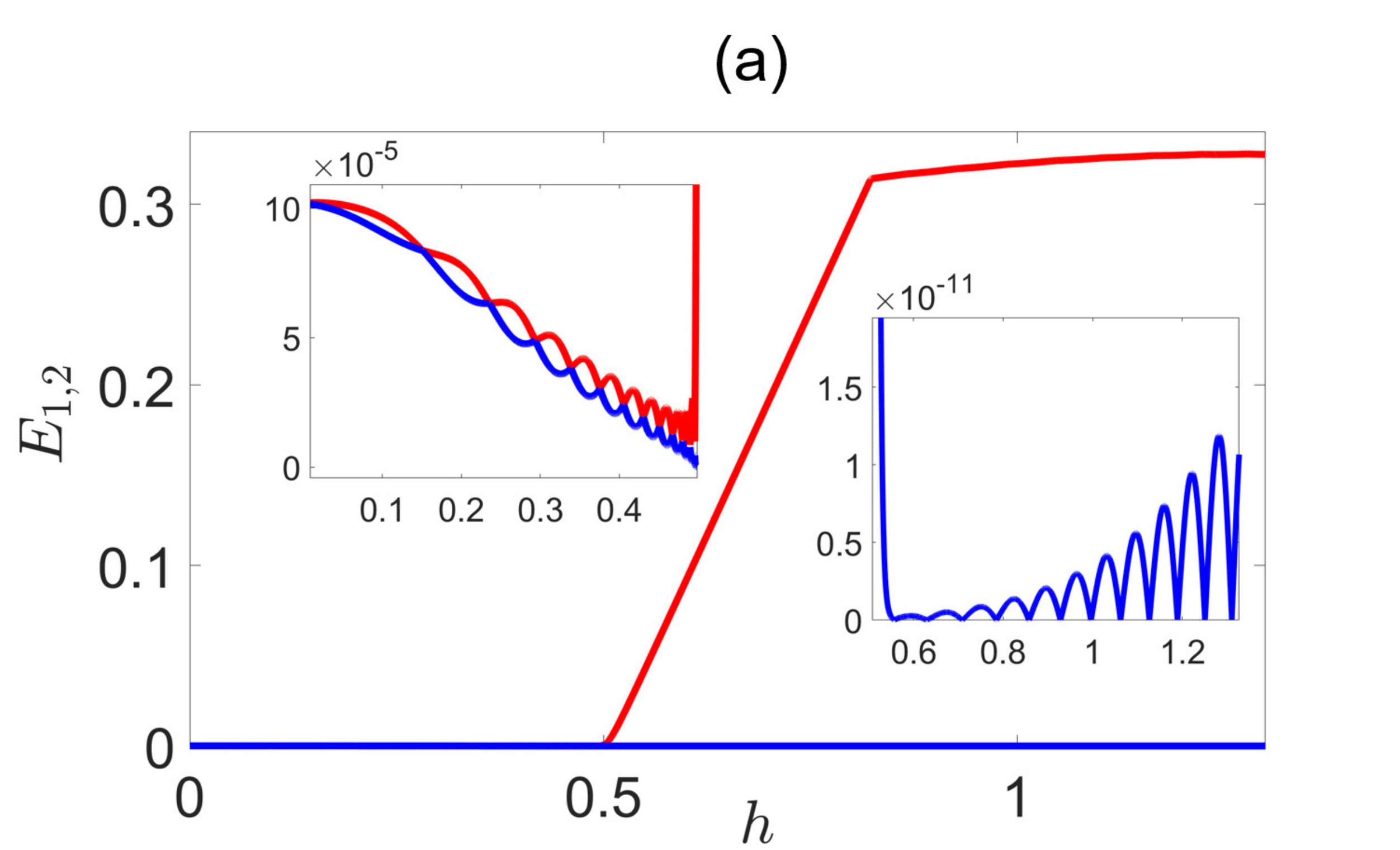}
	\includegraphics[width=0.5\textwidth]{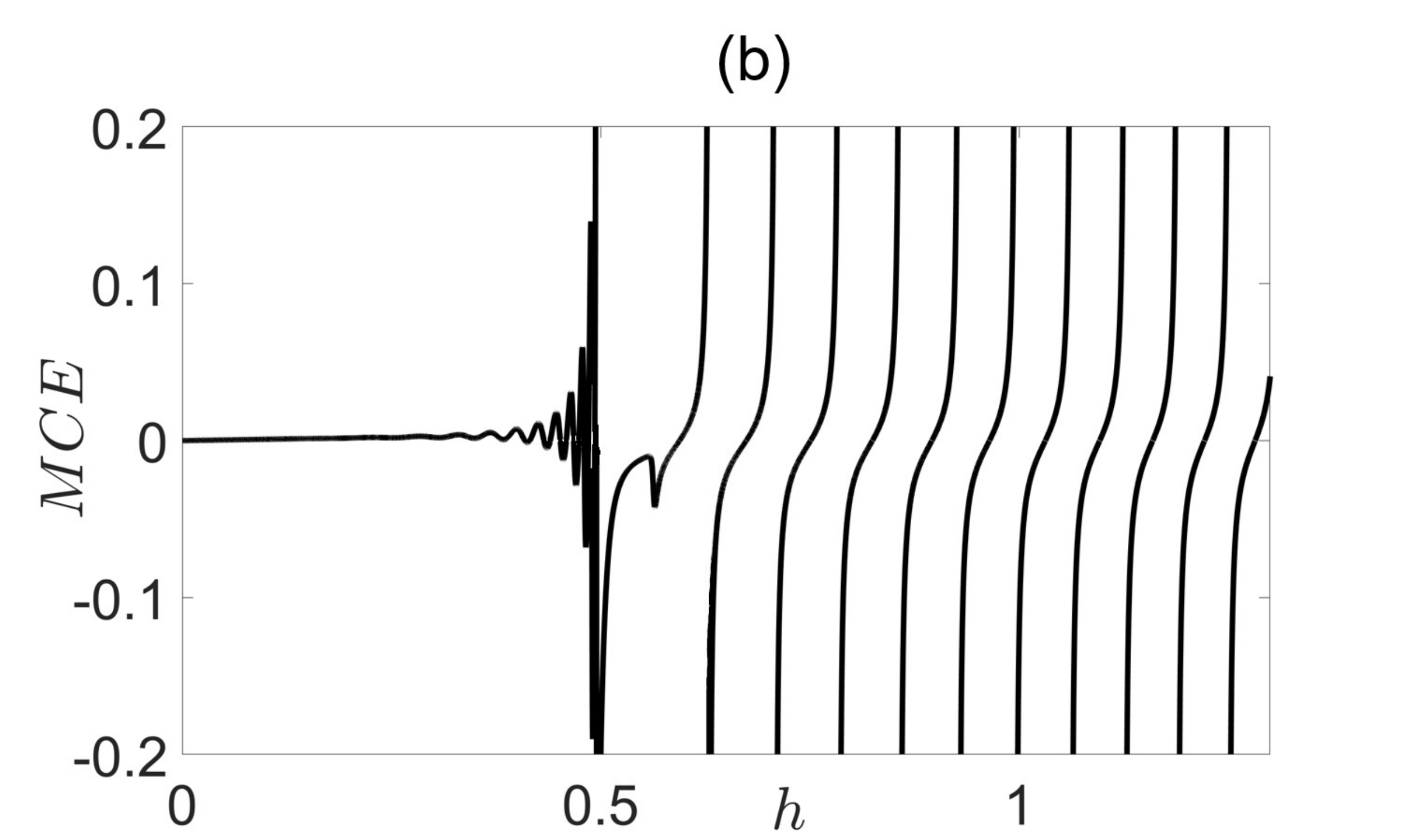}
	\caption{\label{MCE_U0} The magnetic-field dependencies of $E_{1,2}$ (a) and MCE (b) for $\mu=-1.5$. Insets of (a): left and right plots circumstantially depict $E_{1,2}$ in regions of $h$ where the Majorana double and single modes occur, respectively. Parameters: $\Delta=-0.3$, $\Delta_{1}=0.2$, $\alpha=1.5$, $U=0$, $V=0$, $T=10^{-3}$.}
\end{figure}
In the case of non-interacting or weakly interacting fermions the MCE and ECE can be expressed via the $u,v$ Bogoliubov coefficients \cite{valkov-17},
\begin{eqnarray} \label{MCE2}
&~&\partial\langle {N} \rangle /\partial T
=\frac{1}{2T^{2}}\sum_{j=1}^{2N} A_{j}E_{j}f(E_{j})\left( 1- f(E_{j}) \right);\nonumber\\
&~&\partial\langle {M} \rangle /\partial T =\frac{1}{2T^{2}}\sum_{j=1}^{2N} B_{j}E_{j}f(E_{j})\left( 1- f(E_{j}) \right);\nonumber\\
&~&C(T) = \frac{1}{T^{2}}\sum_{j=1}^{2N} E^{2}_{j}f(E_{j})\left( 1- f(E_{j}) \right);
\end{eqnarray}
where
\begin{eqnarray} \label{MCE4}
	&~&A_{j} = \sum_{f=1,{\sigma}}^{N}\left( {|u_{jf\sigma}|^2-|v_{jf\sigma}|^{2}} \right); \nonumber\\
	&~&B_{j} = \sum_{f=1,{\sigma}}^{N}\eta_{\sigma}\left( {|u_{jf\sigma}|^2-|v_{jf\sigma}|^{2}} \right).
\end{eqnarray}
Here $f\left({E_j}/{T}\right)$ is the Fermi-Dirac function. It is seen from \eqref{MCE_ECE}-\eqref{MCE4} that for a single edge state with the energy in the gap $E_{1}\lesssim T \ll E_{j'}$ ($j'>1$) the divergences of caloric effects occur if $E_{1}=0$. Such a situation realizes in the D-class wires for which the oscillations of ground-state fermionic parity occur. Next, if there are a few edge excitations such that $E_{j}\lesssim T \ll E_{j'}$ ($j'>j$) and the corresponding many-particle excited states are changed but not the ground one that the mentioned anomalies of MCE and ECE are not observed. 

Thus, the following relation between the behavior of caloric functions and energy spectrum of the system takes place: the MCE and ECE change signs under quantum transitions and diverge (not diverge) in the transition points if the ground (excited) state is changed. Both scenarios are able to appear in the BDI-class wire. In the parametric region where the MSMs emerge the cascade of transitions accompanied by the switches of ground-state fermionic parity occurs. In the MDM area the set of quantum transitions realizes as well. However, the ground state remains the same but the multiple replacement of many-particle states belonging to the dual-parity sector of the Hilbert space emerges.

The described effects are displayed in Figure \ref{MCE_U0}. The plot \ref{MCE_U0}a includes the magnetic-field dependencies of the first two elementary excitations $E_{1,2}$ without Coulomb interactions. In Fig. \ref{MCE_U0}b the MCE as a function of magnetic field calculated by formulas \eqref{MCE_ECE}-\eqref{MCE4} at $U=V=0$ is demonstrated. This quantity changes its sign and periodically diverges at the fields where $E_{1}\to0$, $E_{2}\neq0$. In opposite, the divergences disappear in the MDM region and the MCE oscillates with finite amplitude. Thus, the numerics fully support the proposed correspondence between the behavior of caloric functions and energy spectrum.

\begin{figure}[htbp]
	\includegraphics[width=0.5\textwidth]{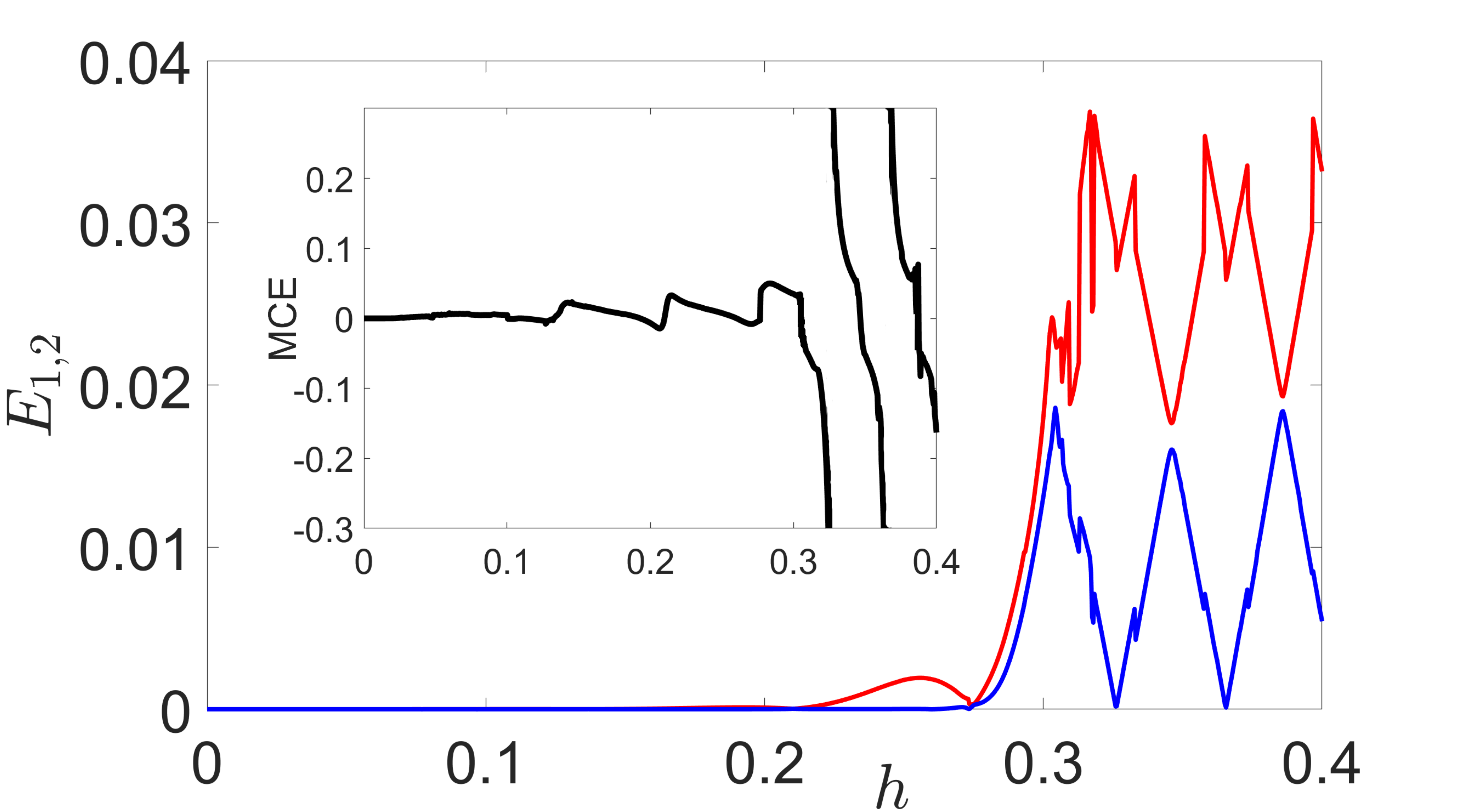}
	\caption{\label{MCE_U10} The magnetic-field dependencies of $E_{1,2}$ for $\Delta=-0.5$, $U=10$, $\mu=-0.75$. Inset: MCE versus $h$. The other parameters are taken the same as in Fig. \ref{MCE_U0}.}
\end{figure}

Strictly speaking, in the strongly correlated regime the expressions \eqref{MCE_ECE}-\eqref{MCE4} are inapplicable. An accurate analysis of caloric functions in such a situation appeals to the finite-temperature DMRG approach that goes beyond the scope of current study. However, a qualitative evaluation of caloric effects in the low-temperature limit ($T$ is much lesser than the bulk gap) can be provided via the following thermodynamic relations: 
\begin{eqnarray}
\label{M_N_C}
&~&\partial\langle \hat{M} \rangle /\partial T= \langle \hat{M}\cdot \mathscr{H} \rangle - \langle \hat{M} \rangle \cdot \langle \mathscr{H} \rangle;\nonumber\\
&~&\partial\langle \hat{N} \rangle /\partial T= \langle \hat{N}\cdot \mathscr{H} \rangle - \langle \hat{N} \rangle \cdot \langle \mathscr{H} \rangle;\nonumber\\
&~& C(T)=\langle \mathscr{H}^2 \rangle - \langle \mathscr{H} \rangle^{2}.
\end{eqnarray}
Since the presence of two edge states results in the four-fold degeneracy of entanglement spectrum that the equilibrium averages can be calculated using first four many-body states (two from each parity sector),
\begin{eqnarray}
\label{Nav_Mav}
&~&\langle \hat{N} \rangle = Sp\left(\hat{N} \cdot \rho \right);~~
\langle \hat{M} \rangle = Sp\left(\hat{M} \cdot \rho \right);\nonumber\\
&~&\rho = \left(\frac{1}{\tilde{Z}}\right) \sum_{j=1,2}\sum_{P=ev,od} e^{-E^{P}_{j}/T}\cdot|\Psi^{P}_{j} \rangle\langle  \Psi^{P}_{j} |;\nonumber\\
&~&\tilde{Z} = \sum_{j=1,2}\sum_{P=ev,od} e^{-E^{P}_{j}/T}.
\end{eqnarray}
This approximation confirms the correspondence between the behavior of caloric functions and spectrum of elementary excitations that is demonstrated in Figure~\ref{MCE_U10}. Namely, there are the MCE oscillations in the MDM area, $h<0.3$, and the series of anomalies in the former MSM region, $h>0.3$ (see the inset of Fig.~\ref{MCE_U10}). As it was already observed the latter appears due to the significant electron-electron interactions leading to reduction of the gap between $E_{1}$ and $E_{2}$. Then the spatial distribution of the lowest state becomes bulk-like even though $E_{1}$ is still periodically equal to zero (as well as true MSM). Thus, the highlighted properties are stable against the Coulomb correlations. Note that such measurements can be supplemented by probing of the spin polarization of the wire as a whole which provides the information about the MM norm.

\section{\label{sec5}Summary and conclusion}

In the present article the effect of Coulomb correlations on topological phases of the 1D BDI-class wire was analyzed employing the DMRG method. To probe and display the MM emergence in the system we generalized the MP concept which had been introduced earlier for the D-class noninteracting structures. The numerics revealed that the MP behavior is in agreement with the entanglement-spectrum degeneracy, that has topological nature, in a wide range of values of the Coulomb interactions. In the noninteracting case the equivalence between the MP and $N_{BDI}$ topological invariant was observed as well. 

The $tUV$-DMRG calculations operating with the initial Hamiltonian \eqref{Ham_Fermi} showed the features which had been already obtained for the D-class systems at weak correlations. In particular, on the $\mu-h$ phase diagram the left parabola with the MSMs inside is stretched to the right and its minimum is shifted right and down. The Mott-Hubbard gap, where the phase is trivial, between the left and right parabolas increases while the on-site Coulomb interaction rises. Consequently, in the case of strong electron correlations the left (right) parabola is located in the lower (upper) Hubbard subband. 

In general, when $U,~V=0$ the BDI-class wire is characterized by the presence of the MDM region around the left parabola if $|\Delta|<2|\Delta_{1}|$ and $\Delta<0,~\Delta_{1}>0$. These conditions directly correspond to the existence of nodal points of SC order parameter. Then it was demonstrated that the interactions can induce two MMs at each edge even in the opposite situation of $|\Delta|>2|\Delta_{1}|$. The effect is attributed to the suppression of on-site SC pairing due to the increasing correlations. Additionally, in case of strong repulsion the MDM-to-MSM transformation was revealed. The norm of such interaction-induced MMs significantly deviates from 1 at high $U$ if the concentrations of spin-up and spin-down carriers are commensurable. In turn, the MSMs survive inside the parabola mainly in the pronounced spin-polarized regime, $n_{\sigma}\gg n_{\bar{\sigma}}$, when they have the close-to-unity norm.

To improve the convergence and speed of DMRG-numerics in the strongly correlated regime we derived the $t-J^{*}-V$-model by integrating out all the states with two electrons on one site in the wire utilizing projection-operator technique. The resulting Hamiltonian includes the effective interactions related to the processes of both standard hopping and Rashba spin-orbit coupling (as well as their combination). Note that if $\Delta_{1}=0$ that the acquired model becomes applicable for the strongly-correlated D-class wire. The comparison between $tUV$- and $tJV$-DMRG data showed partial agreement as the three-center terms were not included in the effective Hamiltonian. The last algorithm was used to obtain the topological-phase diagrams where the mentioned induction and suppression of MMs were clearly seen. Additionally, we found that the inter-site Coulomb interactions result in the extra reduction of MSM and MDM areas. 

In order to demonstrate the dramatic impact of spin and charge fluctuations on the observed effects the simplest Hubbard-I approximation for the $t$-model was considered analytically. In this case the MSM and MDM regions were not affected by $U$ and their norm equals 1 everywhere on the phase diagram even though $U \to \infty$. On the other hand, using the calculations of entanglement spectrum degeneracy in the $t$-DMRG approach it was revealed that both MSMs and MDMs are completely destroyed in the $U \to \infty$ limit.

We also discussed the possibility to probe the MSMs and MDMs via the features of caloric functions. It was shown that in the MSM area the MCE changes its sign and periodically diverges whereas in the case of MDMs this function oscillates with finite amplitude. In the former situation the anomalies appear at the fields where the ground state changes parity while there is no such effect if the MDMs emerge. Using the DMRG data we argue that these features are able to persist in the strongly correlated regime.

\begin{acknowledgments}
We acknowledge fruitful discussions with V. V. Valkov and V. A. Mitskan. The reported study was funded by the RAS Presidium programs for fundamental research Nos. 12 and 32, Russian Foundation for Basic Research (Projects No. 18-32-00443 and No. 19-02-00348), Government of Krasnoyarsk Territory, Krasnoyarsk Regional Fund of Science to the research project: “Coulomb interactions in the problem of Majorana modes in low-dimensional systems with nontrivial topology” (Grant No. 19-42-240011). S.V.A. and A.O.Z. are grateful to the Council of the President of the Russian Federation for Support of Young Scientists and Leading Scientific Schools, Projects No. MK-1641.2020.2 and No. MK-3594.2018.2. S.V.A. acknowledges the support from the Foundation for the Advancement of Theoretical Physics and Mathematics ”BASIS” (Grant No. 18-46-007).
\end{acknowledgments}

\appendix
\section{\label{apxA} The generalized mean-field description of BDI-class wire}

In the Appendix \ref{apxA} the GMF approach, which is useful to probe the effects of weak Coulomb interactions on the topological phases of 1D structures, is discussed for the BDI-class wire. The effective BdG Hamiltonian is defined as
\begin{eqnarray}
\label{Ham_BdG}
{\mathscr{H}} & = & \frac{1}{2}\cdot{{{\bf C} }}^+  \cdot H \cdot {\bf C },
\\
\label{Ham_BdG_MF}
H & = & \left( {\begin{array}{*{20}{c}}
	A_{\uparrow\uparrow}&A_{\uparrow\downarrow}&B_{\uparrow\uparrow}&B_{\uparrow\downarrow}\\
	A_{\uparrow\downarrow}^{+}&A_{\downarrow\downarrow}&-B^{T}_{\uparrow\downarrow}&B_{\downarrow\downarrow}\\
	-B^{*}_{\uparrow\uparrow}&-B^{*}_{\uparrow\downarrow}&-A^{*}_{\uparrow\uparrow}&-A^{*}_{\uparrow\downarrow}\\
	B^{+}_{\uparrow\downarrow}&-B^{*}_{\downarrow\downarrow}&-A_{\uparrow\downarrow}^{T}&-A^{*}_{\downarrow\downarrow}\\
	\end{array}} \right),
\end{eqnarray}
where ${\bf C}^{+} =\left({\bf a}^{+}_{\uparrow}, {\bf a}^{+}_{\downarrow}, {\bf a}^{T}_{\uparrow},    {\bf a}^{T}_{\downarrow} \right),$  ${\bf{a}}_{\sigma} = (a_{1\sigma},\ldots,a_{N\sigma})^{T}$.
The matrices $A_{\sigma,\sigma'}$, $B_{\sigma,\sigma'}$ contain the following nonzero components
($ A_{\sigma\sigma}  =  A^{+}_{\sigma\sigma} $, $B_{\sigma\sigma}=-B_{\sigma\sigma}^{T}$):
\begin{eqnarray}
\label{A_B_int}
&~&\left( A_{\sigma \sigma} \right)_{f,f} = -\mu - \sigma h + U\langle a^{+}_{f\bar{\sigma}}a_{f\bar{\sigma}} \rangle + \nonumber\\
&+&V\cdot\left( \sum_{\sigma'} \langle a^{+}_{f-1,\sigma'}a_{f-1,\sigma'} \rangle + \langle a^{+}_{f+1,\sigma'}a_{f+1,\sigma'} \rangle \right);\nonumber\\
&~&\left( A_{\sigma \sigma} \right)_{f+1,f} = -\frac{t}{2} - V\langle a^{+}_{f\sigma}a_{f+1,\sigma} \rangle;\nonumber\\
&~&\left( A_{\uparrow \downarrow} \right)_{f,f} =-U\langle a^{+}_{f\downarrow}a_{f\uparrow} \rangle; \nonumber\\
&~&\left( A_{\uparrow \downarrow} \right)_{f,f+1} = -\frac{\alpha}{2} - V\langle a^{+}_{f+1\downarrow}a_{f\uparrow}\rangle;\nonumber\\
&~&\left( A_{\uparrow \downarrow} \right)_{f+1,f} = \frac{\alpha}{2} - V\langle a^{+}_{f\downarrow}a_{f+1\uparrow}\rangle; \nonumber\\
&~&\left( B_{\sigma \sigma} \right)_{f+1,f} = -V\langle a_{f+1\sigma}a_{f\sigma}\rangle;\nonumber\\
&~&\left( B_{\uparrow \downarrow} \right)_{f,f} = - \Delta^{*} +U \langle a_{f\downarrow}a_{f\uparrow}\rangle;\nonumber\\
&~&\left( B_{\uparrow \downarrow} \right)_{f,f+1} =-\Delta_{1} + V \langle a_{f+1\downarrow}a_{f\uparrow}\rangle;\nonumber\\
&~&\left( B_{\uparrow \downarrow} \right)_{f+1,f} =-\Delta_{1} -V \langle a_{f+1\uparrow}a_{f\downarrow}\rangle.
\end{eqnarray}
It is seen from \eqref{A_B_int} that there are both renormalized elements and new ones induced by nonzero $U,~V$.

The eigenvectors ${\bf{Y}}_{j} = \left( {\bf{u}}_{j\uparrow}, {\bf{u}}_{j\downarrow}, {\bf{v}}^{*}_{j\uparrow}, {\bf{v}}^{*}_{j\downarrow} \right)^{T}$ of
BdG Hamiltonian (\ref{Ham_BdG_MF}) describe the
electron- and hole-like wave functions of the states
with excitation energy $\varepsilon_{j}$. The averages in the expressions \eqref{A_B_int} are 
nonlinearly related to the sought coefficients of transformation:
\begin{eqnarray}
\label{Corr_func}
\langle a^{+}_{f\sigma}a_{f'\sigma'} \rangle &=&
\sum_{j=1}^{2N}\Big[ u_{jf\sigma}u^{*}_{jf'\sigma'}f\left(\frac{\varepsilon_j}{T}\right) + \nonumber\\
&+&v_{jf\sigma}v^{*}_{jf'\sigma'}\left( 1 - f\left(\frac{\varepsilon_j}{T}\right) \right)\Big];
\end{eqnarray}
\begin{eqnarray}
\label{Corr_func}
\langle a^{+}_{f\sigma}a^{+}_{f'\sigma'} \rangle &=&
\sum_{j=1}^{2N}\Big[ u_{jf\sigma}v_{jf'\sigma'}f\left(\frac{\varepsilon_j}{T}\right) + \nonumber\\
&+&v_{jf\sigma}u_{jf'\sigma'}\left( 1 - f\left(\frac{\varepsilon_j}{T}\right) \right)\Big].
\end{eqnarray}
where $f(x)$ is the Fermi-Dirac function. Analysis of the MP can be carried out using the relation between the quasiparticle operators $\alpha_{1,2}$ and self-adjoint Majorana operators, $\gamma_{Af\sigma}=\frac{1}{2}\left(a_{f\sigma} +
a_{f\sigma}^{+}\right)$, $\gamma_{Bf\sigma}=\frac{i}{2}\left(a_{f\sigma} -
a_{f\sigma}^{+}\right)$,
\begin{eqnarray}
\label{A_Majorana}
\alpha_{j} = \frac{1}{2}\sum^{N}_{f=1;\sigma}\left( {w_{jf\sigma}\gamma_{Af\sigma} + z_{jf\sigma}\gamma_{Bf\sigma}} \right).
\end{eqnarray}

\section{\label{apxB} Hubbard-I approximation in the limit $U \to \infty$}

Here the Majorana quasiparticles in the limit $U \to \infty$ are determined employing the Hubbard-I approximation. To achieve it we solve the system of equations for the Zubarev's Green functions on different sites which is written as:
\begin{eqnarray}
\label{Sys_GF}
\left( {\begin{array}{*{20}{c}}
	{{\omega - {\tilde{A}}_{\uparrow \uparrow}}}&{-{\tilde{A}}_{\uparrow \downarrow}}&\hat{0}&{{-{\tilde{B}}_{\uparrow \downarrow}}}\\
	{-{\tilde{A}}_{\downarrow \uparrow}}&{{\omega - {\tilde{A}}_{\downarrow \downarrow}}}&{{-{\tilde{B}}_{\downarrow \uparrow}}}&\hat{0}\\
	\hat{0}&{{\tilde{B}}^*_{\uparrow \downarrow}}&{ \omega +  {\tilde{A}}_{\uparrow \uparrow}}& {\tilde{A}}_{\uparrow \downarrow}\\
	{\tilde{B}}^*_{\downarrow \uparrow}&\hat{0}&{{\tilde{A}}_{\downarrow \uparrow}}&{\omega + {\tilde{A}}_{\downarrow \downarrow}}
	\end{array}} \right) \cdot \nonumber \\
\cdot \left[ {\begin{array}{*{20}{c}}
	\left\langle \left\langle \hat{X}^{0 \uparrow} | X_{f'}^{\uparrow 0}  \right\rangle \right\rangle_{\omega}\\
	\left\langle \left\langle \hat{X}^{0 \downarrow} | X_{f'}^{\uparrow 0}  \right\rangle \right\rangle_{\omega}\\
	\left\langle \left\langle \hat{X}^{\uparrow 0} | X_{f'}^{\uparrow 0}  \right\rangle \right\rangle_{\omega}\\
	\left\langle \left\langle \hat{X}^{\downarrow 0} | X_{f'}^{\uparrow 0}  \right\rangle \right\rangle_{\omega}
	\end{array}} \right] = \left[ {\begin{array}{*{20}{c}}
	\hat{\delta}_{\uparrow}\\
	\hat{0}\\
	\hat{0}\\
	\hat{0}
	\end{array}} \right],
\end{eqnarray}
where $N \times N$ matrices $\tilde{A}_{\sigma \sigma}$, $\tilde{A}_{\sigma \bar{\sigma}}$, and $\tilde{B}_{\sigma \bar{\sigma}}$ are given by
\begin{eqnarray}
\label{Ham_Matrices}
\tilde{A}_{\sigma \sigma} &=& \left( {\begin{array}{*{20}{c}}
	{{\xi_{1 \sigma}}}&{z_{1\bar{\sigma}}t}&0&0\\
	{z_{2\bar{\sigma}}t}& \ddots & \ddots &0\\
	0& \ddots & \ddots &{z_{N-1,\bar{\sigma}}t}\\
	0&0&{z_{N\bar{\sigma}}t}&{\xi_{N \sigma}}
	\end{array}} \right), \\
\tilde{A}_{\sigma \bar{\sigma}} &=& \left( {\begin{array}{*{20}{c}}
	0&{-z_{1\bar{\sigma}}\alpha}&0&0\\
	{z_{2\bar{\sigma}}\alpha}& \ddots & \ddots &0\\
	0& \ddots & \ddots &{-z_{N-1,\bar{\sigma}}\alpha}\\
	0&0&{z_{N\bar{\sigma}}\alpha}&0
	\end{array}} \right),
\\ \tilde{B}_{\sigma \bar{\sigma}} &=& \left( {\begin{array}{*{20}{c}}
	0&{-z_{1\bar{\sigma}}\Delta_1^*}&0&0\\
	{-z_{2\bar{\sigma}}\Delta_1^*}& \ddots & \ddots &0\\
	0& \ddots & \ddots &{-z_{N-1,\bar{\sigma}}\Delta_1^*}\\
	0&0&{-z_{N\bar{\sigma}}\Delta_1^*}&0
	\end{array}} \right). \nonumber\\
\end{eqnarray}
In \eqref{Sys_GF} $\hat{\delta}_{\uparrow}$ is a vector-column of size $N$,
\begin{eqnarray}
\hat{\delta}_{\uparrow} & = & \left( z_{1 \downarrow} \delta_{1f'}, z_{2 \downarrow} \delta_{2f'}, \dots, z_{N \downarrow} \delta_{Nf'}  \right)',
\end{eqnarray}
where $\xi_{f \sigma} = \xi_{\sigma} - V \left( \langle n_{f+1} \rangle + \langle n_{f-1} \rangle \right) $, $z_{f \sigma} = 1 - \langle n_{f \sigma} \rangle$ is the site-dependent Hubbard renormalization factor, $\delta_{f f'}$ is the Kronecker symbol.

The low-energy quasiparticle Green function can be formally represented
in the form
\begin{eqnarray}
\label{GRFM} \left( \omega - \varepsilon_{j} \right)
\left\langle \left\langle \alpha_j | X_{f'}^{\uparrow 0}  \right\rangle \right\rangle_{\omega} =
z_{f' \downarrow} \left( S^{\dag} \right)_{jf'},
\end{eqnarray}
where $\varepsilon_{j}$ are branches of the excitation spectrum
with $j = 1,2, \dots N$, and $S$ is a transformation matrix diagonalizing the system-of-equation matrix. The factors $z_{f \sigma}$, energy spectrum and Green functions are obtained self-consistently using the relation:
\begin{eqnarray}
\label{nfs_sfc}
\langle n_{f \sigma} \rangle & = & \sum_{j = 1}^{2N} \frac{1}{2 \varepsilon_j \prod_{i \ne j} \left( \varepsilon_j^2 - \varepsilon_i^2  \right)} \left\{ \frac{g_{f \sigma}(\varepsilon_j)}{\exp(\varepsilon_j/T) + 1} - \right.
\nonumber \\
& - & \left.  \frac{g_{f \sigma}(-\varepsilon_j)}{\exp(-\varepsilon_j/T) + 1} \right\},
\end{eqnarray}
where $g_{f \sigma}$ are numerators of the Green functions $\left\langle \left\langle X_f^{0 \sigma} | X_{f}^{\sigma 0}  \right\rangle \right\rangle_{\omega}$ which are numerically found from the Eq. \eqref{Sys_GF} for
$\sigma = \uparrow$. The Green functions containing $X_{f}^{\downarrow 0}$ as the second operator can be received in the similar manner.

The relation \eqref{GRFM} makes it possible
to determine the operators of elementary excitations
in terms of the Hubbard fermion operators:
\begin{eqnarray}
\label{DEFUV} \alpha_{j} & = & \sum_{f=1}^{N} \sum_{\sigma} \left(\tilde{u}_{jf \sigma}
X_{f}^{0 \sigma} + \tilde{v}_{jf \sigma} X_f^{\sigma 0}\right).
\end{eqnarray}
Using the Majorana operators in the atomic representation~\eqref{gammaX}
the quasiparticle operator can be presented in the form
\begin{eqnarray}
\label{DEFUV3} \tilde{\alpha}_{j} & = & \frac{1}{2} \sum_{f=1}^{N} \sum_{\sigma} \left(\tilde{w}_{jf \sigma}
\tilde{\gamma}_{A f \sigma} + i\tilde{z}_{jf \sigma} \tilde{\gamma}_{B f \sigma}\right),
\end{eqnarray}
The difference from the conventional definition \eqref{A_Majorana} is that the operators $\tilde{\gamma}_{A f \sigma}$ and $\tilde{\gamma}_{B f \sigma}$ involve not only one-fermion but also three-fermion summand since $X^{0\sigma}_{f}=a_{f\sigma}\left(1-a_{f\bar{\sigma}}^+ a_{f\bar{\sigma}}\right)$.
The coefficients $\tilde{w}_{jf \sigma} = \tilde{u}_{jf \sigma} + \tilde{v}_{jf \sigma}$, $\tilde{z}_{jf \sigma} = \tilde{u}_{jf \sigma} - \tilde{v}_{jf \sigma}$ are sought Majorana-type coefficients. Now one can calculate the MP using the definition \eqref{MP_def}.

\bibliography{Majorana}

\providecommand{\noopsort}[1]{}\providecommand{\singleletter}[1]{#1}%
\begin{thebibliography}{60}%
\makeatletter
\providecommand \@ifxundefined [1]{%
 \@ifx{#1\undefined}
}%
\providecommand \@ifnum [1]{%
 \ifnum #1\expandafter \@firstoftwo
 \else \expandafter \@secondoftwo
 \fi
}%
\providecommand \@ifx [1]{%
 \ifx #1\expandafter \@firstoftwo
 \else \expandafter \@secondoftwo
 \fi
}%
\providecommand \natexlab [1]{#1}%
\providecommand \enquote  [1]{``#1''}%
\providecommand \bibnamefont  [1]{#1}%
\providecommand \bibfnamefont [1]{#1}%
\providecommand \citenamefont [1]{#1}%
\providecommand \href@noop [0]{\@secondoftwo}%
\providecommand \href [0]{\begingroup \@sanitize@url \@href}%
\providecommand \@href[1]{\@@startlink{#1}\@@href}%
\providecommand \@@href[1]{\endgroup#1\@@endlink}%
\providecommand \@sanitize@url [0]{\catcode `\\12\catcode `\$12\catcode
  `\&12\catcode `\#12\catcode `\^12\catcode `\_12\catcode `\%12\relax}%
\providecommand \@@startlink[1]{}%
\providecommand \@@endlink[0]{}%
\providecommand \url  [0]{\begingroup\@sanitize@url \@url }%
\providecommand \@url [1]{\endgroup\@href {#1}{\urlprefix }}%
\providecommand \urlprefix  [0]{URL }%
\providecommand \Eprint [0]{\href }%
\providecommand \doibase [0]{http://dx.doi.org/}%
\providecommand \selectlanguage [0]{\@gobble}%
\providecommand \bibinfo  [0]{\@secondoftwo}%
\providecommand \bibfield  [0]{\@secondoftwo}%
\providecommand \translation [1]{[#1]}%
\providecommand \BibitemOpen [0]{}%
\providecommand \bibitemStop [0]{}%
\providecommand \bibitemNoStop [0]{.\EOS\space}%
\providecommand \EOS [0]{\spacefactor3000\relax}%
\providecommand \BibitemShut  [1]{\csname bibitem#1\endcsname}%
\let\auto@bib@innerbib\@empty
\bibitem [{\citenamefont {Read}\ and\ \citenamefont {Green}(2000)}]{read-00}%
  \BibitemOpen
  \bibfield  {author} {\bibinfo {author} {\bibfnamefont {N.}~\bibnamefont
  {Read}}\ and\ \bibinfo {author} {\bibfnamefont {D.}~\bibnamefont {Green}},\
  }\href {\doibase 10.1103/PhysRevB.61.10267} {\bibfield  {journal} {\bibinfo
  {journal} {Phys.\ Rev.\ B}\ }\textbf {\bibinfo {volume} {61}},\ \bibinfo
  {pages} {10267} (\bibinfo {year} {2000})}\BibitemShut {NoStop}%
\bibitem [{\citenamefont {Kitaev}(2001)}]{kitaev-01}%
  \BibitemOpen
  \bibfield  {author} {\bibinfo {author} {\bibfnamefont {A.~Y.}\ \bibnamefont
  {Kitaev}},\ }\href {\doibase http://dx.doi.org/10.1070/1063-7869/44/10S/S29}
  {\bibfield  {journal} {\bibinfo  {journal} {Phys.\ Usp.}\ }\textbf {\bibinfo
  {volume} {44}},\ \bibinfo {pages} {131} (\bibinfo {year} {2001})}\BibitemShut
  {NoStop}%
\bibitem [{\citenamefont {Ivanov}(2001)}]{ivanov-01}%
  \BibitemOpen
  \bibfield  {author} {\bibinfo {author} {\bibfnamefont {D.~A.}\ \bibnamefont
  {Ivanov}},\ }\href {\doibase 10.1103/PhysRevLett.86.268} {\bibfield
  {journal} {\bibinfo  {journal} {Phys.\ Rev.\ Lett.}\ }\textbf {\bibinfo
  {volume} {86}},\ \bibinfo {pages} {268} (\bibinfo {year} {2001})}\BibitemShut
  {NoStop}%
\bibitem [{\citenamefont {Kitaev}(2003)}]{kitaev-03}%
  \BibitemOpen
  \bibfield  {author} {\bibinfo {author} {\bibfnamefont {A.~Y.}\ \bibnamefont
  {Kitaev}},\ }\href {\doibase http://dx.doi.org/10.1016/S0003-4916(02)00018-0}
  {\bibfield  {journal} {\bibinfo  {journal} {Ann.\ Phys.}\ }\textbf {\bibinfo
  {volume} {303}},\ \bibinfo {pages} {2} (\bibinfo {year} {2003})}\BibitemShut
  {NoStop}%
\bibitem [{\citenamefont {Lutchyn}\ \emph {et~al.}(2010)\citenamefont
  {Lutchyn}, \citenamefont {Sau},\ and\ \citenamefont {Sarma}}]{lutchyn-10}%
  \BibitemOpen
  \bibfield  {author} {\bibinfo {author} {\bibfnamefont {R.~M.}\ \bibnamefont
  {Lutchyn}}, \bibinfo {author} {\bibfnamefont {J.~D.}\ \bibnamefont {Sau}}, \
  and\ \bibinfo {author} {\bibfnamefont {S.~D.}\ \bibnamefont {Sarma}},\ }\href
  {\doibase 10.1103/PhysRevLett.105.077001} {\bibfield  {journal} {\bibinfo
  {journal} {Phys.\ Rev.\ Lett.}\ }\textbf {\bibinfo {volume} {105}},\ \bibinfo
  {pages} {077001} (\bibinfo {year} {2010})}\BibitemShut {NoStop}%
\bibitem [{\citenamefont {Oreg}\ \emph {et~al.}(2010)\citenamefont {Oreg},
  \citenamefont {Refael},\ and\ \citenamefont {von Oppen}}]{oreg-10}%
  \BibitemOpen
  \bibfield  {author} {\bibinfo {author} {\bibfnamefont {Y.}~\bibnamefont
  {Oreg}}, \bibinfo {author} {\bibfnamefont {G.}~\bibnamefont {Refael}}, \ and\
  \bibinfo {author} {\bibfnamefont {F.}~\bibnamefont {von Oppen}},\ }\href
  {\doibase 10.1103/PhysRevLett.105.177002} {\bibfield  {journal} {\bibinfo
  {journal} {Phys.\ Rev.\ Lett.}\ }\textbf {\bibinfo {volume} {105}},\ \bibinfo
  {pages} {177002} (\bibinfo {year} {2010})}\BibitemShut {NoStop}%
\bibitem [{\citenamefont {Mourik}\ \emph {et~al.}(2012)\citenamefont {Mourik},
  \citenamefont {Zuo}, \citenamefont {Frolov}, \citenamefont {Plissard},
  \citenamefont {Bakkers},\ and\ \citenamefont {Kouwenhoven}}]{mourik-12}%
  \BibitemOpen
  \bibfield  {author} {\bibinfo {author} {\bibfnamefont {V.}~\bibnamefont
  {Mourik}}, \bibinfo {author} {\bibfnamefont {K.}~\bibnamefont {Zuo}},
  \bibinfo {author} {\bibfnamefont {S.~M.}\ \bibnamefont {Frolov}}, \bibinfo
  {author} {\bibfnamefont {S.~R.}\ \bibnamefont {Plissard}}, \bibinfo {author}
  {\bibfnamefont {E.~P. A.~M.}\ \bibnamefont {Bakkers}}, \ and\ \bibinfo
  {author} {\bibfnamefont {L.~P.}\ \bibnamefont {Kouwenhoven}},\ }\href
  {\doibase 10.1126/science.1222360} {\bibfield  {journal} {\bibinfo  {journal}
  {Science}\ }\textbf {\bibinfo {volume} {336}},\ \bibinfo {pages} {1003}
  (\bibinfo {year} {2012})}\BibitemShut {NoStop}%
\bibitem [{\citenamefont {Krogstrup}\ \emph {et~al.}(2015)\citenamefont
  {Krogstrup}, \citenamefont {Ziino}, \citenamefont {Chang}, \citenamefont
  {Albrecht}, \citenamefont {Madsen}, \citenamefont {Johnson}, \citenamefont
  {Nygard}, \citenamefont {Marcus},\ and\ \citenamefont
  {Jespersen}}]{krogstrup-15}%
  \BibitemOpen
  \bibfield  {author} {\bibinfo {author} {\bibfnamefont {P.}~\bibnamefont
  {Krogstrup}}, \bibinfo {author} {\bibfnamefont {N.}~\bibnamefont {Ziino}},
  \bibinfo {author} {\bibfnamefont {W.}~\bibnamefont {Chang}}, \bibinfo
  {author} {\bibfnamefont {S.}~\bibnamefont {Albrecht}}, \bibinfo {author}
  {\bibfnamefont {M.}~\bibnamefont {Madsen}}, \bibinfo {author} {\bibfnamefont
  {E.}~\bibnamefont {Johnson}}, \bibinfo {author} {\bibfnamefont
  {J.}~\bibnamefont {Nygard}}, \bibinfo {author} {\bibfnamefont
  {C.}~\bibnamefont {Marcus}}, \ and\ \bibinfo {author} {\bibfnamefont
  {T.}~\bibnamefont {Jespersen}},\ }\href {\doibase 10.1038/NMAT4176}
  {\bibfield  {journal} {\bibinfo  {journal} {Nature Mat.}\ }\textbf {\bibinfo
  {volume} {14}},\ \bibinfo {pages} {400} (\bibinfo {year} {2015})}\BibitemShut
  {NoStop}%
\bibitem [{\citenamefont {Zhang}\ \emph {et~al.}(2018)\citenamefont {Zhang},
  \citenamefont {Liu}, \citenamefont {Gazibegovic}, \citenamefont {Xu},
  \citenamefont {Logan}, \citenamefont {Wang}, \citenamefont {van Loo},
  \citenamefont {Bommer}, \citenamefont {de~Moor}, \citenamefont {Car},
  \citenamefont {het Veld}, \citenamefont {van Veldhoven}, \citenamefont
  {Koelling}, \citenamefont {Verheijen}, \citenamefont {Pendharkar},
  \citenamefont {Pennachio}, \citenamefont {Shojaei}, \citenamefont {Lee},
  \citenamefont {Palmstrm}, \citenamefont {Bakkers}, \citenamefont {Sarma},\
  and\ \citenamefont {Kouwenhoven}}]{zhang-18}%
  \BibitemOpen
  \bibfield  {author} {\bibinfo {author} {\bibfnamefont {H.}~\bibnamefont
  {Zhang}}, \bibinfo {author} {\bibfnamefont {C.-X.}\ \bibnamefont {Liu}},
  \bibinfo {author} {\bibfnamefont {S.}~\bibnamefont {Gazibegovic}}, \bibinfo
  {author} {\bibfnamefont {D.}~\bibnamefont {Xu}}, \bibinfo {author}
  {\bibfnamefont {J.~A.}\ \bibnamefont {Logan}}, \bibinfo {author}
  {\bibfnamefont {G.}~\bibnamefont {Wang}}, \bibinfo {author} {\bibfnamefont
  {N.}~\bibnamefont {van Loo}}, \bibinfo {author} {\bibfnamefont {J.~D.}\
  \bibnamefont {Bommer}}, \bibinfo {author} {\bibfnamefont {M.~W.}\
  \bibnamefont {de~Moor}}, \bibinfo {author} {\bibfnamefont {D.}~\bibnamefont
  {Car}}, \bibinfo {author} {\bibfnamefont {R.~L. M.~O.}\ \bibnamefont {het
  Veld}}, \bibinfo {author} {\bibfnamefont {P.~J.}\ \bibnamefont {van
  Veldhoven}}, \bibinfo {author} {\bibfnamefont {S.}~\bibnamefont {Koelling}},
  \bibinfo {author} {\bibfnamefont {M.~A.}\ \bibnamefont {Verheijen}}, \bibinfo
  {author} {\bibfnamefont {M.}~\bibnamefont {Pendharkar}}, \bibinfo {author}
  {\bibfnamefont {D.~J.}\ \bibnamefont {Pennachio}}, \bibinfo {author}
  {\bibfnamefont {B.}~\bibnamefont {Shojaei}}, \bibinfo {author} {\bibfnamefont
  {J.~S.}\ \bibnamefont {Lee}}, \bibinfo {author} {\bibfnamefont {C.~J.}\
  \bibnamefont {Palmstrm}}, \bibinfo {author} {\bibfnamefont {E.~P. A.~M.}\
  \bibnamefont {Bakkers}}, \bibinfo {author} {\bibfnamefont {S.~D.}\
  \bibnamefont {Sarma}}, \ and\ \bibinfo {author} {\bibfnamefont {L.~P.}\
  \bibnamefont {Kouwenhoven}},\ }\href {\doibase 10.1038/nature26142}
  {\bibfield  {journal} {\bibinfo  {journal} {Nature}\ }\textbf {\bibinfo
  {volume} {556}},\ \bibinfo {pages} {74} (\bibinfo {year} {2018})}\BibitemShut
  {NoStop}%
\bibitem [{\citenamefont {Cayao}\ \emph {et~al.}(2015)\citenamefont {Cayao},
  \citenamefont {Prada}, \citenamefont {San-Jose},\ and\ \citenamefont
  {Aguado}}]{cayao-15}%
  \BibitemOpen
  \bibfield  {author} {\bibinfo {author} {\bibfnamefont {J.}~\bibnamefont
  {Cayao}}, \bibinfo {author} {\bibfnamefont {E.}~\bibnamefont {Prada}},
  \bibinfo {author} {\bibfnamefont {P.}~\bibnamefont {San-Jose}}, \ and\
  \bibinfo {author} {\bibfnamefont {R.}~\bibnamefont {Aguado}},\ }\href
  {\doibase 10.1103/PhysRevB.91.024514} {\bibfield  {journal} {\bibinfo
  {journal} {Phys.\ Rev.\ B}\ }\textbf {\bibinfo {volume} {91}},\ \bibinfo
  {pages} {024514} (\bibinfo {year} {2015})}\BibitemShut {NoStop}%
\bibitem [{\citenamefont {Liu}\ \emph {et~al.}(2017)\citenamefont {Liu},
  \citenamefont {Sau}, \citenamefont {Stanescu},\ and\ \citenamefont
  {Sarma}}]{liu-17b}%
  \BibitemOpen
  \bibfield  {author} {\bibinfo {author} {\bibfnamefont {C.-X.}\ \bibnamefont
  {Liu}}, \bibinfo {author} {\bibfnamefont {J.~D.}\ \bibnamefont {Sau}},
  \bibinfo {author} {\bibfnamefont {T.~D.}\ \bibnamefont {Stanescu}}, \ and\
  \bibinfo {author} {\bibfnamefont {S.~D.}\ \bibnamefont {Sarma}},\ }\href
  {\doibase 10.1103/PhysRevB.96.075161} {\bibfield  {journal} {\bibinfo
  {journal} {Phys.\ Rev.\ B}\ }\textbf {\bibinfo {volume} {96}},\ \bibinfo
  {pages} {075161} (\bibinfo {year} {2017})}\BibitemShut {NoStop}%
\bibitem [{\citenamefont {Moore}\ \emph {et~al.}(2018)\citenamefont {Moore},
  \citenamefont {Stanescu},\ and\ \citenamefont {Tewari}}]{moore-18a}%
  \BibitemOpen
  \bibfield  {author} {\bibinfo {author} {\bibfnamefont {C.}~\bibnamefont
  {Moore}}, \bibinfo {author} {\bibfnamefont {T.}~\bibnamefont {Stanescu}}, \
  and\ \bibinfo {author} {\bibfnamefont {S.}~\bibnamefont {Tewari}},\ }\href
  {\doibase 10.1103/PhysRevB.97.165302} {\bibfield  {journal} {\bibinfo
  {journal} {Phys.\ Rev.\ B}\ }\textbf {\bibinfo {volume} {97}},\ \bibinfo
  {pages} {165302} (\bibinfo {year} {2018})}\BibitemShut {NoStop}%
\bibitem [{\citenamefont {Reeg}\ \emph {et~al.}(2018)\citenamefont {Reeg},
  \citenamefont {Dmytruk}, \citenamefont {Chevallier}, \citenamefont {Loss},\
  and\ \citenamefont {Klinovaja}}]{reeg-18}%
  \BibitemOpen
  \bibfield  {author} {\bibinfo {author} {\bibfnamefont {C.}~\bibnamefont
  {Reeg}}, \bibinfo {author} {\bibfnamefont {O.}~\bibnamefont {Dmytruk}},
  \bibinfo {author} {\bibfnamefont {D.}~\bibnamefont {Chevallier}}, \bibinfo
  {author} {\bibfnamefont {D.}~\bibnamefont {Loss}}, \ and\ \bibinfo {author}
  {\bibfnamefont {J.}~\bibnamefont {Klinovaja}},\ }\href {\doibase
  10.1103/PhysRevB.98.245407} {\bibfield  {journal} {\bibinfo  {journal}
  {Phys.\ Rev.\ B}\ }\textbf {\bibinfo {volume} {98}},\ \bibinfo {pages}
  {245407} (\bibinfo {year} {2018})}\BibitemShut {NoStop}%
\bibitem [{\citenamefont {Val'kov}\ and\ \citenamefont
  {Aksenov}(2017{\natexlab{a}})}]{valkov-17b}%
  \BibitemOpen
  \bibfield  {author} {\bibinfo {author} {\bibfnamefont {V.~V.}\ \bibnamefont
  {Val'kov}}\ and\ \bibinfo {author} {\bibfnamefont {S.~V.}\ \bibnamefont
  {Aksenov}},\ }\href {\doibase 10.1063/1.4983331} {\bibfield  {journal}
  {\bibinfo  {journal} {Low Temp.\ Phys.}\ }\textbf {\bibinfo {volume} {43}},\
  \bibinfo {pages} {437} (\bibinfo {year} {2017}{\natexlab{a}})}\BibitemShut
  {NoStop}%
\bibitem [{\citenamefont {Val'kov}\ and\ \citenamefont
  {Aksenov}(2017{\natexlab{b}})}]{valkov-17c}%
  \BibitemOpen
  \bibfield  {author} {\bibinfo {author} {\bibfnamefont {V.~V.}\ \bibnamefont
  {Val'kov}}\ and\ \bibinfo {author} {\bibfnamefont {S.~V.}\ \bibnamefont
  {Aksenov}},\ }\href {\doibase http://dx.doi.org/10.1016/j.jmmm.2016.10.155}
  {\bibfield  {journal} {\bibinfo  {journal} {J.\ Magn.\ Magn.\ Mat.}\ }\textbf
  {\bibinfo {volume} {440}},\ \bibinfo {pages} {112} (\bibinfo {year}
  {2017}{\natexlab{b}})}\BibitemShut {NoStop}%
\bibitem [{\citenamefont {Schnyder}\ \emph {et~al.}(2008)\citenamefont
  {Schnyder}, \citenamefont {Ryu}, \citenamefont {Furusaki},\ and\
  \citenamefont {Ludwig}}]{schnyder-08}%
  \BibitemOpen
  \bibfield  {author} {\bibinfo {author} {\bibfnamefont {A.}~\bibnamefont
  {Schnyder}}, \bibinfo {author} {\bibfnamefont {S.}~\bibnamefont {Ryu}},
  \bibinfo {author} {\bibfnamefont {A.}~\bibnamefont {Furusaki}}, \ and\
  \bibinfo {author} {\bibfnamefont {A.}~\bibnamefont {Ludwig}},\ }\href
  {\doibase 10.1103/PhysRevB.78.195125} {\bibfield  {journal} {\bibinfo
  {journal} {Phys.\ Rev.\ B}\ }\textbf {\bibinfo {volume} {78}},\ \bibinfo
  {pages} {195125} (\bibinfo {year} {2008})}\BibitemShut {NoStop}%
\bibitem [{\citenamefont {Kitaev}(2009)}]{kitaev-09}%
  \BibitemOpen
  \bibfield  {author} {\bibinfo {author} {\bibfnamefont {A.}~\bibnamefont
  {Kitaev}},\ }\href {\doibase 10.1063/1.3149495} {\bibfield  {journal}
  {\bibinfo  {journal} {AIP Conf.\ Proc.}\ }\textbf {\bibinfo {volume}
  {1134}},\ \bibinfo {pages} {22} (\bibinfo {year} {2009})}\BibitemShut
  {NoStop}%
\bibitem [{\citenamefont {Sarma}\ \emph {et~al.}(2015)\citenamefont {Sarma},
  \citenamefont {Freedman},\ and\ \citenamefont {Nayak}}]{dassarma-15}%
  \BibitemOpen
  \bibfield  {author} {\bibinfo {author} {\bibfnamefont {S.~D.}\ \bibnamefont
  {Sarma}}, \bibinfo {author} {\bibfnamefont {M.}~\bibnamefont {Freedman}}, \
  and\ \bibinfo {author} {\bibfnamefont {C.}~\bibnamefont {Nayak}},\ }\href
  {\doibase 10.1038/npjqi.2015.1} {\bibfield  {journal} {\bibinfo  {journal}
  {Quant.\ Inf.}\ }\textbf {\bibinfo {volume} {1}},\ \bibinfo {pages} {15001}
  (\bibinfo {year} {2015})}\BibitemShut {NoStop}%
\bibitem [{\citenamefont {Sato}\ \emph {et~al.}(2019)\citenamefont {Sato},
  \citenamefont {Matsuo}, \citenamefont {Hsu}, \citenamefont {Stano},
  \citenamefont {Ueda}, \citenamefont {Takeshige}, \citenamefont {Kamata},
  \citenamefont {Lee}, \citenamefont {Shojaei}, \citenamefont {Wickramasinghe},
  \citenamefont {Shabani}, \citenamefont {Palmstrom}, \citenamefont {Tokura},
  \citenamefont {Loss},\ and\ \citenamefont {Tarucha}}]{sato-19}%
  \BibitemOpen
  \bibfield  {author} {\bibinfo {author} {\bibfnamefont {Y.}~\bibnamefont
  {Sato}}, \bibinfo {author} {\bibfnamefont {S.}~\bibnamefont {Matsuo}},
  \bibinfo {author} {\bibfnamefont {C.-H.}\ \bibnamefont {Hsu}}, \bibinfo
  {author} {\bibfnamefont {P.}~\bibnamefont {Stano}}, \bibinfo {author}
  {\bibfnamefont {K.}~\bibnamefont {Ueda}}, \bibinfo {author} {\bibfnamefont
  {Y.}~\bibnamefont {Takeshige}}, \bibinfo {author} {\bibfnamefont
  {H.}~\bibnamefont {Kamata}}, \bibinfo {author} {\bibfnamefont {J.~S.}\
  \bibnamefont {Lee}}, \bibinfo {author} {\bibfnamefont {B.}~\bibnamefont
  {Shojaei}}, \bibinfo {author} {\bibfnamefont {K.}~\bibnamefont
  {Wickramasinghe}}, \bibinfo {author} {\bibfnamefont {J.}~\bibnamefont
  {Shabani}}, \bibinfo {author} {\bibfnamefont {C.}~\bibnamefont {Palmstrom}},
  \bibinfo {author} {\bibfnamefont {Y.}~\bibnamefont {Tokura}}, \bibinfo
  {author} {\bibfnamefont {D.}~\bibnamefont {Loss}}, \ and\ \bibinfo {author}
  {\bibfnamefont {S.}~\bibnamefont {Tarucha}},\ }\href {\doibase
  10.1103/PhysRevB.99.155304} {\bibfield  {journal} {\bibinfo  {journal}
  {Phys.\ Rev.\ B}\ }\textbf {\bibinfo {volume} {99}},\ \bibinfo {pages}
  {155304} (\bibinfo {year} {2019})}\BibitemShut {NoStop}%
\bibitem [{\citenamefont {Fidkowski}\ and\ \citenamefont
  {Kitaev}(2010)}]{fidkowski-10}%
  \BibitemOpen
  \bibfield  {author} {\bibinfo {author} {\bibfnamefont {L.}~\bibnamefont
  {Fidkowski}}\ and\ \bibinfo {author} {\bibfnamefont {A.}~\bibnamefont
  {Kitaev}},\ }\href {\doibase 10.1103/PhysRevB.81.134509} {\bibfield
  {journal} {\bibinfo  {journal} {Phys.\ Rev.\ B}\ }\textbf {\bibinfo {volume}
  {81}},\ \bibinfo {pages} {134509} (\bibinfo {year} {2010})}\BibitemShut
  {NoStop}%
\bibitem [{\citenamefont {Wang}\ and\ \citenamefont {Senthil}(2014)}]{wang-14}%
  \BibitemOpen
  \bibfield  {author} {\bibinfo {author} {\bibfnamefont {C.}~\bibnamefont
  {Wang}}\ and\ \bibinfo {author} {\bibfnamefont {T.}~\bibnamefont {Senthil}},\
  }\href@noop {} {\bibfield  {journal} {\bibinfo  {journal} {Phys.\ Rev.\ B}\
  }\textbf {\bibinfo {volume} {89}},\ \bibinfo {pages} {195124} (\bibinfo
  {year} {2014})}\BibitemShut {NoStop}%
\bibitem [{\citenamefont {Katsura}\ \emph {et~al.}(2015)\citenamefont
  {Katsura}, \citenamefont {Schuricht},\ and\ \citenamefont
  {Takahashi}}]{katsura-15}%
  \BibitemOpen
  \bibfield  {author} {\bibinfo {author} {\bibfnamefont {H.}~\bibnamefont
  {Katsura}}, \bibinfo {author} {\bibfnamefont {D.}~\bibnamefont {Schuricht}},
  \ and\ \bibinfo {author} {\bibfnamefont {M.}~\bibnamefont {Takahashi}},\
  }\href@noop {} {\bibfield  {journal} {\bibinfo  {journal} {Phys.\ Rev.\ B}\
  }\textbf {\bibinfo {volume} {92}},\ \bibinfo {pages} {115137} (\bibinfo
  {year} {2015})}\BibitemShut {NoStop}%
\bibitem [{\citenamefont {Kells}(2015)}]{kells-15}%
  \BibitemOpen
  \bibfield  {author} {\bibinfo {author} {\bibfnamefont {G.}~\bibnamefont
  {Kells}},\ }\href@noop {} {\bibfield  {journal} {\bibinfo  {journal} {Phys.\
  Rev.\ B}\ }\textbf {\bibinfo {volume} {92}},\ \bibinfo {pages} {081401(R)}
  (\bibinfo {year} {2015})}\BibitemShut {NoStop}%
\bibitem [{\citenamefont {Miao}\ \emph {et~al.}(2017)\citenamefont {Miao},
  \citenamefont {Jin}, \citenamefont {Zhang},\ and\ \citenamefont
  {Zhou}}]{miao-17}%
  \BibitemOpen
  \bibfield  {author} {\bibinfo {author} {\bibfnamefont {J.-J.}\ \bibnamefont
  {Miao}}, \bibinfo {author} {\bibfnamefont {H.-K.}\ \bibnamefont {Jin}},
  \bibinfo {author} {\bibfnamefont {F.-C.}\ \bibnamefont {Zhang}}, \ and\
  \bibinfo {author} {\bibfnamefont {Y.}~\bibnamefont {Zhou}},\ }\href@noop {}
  {\bibfield  {journal} {\bibinfo  {journal} {Phys.\ Rev.\ Lett.}\ }\textbf
  {\bibinfo {volume} {118}},\ \bibinfo {pages} {267701} (\bibinfo {year}
  {2017})}\BibitemShut {NoStop}%
\bibitem [{\citenamefont {White}(1992)}]{white-92}%
  \BibitemOpen
  \bibfield  {author} {\bibinfo {author} {\bibfnamefont {S.~R.}\ \bibnamefont
  {White}},\ }\href@noop {} {\bibfield  {journal} {\bibinfo  {journal} {Phys.\
  Rev.\ Lett.}\ }\textbf {\bibinfo {volume} {69}},\ \bibinfo {pages} {2863}
  (\bibinfo {year} {1992})}\BibitemShut {NoStop}%
\bibitem [{\citenamefont {White}(1993)}]{white-93}%
  \BibitemOpen
  \bibfield  {author} {\bibinfo {author} {\bibfnamefont {S.~R.}\ \bibnamefont
  {White}},\ }\href@noop {} {\bibfield  {journal} {\bibinfo  {journal} {Phys.\
  Rev.\ B}\ }\textbf {\bibinfo {volume} {48}},\ \bibinfo {pages} {10345}
  (\bibinfo {year} {1993})}\BibitemShut {NoStop}%
\bibitem [{\citenamefont {Stoudenmire}\ \emph {et~al.}(2011)\citenamefont
  {Stoudenmire}, \citenamefont {Alicea}, \citenamefont {Starykh},\ and\
  \citenamefont {Fisher}}]{stoudenmire-11}%
  \BibitemOpen
  \bibfield  {author} {\bibinfo {author} {\bibfnamefont {E.}~\bibnamefont
  {Stoudenmire}}, \bibinfo {author} {\bibfnamefont {J.}~\bibnamefont {Alicea}},
  \bibinfo {author} {\bibfnamefont {O.}~\bibnamefont {Starykh}}, \ and\
  \bibinfo {author} {\bibfnamefont {M.}~\bibnamefont {Fisher}},\ }\href
  {\doibase 10.1103/PhysRevB.84.014503} {\bibfield  {journal} {\bibinfo
  {journal} {Phys.\ Rev.\ B}\ }\textbf {\bibinfo {volume} {84}},\ \bibinfo
  {pages} {014503} (\bibinfo {year} {2011})}\BibitemShut {NoStop}%
\bibitem [{\citenamefont {Thomale}\ \emph {et~al.}(2013)\citenamefont
  {Thomale}, \citenamefont {Rachel},\ and\ \citenamefont
  {Schmitteckert}}]{thomale-13}%
  \BibitemOpen
  \bibfield  {author} {\bibinfo {author} {\bibfnamefont {R.}~\bibnamefont
  {Thomale}}, \bibinfo {author} {\bibfnamefont {S.}~\bibnamefont {Rachel}}, \
  and\ \bibinfo {author} {\bibfnamefont {P.}~\bibnamefont {Schmitteckert}},\
  }\href {\doibase 10.1103/PhysRevB.88.161103} {\bibfield  {journal} {\bibinfo
  {journal} {Phys.\ Rev.\ B}\ }\textbf {\bibinfo {volume} {88}},\ \bibinfo
  {pages} {161103(R)} (\bibinfo {year} {2013})}\BibitemShut {NoStop}%
\bibitem [{\citenamefont {Haim}\ \emph {et~al.}(2014)\citenamefont {Haim},
  \citenamefont {Keselman}, \citenamefont {Berg},\ and\ \citenamefont
  {Oreg}}]{haim-14}%
  \BibitemOpen
  \bibfield  {author} {\bibinfo {author} {\bibfnamefont {A.}~\bibnamefont
  {Haim}}, \bibinfo {author} {\bibfnamefont {A.}~\bibnamefont {Keselman}},
  \bibinfo {author} {\bibfnamefont {E.}~\bibnamefont {Berg}}, \ and\ \bibinfo
  {author} {\bibfnamefont {Y.}~\bibnamefont {Oreg}},\ }\href@noop {} {\bibfield
   {journal} {\bibinfo  {journal} {Phys.\ Rev.\ B}\ }\textbf {\bibinfo {volume}
  {89}},\ \bibinfo {pages} {220504(R)} (\bibinfo {year} {2014})}\BibitemShut
  {NoStop}%
\bibitem [{\citenamefont {Gergs}\ \emph {et~al.}(2016)\citenamefont {Gergs},
  \citenamefont {Fritz},\ and\ \citenamefont {Schuricht}}]{gergs-16}%
  \BibitemOpen
  \bibfield  {author} {\bibinfo {author} {\bibfnamefont {N.~M.}\ \bibnamefont
  {Gergs}}, \bibinfo {author} {\bibfnamefont {L.}~\bibnamefont {Fritz}}, \ and\
  \bibinfo {author} {\bibfnamefont {D.}~\bibnamefont {Schuricht}},\ }\href
  {\doibase 10.1103/PhysRevB.93.075129} {\bibfield  {journal} {\bibinfo
  {journal} {Phys.\ Rev.\ B}\ }\textbf {\bibinfo {volume} {93}},\ \bibinfo
  {pages} {075129} (\bibinfo {year} {2016})}\BibitemShut {NoStop}%
\bibitem [{\citenamefont {Wong}\ and\ \citenamefont {Law}(2012)}]{wong-12}%
  \BibitemOpen
  \bibfield  {author} {\bibinfo {author} {\bibfnamefont {C.}~\bibnamefont
  {Wong}}\ and\ \bibinfo {author} {\bibfnamefont {K.}~\bibnamefont {Law}},\
  }\href {\doibase 10.1103/PhysRevB.86.184516} {\bibfield  {journal} {\bibinfo
  {journal} {Phys.\ Rev.\ B}\ }\textbf {\bibinfo {volume} {86}},\ \bibinfo
  {pages} {184516} (\bibinfo {year} {2012})}\BibitemShut {NoStop}%
\bibitem [{\citenamefont {Tanaka}\ and\ \citenamefont
  {Kashiwaya}(1995)}]{tanaka-95}%
  \BibitemOpen
  \bibfield  {author} {\bibinfo {author} {\bibfnamefont {Y.}~\bibnamefont
  {Tanaka}}\ and\ \bibinfo {author} {\bibfnamefont {S.}~\bibnamefont
  {Kashiwaya}},\ }\href {\doibase 10.1103/PhysRevLett.74.3451} {\bibfield
  {journal} {\bibinfo  {journal} {Phys. Rev. Lett.}\ }\textbf {\bibinfo
  {volume} {74}},\ \bibinfo {pages} {3451} (\bibinfo {year}
  {1995})}\BibitemShut {NoStop}%
\bibitem [{\citenamefont {Martin}\ and\ \citenamefont
  {Annett}(1998)}]{martin-98}%
  \BibitemOpen
  \bibfield  {author} {\bibinfo {author} {\bibfnamefont {A.~M.}\ \bibnamefont
  {Martin}}\ and\ \bibinfo {author} {\bibfnamefont {J.~F.}\ \bibnamefont
  {Annett}},\ }\href {\doibase 10.1103/PhysRevB.57.8709} {\bibfield  {journal}
  {\bibinfo  {journal} {Phys. Rev. B}\ }\textbf {\bibinfo {volume} {57}},\
  \bibinfo {pages} {8709} (\bibinfo {year} {1998})}\BibitemShut {NoStop}%
\bibitem [{\citenamefont {Belzig}\ \emph {et~al.}(1998)\citenamefont {Belzig},
  \citenamefont {Bruder},\ and\ \citenamefont {Sigrist}}]{belzig-98}%
  \BibitemOpen
  \bibfield  {author} {\bibinfo {author} {\bibfnamefont {W.}~\bibnamefont
  {Belzig}}, \bibinfo {author} {\bibfnamefont {C.}~\bibnamefont {Bruder}}, \
  and\ \bibinfo {author} {\bibfnamefont {M.}~\bibnamefont {Sigrist}},\ }\href
  {\doibase 10.1103/PhysRevLett.80.4285} {\bibfield  {journal} {\bibinfo
  {journal} {Phys. Rev. Lett.}\ }\textbf {\bibinfo {volume} {80}},\ \bibinfo
  {pages} {4285} (\bibinfo {year} {1998})}\BibitemShut {NoStop}%
\bibitem [{\citenamefont {Hogan-O'Neill}\ \emph {et~al.}(1999)\citenamefont
  {Hogan-O'Neill}, \citenamefont {Martin},\ and\ \citenamefont
  {Annett}}]{hogan-99}%
  \BibitemOpen
  \bibfield  {author} {\bibinfo {author} {\bibfnamefont {J.~J.}\ \bibnamefont
  {Hogan-O'Neill}}, \bibinfo {author} {\bibfnamefont {A.~M.}\ \bibnamefont
  {Martin}}, \ and\ \bibinfo {author} {\bibfnamefont {J.~F.}\ \bibnamefont
  {Annett}},\ }\href {\doibase 10.1103/PhysRevB.60.3568} {\bibfield  {journal}
  {\bibinfo  {journal} {Phys. Rev. B}\ }\textbf {\bibinfo {volume} {60}},\
  \bibinfo {pages} {3568} (\bibinfo {year} {1999})}\BibitemShut {NoStop}%
\bibitem [{\citenamefont {Gangadharaiah}\ \emph {et~al.}(2011)\citenamefont
  {Gangadharaiah}, \citenamefont {Braunecker}, \citenamefont {Simon},\ and\
  \citenamefont {Loss}}]{gangadharaiah-11}%
  \BibitemOpen
  \bibfield  {author} {\bibinfo {author} {\bibfnamefont {S.}~\bibnamefont
  {Gangadharaiah}}, \bibinfo {author} {\bibfnamefont {B.}~\bibnamefont
  {Braunecker}}, \bibinfo {author} {\bibfnamefont {P.}~\bibnamefont {Simon}}, \
  and\ \bibinfo {author} {\bibfnamefont {D.}~\bibnamefont {Loss}},\ }\href@noop
  {} {\bibfield  {journal} {\bibinfo  {journal} {Phys.\ Rev.\ Lett.}\ }\textbf
  {\bibinfo {volume} {107}},\ \bibinfo {pages} {036801} (\bibinfo {year}
  {2011})}\BibitemShut {NoStop}%
\bibitem [{\citenamefont {Lutchyn}\ and\ \citenamefont
  {Fisher}(2011)}]{lutchyn-11}%
  \BibitemOpen
  \bibfield  {author} {\bibinfo {author} {\bibfnamefont {R.}~\bibnamefont
  {Lutchyn}}\ and\ \bibinfo {author} {\bibfnamefont {M.}~\bibnamefont
  {Fisher}},\ }\href {\doibase 10.1103/PhysRevB.84.214528} {\bibfield
  {journal} {\bibinfo  {journal} {Phys.\ Rev.\ B}\ }\textbf {\bibinfo {volume}
  {84}},\ \bibinfo {pages} {214528} (\bibinfo {year} {2011})}\BibitemShut
  {NoStop}%
\bibitem [{\citenamefont {Klinovaja}\ and\ \citenamefont
  {Loss}(2014)}]{klinovaja-14}%
  \BibitemOpen
  \bibfield  {author} {\bibinfo {author} {\bibfnamefont {J.}~\bibnamefont
  {Klinovaja}}\ and\ \bibinfo {author} {\bibfnamefont {D.}~\bibnamefont
  {Loss}},\ }\href {\doibase 10.1103/PhysRevB.90.045118} {\bibfield  {journal}
  {\bibinfo  {journal} {Phys.\ Rev.\ B}\ }\textbf {\bibinfo {volume} {90}},\
  \bibinfo {pages} {045118} (\bibinfo {year} {2014})}\BibitemShut {NoStop}%
\bibitem [{\citenamefont {Zaitsev}(1975)}]{zaitsev-75}%
  \BibitemOpen
  \bibfield  {author} {\bibinfo {author} {\bibfnamefont {R.~O.}\ \bibnamefont
  {Zaitsev}},\ }\href@noop {} {\bibfield  {journal} {\bibinfo  {journal} {J.\
  Theor.\ Exp.\ Phys.}\ }\textbf {\bibinfo {volume} {68}},\ \bibinfo {pages}
  {207} (\bibinfo {year} {1975})}\BibitemShut {NoStop}%
\bibitem [{\citenamefont {Izyumov}\ \emph {et~al.}(1992)\citenamefont
  {Izyumov}, \citenamefont {Letfulov}, \citenamefont {Shipitsyn}, \citenamefont
  {Bartkowiak},\ and\ \citenamefont {Chao}}]{izyumov-92}%
  \BibitemOpen
  \bibfield  {author} {\bibinfo {author} {\bibfnamefont {Y.~A.}\ \bibnamefont
  {Izyumov}}, \bibinfo {author} {\bibfnamefont {B.~M.}\ \bibnamefont
  {Letfulov}}, \bibinfo {author} {\bibfnamefont {E.~V.}\ \bibnamefont
  {Shipitsyn}}, \bibinfo {author} {\bibfnamefont {M.}~\bibnamefont
  {Bartkowiak}}, \ and\ \bibinfo {author} {\bibfnamefont {K.~A.}\ \bibnamefont
  {Chao}},\ }\href {\doibase 10.1103/PhysRevB.46.15697} {\bibfield  {journal}
  {\bibinfo  {journal} {Phys. Rev. B}\ }\textbf {\bibinfo {volume} {46}},\
  \bibinfo {pages} {15697} (\bibinfo {year} {1992})}\BibitemShut {NoStop}%
\bibitem [{\citenamefont {Kikoin}\ and\ \citenamefont
  {Avishai}(2001)}]{kikoin-01}%
  \BibitemOpen
  \bibfield  {author} {\bibinfo {author} {\bibfnamefont {K.}~\bibnamefont
  {Kikoin}}\ and\ \bibinfo {author} {\bibfnamefont {Y.}~\bibnamefont
  {Avishai}},\ }\href {\doibase 10.1103/PhysRevLett.86.2090} {\bibfield
  {journal} {\bibinfo  {journal} {Phys. Rev. Lett.}\ }\textbf {\bibinfo
  {volume} {86}},\ \bibinfo {pages} {2090} (\bibinfo {year}
  {2001})}\BibitemShut {NoStop}%
\bibitem [{\citenamefont {Ovchinnikov}\ and\ \citenamefont
  {Valkov}(2004)}]{valkov-04}%
  \BibitemOpen
  \bibfield  {author} {\bibinfo {author} {\bibfnamefont {S.~G.}\ \bibnamefont
  {Ovchinnikov}}\ and\ \bibinfo {author} {\bibfnamefont {V.~V.}\ \bibnamefont
  {Valkov}},\ }\href@noop {} {\emph {\bibinfo {title} {Hubbard Operators in the
  Theory of Strongly Correlated Electrons}}}\ (\bibinfo  {publisher} {Imperial
  College Press, London},\ \bibinfo {year} {2004})\BibitemShut {NoStop}%
\bibitem [{\citenamefont {Izyumov}(1997)}]{izyumov-97}%
  \BibitemOpen
  \bibfield  {author} {\bibinfo {author} {\bibfnamefont {Y.~A.}\ \bibnamefont
  {Izyumov}},\ }\href {\doibase 10.1070/pu1997v040n05abeh000234} {\bibfield
  {journal} {\bibinfo  {journal} {Physics-Uspekhi}\ }\textbf {\bibinfo {volume}
  {40}},\ \bibinfo {pages} {445} (\bibinfo {year} {1997})}\BibitemShut
  {NoStop}%
\bibitem [{\citenamefont {Anderson}(1987)}]{anderson-87}%
  \BibitemOpen
  \bibfield  {author} {\bibinfo {author} {\bibfnamefont {P.~W.}\ \bibnamefont
  {Anderson}},\ }\href {\doibase 10.1126/science.235.4793.1196} {\bibfield
  {journal} {\bibinfo  {journal} {Science}\ }\textbf {\bibinfo {volume}
  {235}},\ \bibinfo {pages} {1196} (\bibinfo {year} {1987})}\BibitemShut
  {NoStop}%
\bibitem [{\citenamefont {Val'kov}\ \emph {et~al.}(2017)\citenamefont
  {Val'kov}, \citenamefont {Mitskan},\ and\ \citenamefont
  {Shustin}}]{valkov-17}%
  \BibitemOpen
  \bibfield  {author} {\bibinfo {author} {\bibfnamefont {V.~V.}\ \bibnamefont
  {Val'kov}}, \bibinfo {author} {\bibfnamefont {V.~A.}\ \bibnamefont
  {Mitskan}}, \ and\ \bibinfo {author} {\bibfnamefont {M.~S.}\ \bibnamefont
  {Shustin}},\ }\href {\doibase 10.7868/S0370274X17240080} {\bibfield
  {journal} {\bibinfo  {journal} {JETP Lett.}\ }\textbf {\bibinfo {volume}
  {106}},\ \bibinfo {pages} {798} (\bibinfo {year} {2017})}\BibitemShut
  {NoStop}%
\bibitem [{\citenamefont {Shubin}\ and\ \citenamefont
  {Vonsovsky}(1934)}]{shubin-34}%
  \BibitemOpen
  \bibfield  {author} {\bibinfo {author} {\bibfnamefont {S.~P.}\ \bibnamefont
  {Shubin}}\ and\ \bibinfo {author} {\bibfnamefont {S.~V.}\ \bibnamefont
  {Vonsovsky}},\ }\href@noop {} {\bibfield  {journal} {\bibinfo  {journal}
  {Proc.\ R.\ Soc.\ Lond.\ A}\ }\textbf {\bibinfo {volume} {145}},\ \bibinfo
  {pages} {159} (\bibinfo {year} {1934})}\BibitemShut {NoStop}%
\bibitem [{\citenamefont {Vonsovsky}\ and\ \citenamefont
  {Katsnelson}(1979)}]{vonsovsky-79}%
  \BibitemOpen
  \bibfield  {author} {\bibinfo {author} {\bibfnamefont {S.~V.}\ \bibnamefont
  {Vonsovsky}}\ and\ \bibinfo {author} {\bibfnamefont {M.~I.}\ \bibnamefont
  {Katsnelson}},\ }\href@noop {} {\bibfield  {journal} {\bibinfo  {journal} {J.
  Phys. C: Solid State Phys.}\ }\textbf {\bibinfo {volume} {12}},\ \bibinfo
  {pages} {2043} (\bibinfo {year} {1979})}\BibitemShut {NoStop}%
\bibitem [{\citenamefont {Qi}\ \emph {et~al.}(2009)\citenamefont {Qi},
  \citenamefont {Hughes}, \citenamefont {Raghu},\ and\ \citenamefont
  {Zhang}}]{qi-09}%
  \BibitemOpen
  \bibfield  {author} {\bibinfo {author} {\bibfnamefont {X.-L.}\ \bibnamefont
  {Qi}}, \bibinfo {author} {\bibfnamefont {T.~L.}\ \bibnamefont {Hughes}},
  \bibinfo {author} {\bibfnamefont {S.}~\bibnamefont {Raghu}}, \ and\ \bibinfo
  {author} {\bibfnamefont {S.-C.}\ \bibnamefont {Zhang}},\ }\href@noop {}
  {\bibfield  {journal} {\bibinfo  {journal} {Phys.\ Rev.\ Lett.}\ }\textbf
  {\bibinfo {volume} {102}},\ \bibinfo {pages} {187001} (\bibinfo {year}
  {2009})}\BibitemShut {NoStop}%
\bibitem [{\citenamefont {edited~by I.~Peschel}\ \emph
  {et~al.}(1999)\citenamefont {edited~by I.~Peschel}, \citenamefont {Wang},
  \citenamefont {Kaulke},\ and\ \citenamefont {Hallberg}}]{peschel-99}%
  \BibitemOpen
  \bibfield  {author} {\bibinfo {author} {\bibnamefont {edited~by I.~Peschel}},
  \bibinfo {author} {\bibfnamefont {X.}~\bibnamefont {Wang}}, \bibinfo {author}
  {\bibfnamefont {M.}~\bibnamefont {Kaulke}}, \ and\ \bibinfo {author}
  {\bibfnamefont {K.}~\bibnamefont {Hallberg}},\ }\href@noop {} {\emph
  {\bibinfo {title} {Density Matrix Renormalization—A New Numerical Method in
  Physics, Lecture Notes in Physics Vol. 528}}}\ (\bibinfo  {publisher}
  {Springer, Berlin, Heidelberg},\ \bibinfo {year} {1999})\BibitemShut
  {NoStop}%
\bibitem [{\citenamefont {Sedlmayr}\ and\ \citenamefont
  {Bena}(2015)}]{sedlmayr-15}%
  \BibitemOpen
  \bibfield  {author} {\bibinfo {author} {\bibfnamefont {N.}~\bibnamefont
  {Sedlmayr}}\ and\ \bibinfo {author} {\bibfnamefont {C.}~\bibnamefont
  {Bena}},\ }\href {\doibase 10.1103/PhysRevB.92.115115} {\bibfield  {journal}
  {\bibinfo  {journal} {Phys.\ Rev.\ B}\ }\textbf {\bibinfo {volume} {92}},\
  \bibinfo {pages} {115115} (\bibinfo {year} {2015})}\BibitemShut {NoStop}%
\bibitem [{\citenamefont {Sedlmayr}\ \emph {et~al.}(2016)\citenamefont
  {Sedlmayr}, \citenamefont {Aguiar-Hualde},\ and\ \citenamefont
  {Bena}}]{sedlmayr-16}%
  \BibitemOpen
  \bibfield  {author} {\bibinfo {author} {\bibfnamefont {N.}~\bibnamefont
  {Sedlmayr}}, \bibinfo {author} {\bibfnamefont {J.~M.}\ \bibnamefont
  {Aguiar-Hualde}}, \ and\ \bibinfo {author} {\bibfnamefont {C.}~\bibnamefont
  {Bena}},\ }\href {\doibase 10.1103/PhysRevB.93.155425} {\bibfield  {journal}
  {\bibinfo  {journal} {Phys.\ Rev.\ B}\ }\textbf {\bibinfo {volume} {93}},\
  \bibinfo {pages} {155425} (\bibinfo {year} {2016})}\BibitemShut {NoStop}%
\bibitem [{\citenamefont {Turner}\ \emph {et~al.}(2011)\citenamefont {Turner},
  \citenamefont {Pollmann},\ and\ \citenamefont {Berg}}]{turner-11}%
  \BibitemOpen
  \bibfield  {author} {\bibinfo {author} {\bibfnamefont {A.~M.}\ \bibnamefont
  {Turner}}, \bibinfo {author} {\bibfnamefont {F.}~\bibnamefont {Pollmann}}, \
  and\ \bibinfo {author} {\bibfnamefont {E.}~\bibnamefont {Berg}},\ }\href
  {\doibase 10.1103/PhysRevB.83.075102} {\bibfield  {journal} {\bibinfo
  {journal} {Phys.\ Rev.\ B}\ }\textbf {\bibinfo {volume} {83}},\ \bibinfo
  {pages} {075102} (\bibinfo {year} {2011})}\BibitemShut {NoStop}%
\bibitem [{\citenamefont {Kukharenko}(1975)}]{kukharenko-75}%
  \BibitemOpen
  \bibfield  {author} {\bibinfo {author} {\bibfnamefont {B.~G.}\ \bibnamefont
  {Kukharenko}},\ }\href@noop {} {\bibfield  {journal} {\bibinfo  {journal}
  {JETP}\ }\textbf {\bibinfo {volume} {42}},\ \bibinfo {pages} {321} (\bibinfo
  {year} {1975})}\BibitemShut {NoStop}%
\bibitem [{\citenamefont {Val'kov}\ and\ \citenamefont
  {Val'kova}(1991)}]{valkov-91}%
  \BibitemOpen
  \bibfield  {author} {\bibinfo {author} {\bibfnamefont {V.~V.}\ \bibnamefont
  {Val'kov}}\ and\ \bibinfo {author} {\bibfnamefont {T.~A.}\ \bibnamefont
  {Val'kova}},\ }\href@noop {} {\bibfield  {journal} {\bibinfo  {journal}
  {JETP}\ }\textbf {\bibinfo {volume} {72}},\ \bibinfo {pages} {1053} (\bibinfo
  {year} {1991})}\BibitemShut {NoStop}%
\bibitem [{\citenamefont {Val'kov}\ \emph {et~al.}(2019)\citenamefont
  {Val'kov}, \citenamefont {Mitskan},\ and\ \citenamefont
  {Shustin}}]{valkov-19a}%
  \BibitemOpen
  \bibfield  {author} {\bibinfo {author} {\bibfnamefont {V.~V.}\ \bibnamefont
  {Val'kov}}, \bibinfo {author} {\bibfnamefont {V.~A.}\ \bibnamefont
  {Mitskan}}, \ and\ \bibinfo {author} {\bibfnamefont {M.~S.}\ \bibnamefont
  {Shustin}},\ }\href@noop {} {\bibfield  {journal} {\bibinfo  {journal}
  {JETP}\ }\textbf {\bibinfo {volume} {129}},\ \bibinfo {pages} {426} (\bibinfo
  {year} {2019})}\BibitemShut {NoStop}%
\bibitem [{\citenamefont {Koshibae}\ \emph {et~al.}(1993)\citenamefont
  {Koshibae}, \citenamefont {Ohta},\ and\ \citenamefont
  {Maekawa}}]{koshibae-93}%
  \BibitemOpen
  \bibfield  {author} {\bibinfo {author} {\bibfnamefont {W.}~\bibnamefont
  {Koshibae}}, \bibinfo {author} {\bibfnamefont {Y.}~\bibnamefont {Ohta}}, \
  and\ \bibinfo {author} {\bibfnamefont {S.}~\bibnamefont {Maekawa}},\ }\href
  {\doibase 10.1103/PhysRevB.47.3391} {\bibfield  {journal} {\bibinfo
  {journal} {Phys.\ Rev.\ B}\ }\textbf {\bibinfo {volume} {47}},\ \bibinfo
  {pages} {3391} (\bibinfo {year} {1993})}\BibitemShut {NoStop}%
\bibitem [{\citenamefont {Wieckowski}\ and\ \citenamefont
  {Ptok}(2019)}]{wieckowski-19}%
  \BibitemOpen
  \bibfield  {author} {\bibinfo {author} {\bibfnamefont {A.}~\bibnamefont
  {Wieckowski}}\ and\ \bibinfo {author} {\bibfnamefont {A.}~\bibnamefont
  {Ptok}},\ }\href {\doibase 10.1103/PhysRevB.100.144510} {\bibfield  {journal}
  {\bibinfo  {journal} {Phys.\ Rev.\ B}\ }\textbf {\bibinfo {volume} {100}},\
  \bibinfo {pages} {144510} (\bibinfo {year} {2019})}\BibitemShut {NoStop}%
\bibitem [{\citenamefont {Val'kov}\ and\ \citenamefont
  {Zlotnikov}(2019)}]{valkov-19b}%
  \BibitemOpen
  \bibfield  {author} {\bibinfo {author} {\bibfnamefont {V.~V.}\ \bibnamefont
  {Val'kov}}\ and\ \bibinfo {author} {\bibfnamefont {A.~O.}\ \bibnamefont
  {Zlotnikov}},\ }\href {\doibase 10.1134/S0021364019110158} {\bibfield
  {journal} {\bibinfo  {journal} {JETP Letters}\ }\textbf {\bibinfo {volume}
  {109}},\ \bibinfo {pages} {736} (\bibinfo {year} {2019})}\BibitemShut
  {NoStop}%
\bibitem [{\citenamefont {Zhu}\ \emph {et~al.}(2003)\citenamefont {Zhu},
  \citenamefont {Garst}, \citenamefont {Rosch},\ and\ \citenamefont
  {Si}}]{zhu-03}%
  \BibitemOpen
  \bibfield  {author} {\bibinfo {author} {\bibfnamefont {L.}~\bibnamefont
  {Zhu}}, \bibinfo {author} {\bibfnamefont {M.}~\bibnamefont {Garst}}, \bibinfo
  {author} {\bibfnamefont {A.}~\bibnamefont {Rosch}}, \ and\ \bibinfo {author}
  {\bibfnamefont {Q.}~\bibnamefont {Si}},\ }\href@noop {} {\bibfield  {journal}
  {\bibinfo  {journal} {Phys.\ Rev.\ lett.}\ }\textbf {\bibinfo {volume}
  {91}},\ \bibinfo {pages} {066404} (\bibinfo {year} {2003})}\BibitemShut
  {NoStop}%
\bibitem [{\citenamefont {Garst}\ and\ \citenamefont {Rosch}(2005)}]{garst-05}%
  \BibitemOpen
  \bibfield  {author} {\bibinfo {author} {\bibfnamefont {M.}~\bibnamefont
  {Garst}}\ and\ \bibinfo {author} {\bibfnamefont {A.}~\bibnamefont {Rosch}},\
  }\href@noop {} {\bibfield  {journal} {\bibinfo  {journal} {Phys.\ Rev.\ B}\
  }\textbf {\bibinfo {volume} {72}},\ \bibinfo {pages} {205129} (\bibinfo
  {year} {2005})}\BibitemShut {NoStop}%
\end{thebibliography}%

\end{document}